\documentclass[aps,prb,twocolumn,floatfix,showpacs,nofootinbib]{revtex4-1}
\usepackage{graphicx}
\usepackage[caption=false]{subfig}
\usepackage{float}
\usepackage{longtable}
\usepackage{bm}
\usepackage{siunitx}
\usepackage{xfrac}
\usepackage{amssymb,mathtools,mathrsfs,bbold,dsfont}
\usepackage{amstext,braket,amsmath,tensor}  
\usepackage{array} 
\usepackage{siunitx}
\usepackage[usenames,dvipsnames,svgnames,table]{xcolor}
\usepackage[normalem]{ulem}
\usepackage{empheq}
\usepackage{xcolor}
\usepackage{multirow}
\usepackage{amsthm,etoolbox}
\usepackage{cleveref}

\newcommand{\bq}{\boldsymbol{q}}

\newcommand{\Vbb}{\mathbb{V}}

\newcommand{\fbb}{\mathbb{f}}

\newcommand{\Sds}{\mathds{S}}
\newcommand{\Ods}{\mathds{O}}
\newcommand{\Acal}{\mathcal{A}}

\newcommand{\Dcal}{D}
\newcommand{\Ecal}{E}

\newcommand{\Ocal}{\mathcal{O}}
\newcommand{\Pcal}{\mathcal{P}}

\newcommand{\Fcal}{\mathcal{F}}
\newcommand{\Rcal}{\mathcal{R}}

\newcommand{\Fscr}{\mathscr{F}}

\newcommand{\Ical}{\mathcal{I}}
\newcommand{\Tcal}{\mathcal{T}}
\newcommand{\Gcal}{\mathcal{G}}

\newcommand{\bPi}{\boldsymbol{\Pi}}
\newcommand{\bM}{\boldsymbol{M}}

\newcommand{\bG}{\boldsymbol{G}}

\newcommand{\bT}{\boldsymbol{T}}

\newcommand{\bR}{\boldsymbol{R}}

\newcommand{\bl}{\boldsymbol{l}}

\newcommand{\bu}{\boldsymbol{u}}
\newcommand{\bL}{\boldsymbol{L}}
\newcommand{\bD}{\boldsymbol{D}}
\newcommand{\bDcal}{\boldsymbol{\Dcal}}
\newcommand{\bEcal}{\boldsymbol{\Ecal}}

\newcommand{\bRcal}{\boldsymbol{\Rcal}}
\newcommand{\bk}{\boldsymbol{k}}
\newcommand{\bchi}{\boldsymbol{\chi}}

\newcommand{\bUpsilon}{\boldsymbol{\Upsilon}}
\newcommand{\bphi}{\boldsymbol{\phi}}

\newcommand{\bLa}{\boldsymbol{\Lambda}}
\newcommand{\bLambda}{\boldsymbol{\Lambda}}
\newcommand{\bTheta}{\boldsymbol{\Theta}}

\newcommand{\bvarPhi}{\boldsymbol{\varPhi}}
\newcommand{\bvarPsi}{\boldsymbol{\varPsi}}

\newcommand{\bXi}{\boldsymbol{\Xi}}

\newcommand{\dR}{\delta\!R}

\newcommand{\eq}{\scriptscriptstyle{\text{eq}}}

\newcommand{\fcm}{\phi} 
\newcommand{\fcmi}{\overset{\scriptscriptstyle{-1}}{\fcm}}

\newcommand{\fcmtrial}{\Phi} 

\newcommand{\fct}{\overset{\scriptscriptstyle{(3)}}{\phi}}
\newcommand{\fcq}{\overset{\scriptscriptstyle{(4)}}{\phi}}

\newcommand{\avg}[1]{\left\langle{#1}\right\rangle}
\newcommand{\Tr}[1]{\textup{tr}\left[{#1}\right]}

\newcommand{\angs}{\textup{\AA}}

\newcommand{\bPhi}{\boldsymbol{\Phi}}

\newcommand{\Linv}{\overset{\scriptscriptstyle{-1}}{L}{}}
\newcommand{\by}{\boldsymbol{y}}

\newcommand{\bare}{\bar{e}}

\newcommand{\Nat}{N_{\text{a}}}
\newcommand{\Ncell}{N_{\text{c}}}

\newcommand{\hs}{\scriptscriptstyle{\text{hs}}}

\newcommand{\f}{\text{f}}
\renewcommand{\bf}{\text{\textbf{f}}}

\newcommand{\Bavg}[1]{\Bigl\langle #1 \Bigr\rangle}

\newcommand{\lat}{\scriptscriptstyle{\text{lat}}}

\newcommand{\SCHA}{\scriptscriptstyle{(\text{SCHA})}}

\newcommand{\ssn}{\scriptscriptstyle{(n)}}

\newcommand{\ssF}{\scriptscriptstyle{(F)}}
\newcommand{\sst}{\scriptscriptstyle{(3)}}
\newcommand{\ssf}{\scriptscriptstyle{(4)}}

\newcommand{\bsst}{\boldsymbol{\sst}}
\newcommand{\bssf}{\boldsymbol{\ssf}}
\newcommand{\ssi}{\scriptscriptstyle{-1}}

\DeclareMathAlphabet{\mathbbmsl}{U}{bbm}{m}{sl}

\newcommand{\varPsiinv}{\overset{\scriptscriptstyle{-1}}{\varPsi}}

\newcommand{\trialrho}{\tilde{\rho}}
\newcommand{\rhotrial}{\tilde{\rho}}
\renewcommand{\trialrho}{\scalebox{1.1}{$\tilde{\rho}$}}
\renewcommand{\rhotrial}{\scalebox{1.1}{$\tilde{\rho}$}}
\newcommand{\rhotrue}{\scalebox{1.1}{$\rho$}}

\newcommand{\ssz}{\scriptscriptstyle{(0)}}
\DeclareSymbolFont{bbold}{U}{bbold}{m}{n}
\DeclareSymbolFontAlphabet{\mathbbold}{bbold}

\newcommand{\ds}[1]{\displaystyle{#1}}

\newcommand{\ssS}{\scriptscriptstyle{(S)}}
\newcommand{\ssL}{\scriptscriptstyle{(L)}}
\newcommand{\ssT}{\scriptscriptstyle{(T)}}
\newcommand{\ssB}{\scriptscriptstyle{(B)}}

\newcommand{\sso}{\scriptscriptstyle{(1)}}

\newcommand{\Ftrial}{\widetilde{F}}
\newcommand{\Vtrial}{\widetilde{V}}
\newcommand{\tagmedia}{\displaystyle\rhotrial_{\bRcal,\bvarPhi}}
\newcommand{\tagmediaSCHA}{\displaystyle\rhotrial_{\bRcal,\bPhi(\bRcal)}}
\newcommand{\tagmediaSCHAnoR}{\displaystyle\rhotrial_{\bRcal,\bPhi}}
\newcommand{\ssIcal}{\scriptscriptstyle{(\Ical)}}
\renewcommand{\SCHA}{\scriptscriptstyle{(\text{SH})}}
\renewcommand{\SCHA}{\scriptscriptstyle{(\text{S})}}
\newcommand{\Nsym}{N_{\scriptscriptstyle{\text{sym}}}}
\newcommand{\sym}{\scriptscriptstyle{\text{sym}}}
\newcommand{\perm}{\scriptscriptstyle{\text{perm}}}
\newcommand{\ssin}{\scriptscriptstyle{\text{in}}}
\newcommand{\rref}{\scriptscriptstyle{\text{ref}}}
\newcommand{\odd}{\scriptscriptstyle{\text{odd}}}
\newcommand{\even}{\scriptscriptstyle{\text{even}}}

\newcommand{\mathp}{\,.}

\newcommand{\mathv}{\,,}

\begin{document}

\title{Second order structural phase transitions, free energy curvature, and temperature-dependent
anharmonic phonons in the self-consistent harmonic approximation: theory and stochastic implementation}
\author{Raffaello Bianco$^{1,2}$}
\email{raffaello.bianco@roma1.infn.it}
\author{Ion Errea$^{3,4}$}
\author{Lorenzo Paulatto$^{1}$}
\author{Matteo Calandra$^{1}$}
\author{Francesco Mauri$^{2,5}$}

\affiliation{$^1$Institut de min\'eralogie, de physique des mat\'eriaux et de 
cosmochimie (IMPMC), Universit\'e Pierre et Marie Curie (Paris VI), CNRS UMR 
7590, IRD UMR 206, Case 115, 4 place Jussieu, 75252 Paris Cedex 05, France}
\affiliation{$^{2}$ Dipartimento di Fisica, Universit\`a di Roma La Sapienza, 
Piazzale Aldo Moro 5, I-00185 Roma, Italy}
\affiliation{$^3$Fisika Aplikatua 1 Saila, Bilboko Ingeniaritza Eskola, University of the Basque Country (UPV/EHU),
Rafael Moreno ``Pitxitxi'' Pasealekua 3, 48013 Bilbao, Basque Country, Spain}
\affiliation{$^4$Donostia International Physics Center (DIPC),
            Manuel de Lardizabal pasealekua 4, 20018 Donostia-San Sebasti\'an,
	    Basque Country, Spain}
\affiliation{$^{5}$ Graphene Labs, Fondazione Istituto Italiano di Tecnologia, Via Morego, I-16163 Genova, Italy}

\begin{abstract}
The self-consistent harmonic approximation is an effective harmonic theory to calculate the free energy of systems with strongly anharmonic atomic vibrations, and its stochastic implementation has proved to be an efficient method to study, from first-principles, the anharmonic properties of solids. The free energy as a function of average atomic positions (centroids) can be used to study quantum or thermal lattice instability. In particular the centroids are order parameters in second-order structural phase transitions such as, e.g., charge-density-waves or ferroelectric instabilities.  According to Landau's theory, the knowledge of the second derivative of the free energy (i.e. the curvature) with respect to the centroids in a high-symmetry configuration allows the identification of the phase-transition and of the instability modes.  In this work we derive the exact analytic formula for the second derivative of the free energy in the self-consistent harmonic approximation for a generic atomic configuration.
The analytic derivative is expressed in terms of the atomic displacements and forces in a form that can be evaluated by a stochastic technique using importance sampling.  Our approach is particularly suitable for applications based on first-principles density-functional-theory calculations, where the forces on atoms can be obtained with a negligible computational effort compared to total energy determination.
Finally we propose a dynamical extension of the theory to calculate spectral properties of strongly anharmonic phonons, as probed by inelastic scattering processes. We illustrate our method with a numerical application on a toy model that mimics the ferroelectric transition in rock-salt crystals such as SnTe or GeTe.
\end{abstract}

\maketitle

\newtoggle{draft}
\togglefalse{draft}

\newtoggle{tabella_simboli}
\toggletrue{tabella_simboli}

\section{Introduction}

Describing accurately atomic vibrations 
is crucial in many branches of physics and chemistry because 
thermodynamic, transport, and superconducting properties of materials and molecules
as well as the spectra obtained in many spectroscopic techniques depend on  
how atoms vibrate\cite{born-huang}.  
The standard harmonic approximation provides the simplest description
of vibrations, which are also present at 0 K due to the quantum zero-point motion. 
The harmonic approximation is based on the
expansion of the Born-Oppenheimer (BO) energy surface to the second 
order around the ionic equilibrium positions{. It predicts well-defined
non-interacting quasi-particles (phonons) with infinite lifetime
and a temperature-independent energy spectrum.} 
Within this approximation many {physical} effects cannot be described.
For example, finite values of {the} thermal conductivity
and temperature dependent effects, like {the thermal expansion in solids,
cannot be accounted for at the harmonic level}. Therefore, it is
of paramount importance to describe accurately the vibrations of atoms
beyond the harmonic approximation.

Anharmonic effects, i.e.  effects due  to {higher} orders in the energy surface expansion,  
introduce interaction between phonons, thus finite scattering rates and finite 
{lifetimes}. 
Anharmonicity can be treated {at} different levels 
{of theory}. The basic approach is to consider higher order terms
in the potential expansion as a small perturbation of the harmonic potential\cite{PhysRev.128.2589}.
However, the perturbative approach can be used under quite restrictive conditions{:
the displacements of the atoms must be within the range in which the
harmonic approximation is valid so that higher order terms are considerably smaller
than the harmonic potential. Unfortunately, there are
several cases in which a non-perturbative regime is reached. For example
when light atoms such as hydrogen are present~\cite{PhysRevLett.111.177002,PhysRevB.89.064302,PhysRevLett.114.157004,Errea81,
0953-8984-28-49-494001,PhysRevB.82.104504},
or when the system is close to a dynamical instability (a phase transition)
as in ferroelectrics or materials undergoing a charge-density wave (CDW) 
instability~\cite{PhysRevLett.17.753,delaire614,0022-3719-13-19-018,
PhysRevLett.107.107403,PhysRevB.92.140303,PhysRevLett.86.3799,PhysRevB.86.155125,
doi:10.1080/00150199808009159,PhysRevB.80.241108,PhysRevLett.106.196406,PhysRev.140.A863, PhysRevB.63.144109,PhysRevB.14.4321,PhysRevB.92.094107}. 
In these cases, a non-perturbative approach is required in order to account for
anharmonic effects~\cite{Errea237}.} 

Anharmonic effects at {a non-perturbative level are commonly treated within molecular 
dynamics (MD) simulations or methods based on 
them~\cite{PhysRevLett.103.125902,PhysRevLett.112.058501,PhysRevB.42.11276,PhysRevB.87.174110,
PhysRevLett.110.105503,PhysRevLett.112.058501,PhysRevB.84.180301,PhysRevB.87.104111,PhysRevB.88.144301}.
However, these approaches are computationally expensive as long simulation times
are needed to obtain converged renormalized phonon energies and have an intrinsic limitation
because they are based on Newtonian dynamics, 
which limits their application to temperatures above the Debye temperature}. 
This problem can be overcome by path-integral molecular dynamics~\cite{RevModPhys.67.279},
but the {even greater computational cost that the method needs to
incorporate the quantum character of atomic vibrations} makes it challenging. To surmount these difficulties, 
{several methods~\cite{PhysRevLett.106.165501,PhysRevB.92.054301,Needs,PhysRevB.89.064302,
PhysRevLett.111.177002,:/content/aip/journal/jcp/138/4/10.1063/1.4788977,:/content/aip/journal/jcp/137/14/10.1063/1.4754819,
PhysRevB.92.201205}
have been developed, 
mainly inspired by the self-consistent harmonic approximation (SCHA) 
devised by Hooton~\cite{doi:10.1080/14786440408520575}}. 
The main idea of the SCHA is to use a variational principle, the Gibbs-Bogoliubov (GB) principle, 
in order to approximate the free energy of the true ionic Hamiltonian 
with the free energy {calculated with a trial harmonic density matrix for the same system}. 
In particular, in the stochastic self-consistent harmonic approximation (SSCHA)~\cite{PhysRevB.89.064302,PhysRevLett.111.177002}, 
the free energy is explicitly minimized by using a conjugate-gradient algorithm 
with respect to the independent coefficients of 
the trial harmonic potential{. In the SSCHA} 
the free energy and its gradient are evaluated through averages computed 
with stochastic sampling of the configuration space and the importance sampling technique~\cite{PhysRevB.89.064302,PhysRevLett.111.177002}.
In that way the (approximated but {non-perturbative}) 
anharmonic free energy of the system is directly accessible. 
The stochastic approach is particularly suited to be used in 
conjunction with \emph{ab initio} calculations, 
and it has been employed to study thermal anharmonic effects in several 
compounds {such as hydrides and transition metal 
dichalcogenides}~\cite{PhysRevLett.111.177002,PhysRevLett.114.157004,PhysRevB.89.064302,Errea81,
PhysRevB.93.174308,PhysRevB.92.140303,0953-8984-28-49-494001}.

{Within the SCHA the free energy as a
function of the average atomic positions, i.e. the \emph{centroids} positions, 
can be estimated 
for any temperature. These can be used to study 
structural second order phase transitions like, for example, in some ferroelectric 
and CDW phase transitions~\cite{PhysRevLett.17.753,delaire614,0022-3719-13-19-018,
PhysRevLett.107.107403,PhysRevB.92.140303,PhysRevLett.86.3799,PhysRevB.86.155125,
doi:10.1080/00150199808009159,PhysRevB.80.241108,PhysRevLett.106.196406,
PhysRev.140.A863, PhysRevB.63.144109,PhysRevB.14.4321,PhysRevB.92.094107}}.
In general, at any temperature the system is in equilibrium in the configuration which 
{minimizes} the free energy.
{According to Landau's theory~\cite{landau-vol-5}}, 
in a second order phase transition the high temperature free energy minimum is in a high-symmetry phase. 
As the temperature decreases, the 
minimum 
becomes less and less pronounced until {it becomes a saddle point
at {the} transition temperature $T_c$,} and, 
on lowering the temperature further,
the equilibrium position moves continuously towards lower symmetry configurations, where the free 
energy is {smaller} (Cfr.~Fig.~\ref{fig:phase_transition}). 
In this scheme the observable to be studied {as a function of temperature} 
is the second derivative of the free energy with respect to the centroids positions, 
i.e. the {free energy curvature, in the high-symmetry phase.}  
Indeed the Hessian in the high-symmetry phase is positive-definite at high-temperature, but lowering the temperature it
develops {first a null ($T=T_c$) and then negative ($T<T_c$) eigendirection,} 
which indicates the instability distortion that lowers the free energy.

Using the SSCHA {code~\cite{PhysRevLett.111.177002,PhysRevB.89.064302}} it is possible 
to compute the free energy for several centroids positions and, therefore,
to calculate numerically, by finite difference, the curvature in a 
point. This has been done, for example, to study the {quantum} H-bond 
symmetrization in the record superconductor H${}_3$S~\cite{Errea81}.
However, even if legitimate, this `brute force' approach to compute the free energy curvature
{is computationally demanding.
In fact, a careful calculation of the curvature by finite differences requires
small statistical noise, implying a great deal of calls to the total-energy-force
engine used. Moreover, it also
requires SSCHA calculations in the low-symmetry distorted phase, 
which are always more statistically demanding because the number of
free parameters in the trial harmonic Hamiltonian is larger due to the
reduced symmetry.} 


Motivated by these considerations, in this paper we {derive the general exact
analytic expression of the SCHA free energy curvature for a generic atomic configuration.
Our approach is similar to the one proposed by G\"otze and Michel in the
context of elastic constants of anharmonic crystals~\cite{GotzeZfurphys}. 
We also present an expression of the SCHA free energy curvature
that only depends on atomic displacements and forces. The latter is suited
for a stochastic implementation in conjunction with any total-energy-force
engine.  The method we are presenting here allows, thus,} 
to compute the curvature of the SCHA free energy {for 
a given structure straightforwardly
once the GB functional has been minimized within the SSCHA.
Since the method is much more efficient and precise than 
any finite-difference approach~\cite{Errea81}, it is 
especially suited to be used in 
conjunction with first-principles calculations.}

Besides the practical achievements, 
the curvature formula {described here} has also interesting conceptual consequences.
{Since the only physical observable given by the SCHA is the (approximated) free energy,
the effective SCHA quadratic matrix that minimizes the GB functional 
must be understood just as an auxiliary quantity, even if its eigenvalues have been often used
to calculate renormalized anharmonic phonon spectra~\cite{PhysRevLett.111.177002,PhysRevLett.114.157004,PhysRevB.89.064302,Errea81,
PhysRevB.93.174308,PhysRevB.92.140303,0953-8984-28-49-494001,PhysRevLett.106.165501}}. 
A significant parallelism can be traced with the Hartree-Fock approximation.
In that case, the Rayleigh-Ritz functional of the total energy is 
{minimized} with trial Slater determinants describing {non-interacting} fermions.
The energy obtained is an approximation of the true energy of the system{, but, on the contrary,
the corresponding trial non-interacting many-body wave function and related single particle spectrum 
do not have in general a physical meaning 
(except that in some aspects, e.g. see Koopmans' theorem~\cite{1934Phy.....1..104K}). An analogous 
situation occurs with density-function theory (DFT)
and the corresponding non-interacting electrons and energy bands~\cite{PhysRevLett.49.1691}}. In the same way, 
the SCHA matrix (divided by the square root of the masses, in order to have the correct dimensions) cannot be considered
a {generalized dynamical matrix and, therefore, its temperature-dependent eigenstates do not 
represent phonons renormalized by anharmonicity. 
On the contrary, the free energy curvature (in the equilibrium position) divided by the square root of the masses 
defines a proper anharmonic temperature-dependent generalization of the harmonic dynamical matrix, whose eigenstates represent 
temperature-dependent anharmonic phonons.
Indeed, at variance with the SCHA matrix, which is positive-definite by construction, 
the dynamical matrix based on 
the free energy curvature can have negative eigenvalues, and
a softening in its spectrum corresponds to a system instability.}
The free-energy based dynamical matrix is a particularly important tool especially when 
we consider crystalline systems. Indeed in that case, exploiting the lattice periodicity and the
Fourier interpolation technique, it allows to find structural 
instabilities with any modulations in real space by performing calculations
only on a coarse grid of the Brillouin zone.

The theory based on the free energy curvature with respect to the centroids position is `static' 
in the sense that there are no dynamical variables evolving
with time. However, here we also propose a minimal `dynamic' extension of the theory that resembles the 
work by Goldman \emph{et al.}~\cite{PhysRevLett.24.1424}, which allows to study, in a full
non-perturbative way, the spectral properties of anharmonic phonons, and thus allows to have finite 
scattering times and linewidths. This makes the theory able to interpret 
the results of inelastic scattering processes between anharmonic phonons and external incident particles (typically neutrons),
as well as allowing the calculation of the thermal conductivity in strongly 
anharmonic solids where the harmonic approximation breaks down.
Despite the proposed dynamic extension being based on an \emph{ansatz}, 
it is reasonable because it yields good results in two limits: 
at the lowest perturbative level it reduces to the standard perturbation theory result and 
in the static limit it predicts the same instabilities found with the free energy curvature.

The paper is structured as follows. In  section~\ref{Sec:Self-consistent_harmonic_approximation}, 
we present the fundamentals of the SCHA method, we define the SCHA free energy as a function of 
the centroids position, and  we fix the notation used.
In section~\ref{sec:Second_order_phase_transition_and_curvature_of_the_free_energy} 
we show how to analyze structural second order phase transitions through 
the Hessian of the free energy with respect to the centroids position (i.e. the curvature), and
in section~\ref{Derivatives_of_F} we give the explicit expression of the free energy curvature. 
In section~\ref{sec:phonons_in_the_SCHA}, on the basis of the results obtained, 
we describe the temperature-dependent free-energy-based
generalization of the harmonic dynamical matrix. In section~\ref{sec:Diagrammatic_representation},
we express the theory developed so far in a diagrammatic way. 
In section~\ref{sec:Stochastic_implementation}, we show how to implement the curvature formula in a 
stochastic way and, in particular, how to take into account symmetries in order to 
speed up the statistical convergence.
In section~\ref{sec:Perturbative_limit}, we find the lowest order perturbative limit of the results obtained.
%
In section~\ref{sec:Anstaz_for_a_dynamic_theory}, inspired by the results obtained, we propose the \emph{ansatz} 
to obtain a dynamical extension of the theory. Finally, in section~\ref{sec:Numerical_test},  
we perform numerical tests on a toy model based on the ferroelectric transition in 
SnTe, with the double objective of demonstrating the correctness of our findings
and showing the power of the method. 
In section~\ref{sec:conclusions}, we summarize our results and draw some final conclusions.
The paper is completed with several appendices including the proofs of all the
equations given in the manuscript and the details of the toy model.

\allowdisplaybreaks
%
\section{Self-consistent harmonic approximation}
\label{Sec:Self-consistent_harmonic_approximation}
We consider the quantum atomic free energy of crystal lattices and molecules.
For notation clarity, we derive the main results by using a real space formalism 
for both cases. This means that in the case of periodic crystals, we are actually studying a supercell
with periodic boundary conditions. Later, we will explicitly consider 
a crystalline case for the numerical example 
and we will take advantage of translational lattice symmetry.
The dynamic of atomic degrees of freedom is determined by the 
quantum Hamiltonian
\begin{equation}
H=\sum_{s=1}^{\Nat}\sum_{\alpha=1}^3\frac{{p^2_{s,\alpha}}}{2M_s}+V(\bR)\mathv
\label{eq:full_ham}
\end{equation}
where $\Nat$ is the total number of atoms,
$s$ and $\alpha$ are the atom and Cartesian component indices, respectively, $M_s$ is the mass of the $s$-th atom,
$p_{s,\alpha}$ and $R^{s,\alpha}$ are the momentum and position operators, respectively, and $V(\bR)$ is the
Born-Oppenheimer potential,
where the bold letter $\bR$ indicates $R^{s,\alpha}$ in component-free notation.
In what follows, we will use bold letters 
in component-free notation also for other observables and higher order tensors with respect to the
$(s,\alpha)$ index.
Moreover, in order to simplify the notation, we will
use a single composite index $a=(s,\alpha)$ to indicate both
atom and Cartesian indices. Notice that double index can be used also for the masses by
defining $M_{s,\alpha}=M_s$.

Fixed the temperature $T$, the free energy $F$ of the ionic Hamiltonian is given by 
the sum of the total energy and the entropic contribution
\begin{equation}
F = \mathrm{tr} \bigl[ \rhotrue H\bigr] + \frac{1}{\beta} 
      \mathrm{tr} \bigl[ \rhotrue \ln \rhotrue\bigr] \mathv
\label{true-free-energy}
\end{equation}
where  $\beta = (k_BT)^{-1}$ , `tr' is the trace operation on the $\Nat$ 
atom Hilbert space, and 
\begin{equation}
\rhotrue = e^{- \beta H} \big/  \mathrm{tr} [ e^{- \beta H}]
\end{equation}
is the equilibrium density matrix.
In systems comprising many interacting particles, calculating $F$
can represent a complicated task. Nevertheless, a quantum variational principle for the free energy 
can be established.
By replacing the true density matrix $\rhotrue$ in Eq.~\eqref{true-free-energy} with a generic density matrix 
$\trialrho$, we can define the functional 
\begin{equation}
\mathcal{F}\bigl[\trialrho\bigr] = \mathrm{tr} \bigl[ \trialrho H\bigr] + 
      \frac{1}{\beta} 
      \mathrm{tr} \bigl[ \trialrho\ln \trialrho\bigr]\mathv
\label{var-free-energy}
\end{equation}
and the Gibbs-Bogoliubov (GB) variational principle~\cite{0022-3689-1-5-305} states that
\begin{equation}
F\le \mathcal{F}\bigl[\trialrho\bigr]\mathp
\label{gibbs-bogoliubov}
\end{equation}
Obviously the equality holds when $\trialrho=\rhotrue$.

In particular, the SCHA is obtained by restricting {the} trial density {matrix} 
to equilibrium density matrices  
\begin{equation}
\rhotrial _{\bRcal,\bvarPhi}= e^{- \beta \widetilde{H}_{\bRcal,\bvarPhi}} \big/  \mathrm{tr} [ e^{- \beta \widetilde{H}_{\bRcal,\bvarPhi}}]
\label{eq:rhotrial}
\end{equation}
for the same temperature of a general trial harmonic Hamiltonian $\widetilde{H}_{\bRcal,\bvarPhi}$ for 
the same particles~\cite{doi:10.1080/14786440408520575}.
The trial harmonic Hamiltonian is parametrized in terms of
the vector $\bRcal$ of dimension $3\Nat$  and the square positive-definite matrix $\bvarPhi$ of order $3\Nat$ as
\begin{subequations}\begin{align}
&\widetilde{H}_{\bRcal,\bvarPhi}=\sum_a\frac{{p^2_{a}}}{2M_a}+\widetilde{V}_{\bRcal,\bvarPhi}\label{eq:Htrial}\\
&\widetilde{V}_{\bRcal,\bvarPhi}=\frac{1}{2}\sum_{ab}\varPhi_{ab}  (R-\Rcal)^{a} (R-\Rcal)^{b}\label{eq:Vtrial}\mathp
\end{align}\end{subequations}
In what follows we consider only  trial harmonic potentials $\widetilde{V}_{\bRcal,\bvarPhi}$  
that respect the symmetries of the system.

With $\Bavg{\Ods}_{{\displaystyle\rhotrial_{\bRcal,\bvarPhi}}}$
we indicate the average of an observable $\Ods$ with respect to the density matrix $\rhotrial_{\bRcal,\bvarPhi}$:
\begin{equation}
\Bavg{\Ods}_{{\displaystyle\rhotrial_{\bRcal,\bvarPhi}}}=\Tr{\Ods\,\rhotrial_{\bRcal,\bvarPhi}}\mathp
\label{eq:def_avg}
\end{equation}
In what follows it will be relevant to consider observables $\Ods(\bR)$ that are function of the position only.
In that case, the average can be written as
\begin{equation}
\Bavg{\Ods}_{{\displaystyle\rhotrial_{\bRcal,\bvarPhi}}}=\int \Ods(\bR)\,\rhotrial_{\bRcal,\bvarPhi}(\bR)\,d\bR\mathv
\label{eq:avg_pos_operator}
\end{equation}
where $\rhotrial_{\bRcal,\bvarPhi}(\bR)$ is the diagonal part of the density matrix $\rhotrial_{\bRcal,\bvarPhi}$
in coordinate representation~\cite{PhysRevLett.111.177002,PhysRevB.89.064302}:
\allowdisplaybreaks[0]
\begin{align}
&\rhotrial_{\bRcal,\bvarPhi}(\bR)=\sqrt{\text{det}\,(\boldsymbol{\Upsilon}/2\pi)}\nonumber\\
&\qquad\qquad\times\text{exp}\biggl[-\frac{1}{2}\sum_{ab}\Upsilon_{ab}(R-\Rcal)^a(R-\Rcal)^b\biggr]\mathp
\label{eq:dens_prob}
\end{align}
\allowdisplaybreaks
Here `det' indicates the determinant and $\bUpsilon$ is the symmetric matrix associated to $\bvarPhi$ by
\begin{equation}
\Upsilon_{ab}=\sqrt{M_aM_b}\,\sum_{\mu} \frac{2\omega_{\mu}}{(1+2n_\mu)\hbar}\,e^a_{\mu}e^b_{\mu}\mathv
\label{eq:def_Upsilon}
\end{equation}
where $\omega^2_{\mu}$ and $e^a_{\mu}$ are eigenvalues and corresponding eigenvectors of 
$\varPhi_{ab}/\sqrt{M_aM_b}$, and $n_{\mu}=1/(e^{\beta\hbar\omega_{\mu}}-1)$
is the bosonic average occupation number associated to $\omega_{\mu}$.
Note that to have a normalizable distribution $\rhotrial_{\bRcal,\bvarPhi}(\bR)$,
$\bUpsilon$ and thus $\bvarPhi$ must be positive-definite matrices, as specified above.

From Eqs.~\eqref{eq:avg_pos_operator}--\eqref{eq:dens_prob} we see that the average positions of the atoms
for the trial density matrix $\rhotrial _{\bRcal,\bvarPhi}$, namely the `centroids',
coincide with $\bRcal$: 
\begin{equation}
\Bavg{\bR}_{{\displaystyle\rhotrial_{\bRcal,\bvarPhi}}}=\bRcal\mathp
\end{equation}

According to the GB variational principle, the best approximation of the free energy within the SCHA is $F^{\SCHA}$, given by
\begin{equation}
F^{\SCHA}=\min_{\bRcal,\bvarPhi}\Fcal\bigl[\trialrho_{\bRcal,\bvarPhi}\bigr]
=\min_{\bRcal}\Bigl(\min_{\bvarPhi} \Fcal\bigl[{\trialrho}_{\bRcal,\bvarPhi}\bigr]\Bigr)\mathp
\label{eq:minFcal}
\end{equation}
With $F^{\SCHA}(\bRcal)$ we indicate the SCHA free energy for the centroids position $\bRcal$,
\begin{equation}
F^{\SCHA}(\bRcal)=\min_{\bvarPhi} \Fcal\bigl[\trialrho_{\bRcal,\bvarPhi}\bigr]\mathv
\label{eq:F(R)_def_1}
\end{equation}
and with $\bRcal_{\eq}$ we indicate the configuration that minimizes $F^{\SCHA}(\bRcal)$:
\begin{equation}
F^{\SCHA}=\min_{\bRcal}F^{\SCHA}(\bRcal)=F^{\SCHA}(\bRcal_{\eq})\mathp
\label{eq:defin_Req}
\end{equation}
Therefore, $\bRcal_{\eq}$ is the SCHA equilibrium configuration of the centroids at the considered temperature.

Given a configuration $\bRcal$ for the centroids, we define the corresponding SCHA square matrix
$\bPhi(\bRcal)$ as the matrix that minimizes the functional $\Fcal[\tagmedia]$ with respect to $\bvarPhi$.
Therefore, from Eq.~\eqref{eq:F(R)_def_1} we have
\begin{equation}
F^{\SCHA}(\bRcal)=\Fcal\bigl[\rhotrial_{\bRcal,\bPhi(\bRcal)}\bigr]\mathp
\label{eq:F(R)_def_2}
\end{equation}
The SCHA matrix satisfies the following self-consistent equation (see Eq.~\eqref{eq:app_SCHA_rel}):
\begin{equation}
\Phi_{ab}(\bRcal)=\avg{\frac{\partial^2V}{\partial R^a\partial R^b}}_{{\displaystyle\rhotrial_{\bRcal,\bPhi(\bRcal)}}}\mathp
\label{eq:SCHA_matrix_def}
\end{equation}
Notice that, for clarity, we are using two different symbols for the generic trial matrix $\bvarPhi$ (a `slanted' phi),
and for the SCHA matrix $\bPhi(\bRcal)$, the specific trial matrix that minimizes $\Fcal[\tagmedia]$ with respect to $\bvarPhi$
for a given centroids position $\bRcal$.

In the rest of this paper, we will consider exclusively the SCHA approximation for the free energy. Therefore,  in order to simplify the notation, 
in what follows we can safely omit the superscript ${}^{\SCHA}$ without ambiguity:
$F$ will always refer to the SCHA free energy. Moreover, as guide and reference for the reader, we collect in table~\ref{tab:symbols} some 
symbols used in the text with a concise description. Several symbols collected in the table will appear later in the course of the paper.

\iftoggle{tabella_simboli}{
\begin{table*}
\begin{ruledtabular}  
\begin{tabular}{lll}
   Symbol  & Meaning     &     First use    \\
\hline
$\bR$ & Atomic position (canonical variable) & Eq.~\eqref{eq:full_ham} \\
$V(\bR)$ & Potential energy & Eq.~\eqref{eq:full_ham} \\
$\bRcal$ & Centroids position (parameter) & Eq.~\eqref{eq:rhotrial}  \\
$F(\bRcal)$ & SCHA free energy for the centroids $\bRcal$ & Eq.~\eqref{eq:F(R)_def_1}\\
$\bRcal_{\ssz}$ & Minimum point of $V(\bR)$ & Eq.~\eqref{eq:Harmonic_dyn_mat}\\
$\bRcal_{\eq}$ & Minimum point of $F(\bRcal)$ &Eq.~\eqref{eq:defin_Req}\\
$\bvarPhi$& Generic trial harmonic matrix (parameter) & Eq.~\eqref{eq:rhotrial}\\
$\rhotrial _{\bRcal,\bvarPhi}(\bR)$ & Probability distribution of $\bR$ for a given value of the parameters  $\bRcal$, $\bvarPhi$ & Eq.~\eqref{eq:rhotrial}\\
$\bPhi(\bRcal)$& $\bvarPhi$ that minimizes the SCHA free energy functional at a given $\bRcal$& Eq.~\eqref{eq:F(R)_def_2}\\
$\overset{\ssn}{\bPhi}(\bRcal)$ & 
	Average of the $n$-th derivative of $V(\bR)$ with the probability $\rhotrial _{\bRcal,\bPhi(\bRcal)}(\bR)$ & Eq.~\eqref{eq:SCHA_matrix_def_nth} \\
$\bphi$ & Second derivative of $V(\bR)$ in $\bRcal_{\ssz}$& Eq.~\eqref{eq:Ham_harm}\\
$\!\!\!\overset{\,\,\,\ssn}{\bphi}$ & $N$-th derivative of $V(\bR)$ in $\bRcal_{\ssz}$ & Eq.~\eqref{eq:n-th_order_force_constant}\\
$\bD^{\ssF}$& Second derivative of $F(\bRcal)$ in $\bRcal_{\eq}$, divided by the square root of the masses & Eq.~\eqref{eq:SCHA_dyn_mat}\\
$\bD^{\ssS}$& Matrix $\bPhi(\bRcal_{\eq})$ divided by the square root of the masses & Eq.~\eqref{eq:def_D}\\
$\overset{\ssn}{\bD}{}^{\ssS}$&  Tensor $\overset{\ssn}{\bPhi}(\bRcal_{\eq})$ divided by the square root of the masses &Eq.~\eqref{eq:def_D_n}\\
$\bD^{\ssz}$& Matrix $\bphi$ divided by the square root of the masses & Eq.~\eqref{eq:Harmonic_dyn_mat} \\
$\overset{\ssn}{\bD}{}^{\ssz}$& Tensor $\overset{\ssn}{\bphi}$ divided by the square root of the masses & Eq.~\eqref{eq:def_dyn_harm_nth}\\
$\overset{\ssi}{\boldsymbol{\Acal}}$ & Inverse of the matrix $\boldsymbol{\Acal}$ & Eq.~\eqref{eq:relaz_GS_DS}\\
$\bG_{\ssS}$& Green function associated to $\bD^{\ssS}$ & Eq.~\eqref{eq:relaz_GS_DS}\\
$\bG_{\ssz}$& Green function  associated to $\bD^{\ssz}$ & Eq.~\eqref{eq:relaz_G0_D0}\\
\end{tabular}
\end{ruledtabular}
\caption{Collection of some symbols frequently used in the main text.
First column, the symbol used. Second column, a short description of the meaning. Third column,
first labeled equation where the symbol appears.}
\label{tab:symbols}
\end{table*}
}

\section{Structural second order phase transition and curvature of the free energy}
\label{sec:Second_order_phase_transition_and_curvature_of_the_free_energy}
In second order phase transitions involving the position of the atoms,
e.g. in ferroelectric and in charge-density wave phase transitions~\cite{PhysRevLett.17.753,delaire614,0022-3719-13-19-018,
PhysRevLett.107.107403,PhysRevB.92.140303,PhysRevLett.86.3799,PhysRevB.86.155125,
doi:10.1080/00150199808009159,PhysRevB.80.241108,PhysRevLett.106.196406,PhysRev.140.A863, PhysRevB.63.144109,PhysRevB.14.4321,
PhysRevB.92.094107}, we can
use the centroids $\bRcal$ to define the order parameter, which is the
observable measured in diffraction experiments.
The (temperature-dependent) function $F(\bRcal)$ rules the
phase transitions.  At each temperature, the system is in equilibrium in the 
(temperature-dependent) configuration $\bRcal_{\eq}$, where $F(\bRcal)$ has a minimum.
Therefore, in $\bRcal_{\eq}$ the first derivative of $F(\bRcal)$ is zero,
$\partial F/\partial\Rcal^a|_{\bRcal_{\eq}}=0$, and the Hessian matrix of $F(\bRcal)$ (i.e. the curvature), $\partial^2 F/\partial\Rcal^a\partial\Rcal^b|_{\bRcal_{\eq}}$, 
is positive-definite.

\begin{figure}
\centering
\includegraphics[width=1.0\columnwidth]{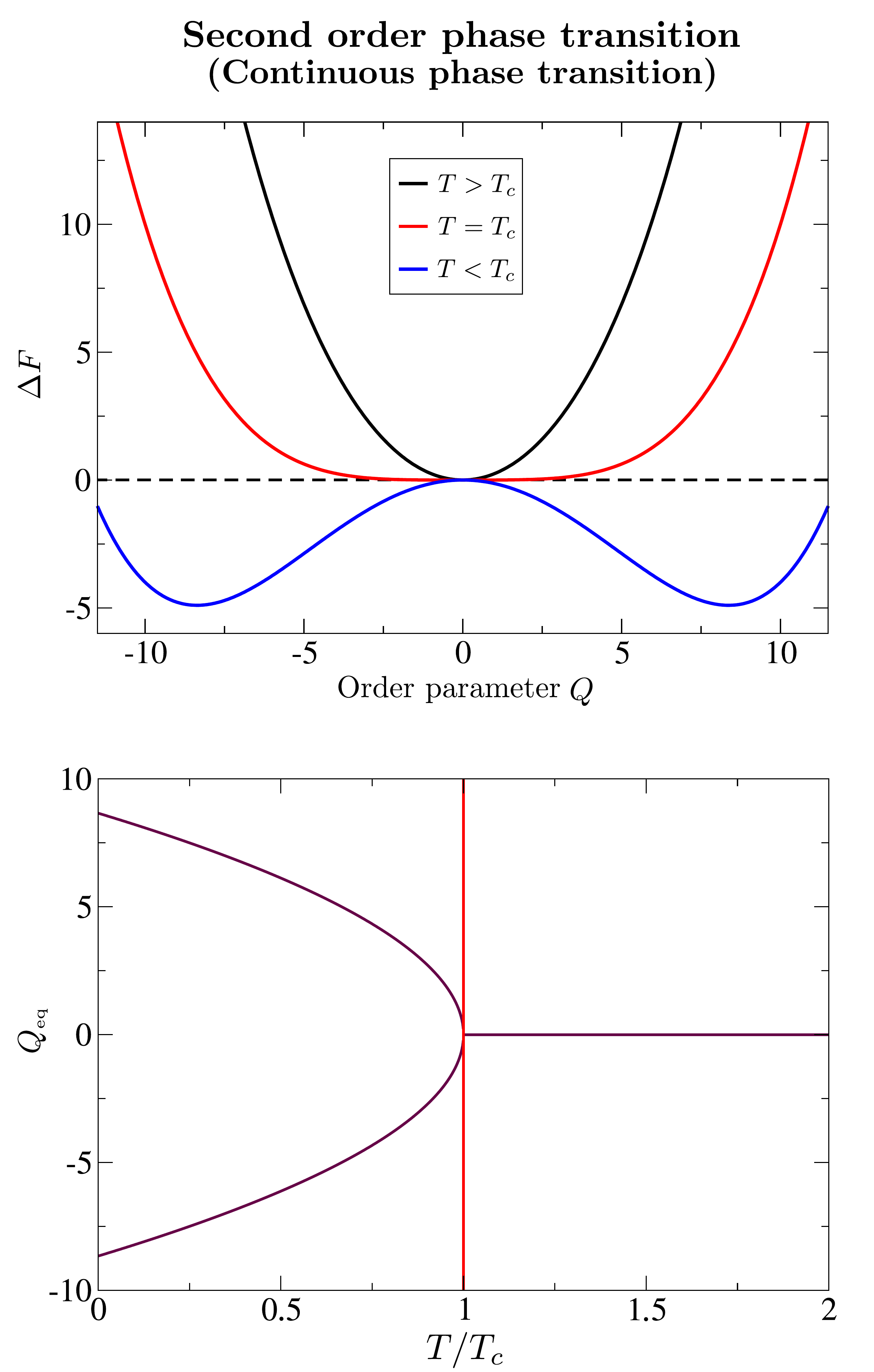}
\caption{(Color online) Example of a second order (i.e. continuous) phase transition, as described by Landau's theory. 
$Q$ is a macroscopic, scalar, order parameter identifying  a system configuration. We consider a situation in which symmetry $Q\rightarrow -Q$ holds. 
$Q=0$ is a high-symmetry phase. $\Delta F(Q)= F(Q)-F(0)$ is the difference between the free energy of phase $Q$ and the
free energy of the high-symmetry phase, at a certain temperature $T$. For each $T$, $\Delta F(Q)$ has minimum in the equilibrium 
configuration $Q_{\eq}(T)$. Plots are in arbitrary units. 
The free energy difference is an even polynomial  $\Delta F(Q)=A_2(T)\,Q^2+A_4(T)\,Q^4+\Ocal(Q^6)$, with
 $A_4(T)>0$ and $A_2(T)$ that decreases from positive to negative values. $T_c$ is the transition temperature.
 At $T>T_c$, the free energy has minimum in $Q_{\eq}(T)=0$, i.e. $A_2(T)$ is positive.
At $T<T_c$, $A_2(T)$ is negative: $Q=0$ becomes a local maximum, whereas the minimum $Q_{\eq}(T)$ acquires two opposite degenerate values, 
different from zero. $Q_{\eq}(T)$ is a continuous function even during the transition. 
Upper panel: variation of the free energy $\Delta F$ as a function of the
order parameter $Q$ for three temperatures $T$, above, below and equal to the transition temperature $T_c$.
Bottom panel: value of the equilibrium order parameter $Q_{\eq}$ as a function of the temperature $T$.}
\label{fig:phase_transition}
\end{figure}

Landau's theory of second order phase transitions~\cite{landau-vol-5} shows that above a certain critical temperature $T_c$
the equilibrium configuration $\bRcal_{\eq}$  is in a high-symmetry phase $\bRcal_{\hs}$.
As $T$ decreases and approaches $T_c$ from above, 
the minimum of $F(\bRcal)$ in  $\bRcal_{\hs}$ becomes less and less pronounced. At $T=T_c$,
$\bRcal_{\hs}$ becomes a saddle point, i.e. the Hessian of $F(\bR)$ in $\bRcal_{\hs}$  develops at least one null eigenvalue,
which becomes negative by lowering further the temperature. At the same time, 
the minimum point $\bRcal_{\eq}(T)$, now depending on temperature, continuously deviates from 
$\bRcal_{\hs}$ to different configurations having lower symmetry.
Since during the phase transition the equilibrium configuration $\bRcal_{\eq}(T)$ remains a continuous function of temperature, 
these are also called ``continuous phase transitions''.
In Fig.~\ref{fig:phase_transition} we show an example of a typical second order phase transition.

In conclusion, at any temperature it is $\partial F/\partial\Rcal^a|_{\bRcal_{\hs}}=0$ and the phase transition is
 characterized by the change of character of the Hessian matrix $\left.\partial^2F/\partial\Rcal^a \partial\Rcal^b\right|_{\bRcal_{\hs}}$:
at $T>T_c$ it  is positive-definite, whereas as $T<T_c$ 
it develops at least one negative eigendirection indicating the distortion which decreases the free energy.
It follows that a method to estimate the transition temperature $T_c$ and the instability modes
of a second order phase transition can be obtained by computing the Hessian of $F(\bRcal)$
in the high-symmetry configuration 
and studying its evolution as a function of temperature.
In the next section we will find explicit formulas for the first and second derivative of $F(\bRcal)$.

\section{Derivatives of $F(\bRcal)$}
\label{Derivatives_of_F}
From the definition of Eq.~\eqref{eq:F(R)_def_2}, we can calculate explicitly the derivatives 
of $F(\bRcal)$ with respect to $\bRcal$.
Here we present only the results, while the derivation is given in appendix~\ref{app:SCHA_proofs}.
%
For the first derivative we have the intuitive result (see~Eq.~\eqref{eq:app_first_der})
\begin{equation}
\frac{\partial F}{\partial\bRcal}=\avg{\frac{\partial V}{\partial \bR}}_{\ds{\rhotrial_{\bRcal,\bPhi(\bRcal)}}}\mathp
\label{eq:first_derivative_F}
\end{equation}
The derivative of the free energy is the average of the potential derivative. In other words,
the forces  on the centroids are equal to the average of the mechanical forces on the atoms.
From Eq.~\eqref{eq:first_derivative_F}, for the equilibrium position $\bRcal_{\eq}$
defined in Eq.~\eqref{eq:defin_Req} it is
\begin{equation}
0=\avg{\frac{\partial V}{\partial \bR}}_{\ds{\rhotrial_{\bRcal_{\eq},\bPhi(\bRcal_{\eq})}}}\mathp
\label{eq:first_derivative_F_in_eq}
\end{equation}

For what follows it is convenient to define the $n$-th order SCHA tensor, which generalizes Eq.~\eqref{eq:SCHA_matrix_def} 
to higher orders:
\begin{equation}
\overset{\ssn}{\Phi}_{a_1\cdots a_n}(\bRcal)=\avg{\frac{\partial^nV}{\partial R^{a_1}\cdots\partial R^{a_n}}}_{{\displaystyle\rhotrial_{\bRcal,\bPhi(\bRcal)}}}\mathp
\label{eq:SCHA_matrix_def_nth}
\end{equation}
Notice that we did not use the superscript $\scriptstyle{(2)}$ for the square SCHA matrix.
The $n$-th order SCHA tensor has the same properties of the $n$-th order force constant (i.e. the $n$-th derivative of the potential). Notably,
it is invariant with respect to permutation of indices; it is invariant with respect to
all the symmetry operations (including lattice translations 
in a crystal) associated to the configuration $\bRcal$~\cite{RevModPhys.40.1};
and it satisfies the acoustic sum rule (ASR), i.e.
the sum over any atom index vanishes (see~Eq.~\eqref{eq:app_def_ASR}).

Deriving a second time Eq.~\eqref{eq:first_derivative_F} with respect to $\bRcal$ we obtain
(see~Eqs.~\eqref{eq:app_der_int_0}--\eqref{eq:app_Theta_equation} )
\begin{align}
&\frac{\partial^2 F}{\partial \Rcal^a \partial \Rcal^b}=\Phi_{ab}+\sum_{c_1c_2c_3c_4}\overset{\scriptscriptstyle{(3)}}{\Phi}_{ac_1c_2}\Lambda^{c_1c_2c_3c_4}\overset{\scriptscriptstyle{(3)}}{\Phi}_{c_3c_4b}\nonumber\\
&\,\,\,+\sum_{\substack{c_1c_2c_3c_4\\d_1d_2d_3d_4}}\overset{\scriptscriptstyle{(3)}}{\Phi}_{ac_1c_2}\Lambda^{c_1c_2c_3c_4}\Theta_{c_3c_4d_1d_2}\Lambda^{d_1d_2d_3d_4}\overset{\scriptscriptstyle{(3)}}{\Phi}_{d_3d_4b}\mathv
\label{eq:sec_der_F_1}
\end{align}
where
\begin{align}
\Lambda^{abcd}&=-\frac{\hbar^2}{8}\sum_{\nu\mu}
\frac{F(0,\omega_{\mu},\omega_{\nu})}{\omega_{\mu}\omega_{\nu}}\nonumber\\
&\mkern72mu\times\frac{e^a_{\nu}}{\sqrt{M_a}}\frac{e^b_{\mu}}{\sqrt{M_b}}\frac{e^c_{\nu}}{\sqrt{M_c}}\frac{e^d_{\mu}}{\sqrt{M_d}}\mathp
\label{eq:main_lambda_1}
\end{align}
Here $e_\mu^a$ and $\omega_\mu^2$ are eigenvectors and eigenvalues of $\Phi_{ab}/\sqrt{M_a M_b}$, respectively, and
\allowdisplaybreaks[0]
\begin{align}
&F(0,\omega_\nu,\omega_\mu)=\phantom{\Biggl\{}\nonumber\\
&\quad\left\{
\begin{aligned}
&\frac{2}{\hbar}\left[\frac{2n_\nu+1}{2\omega_\nu}-\frac{dn_\nu}{d\omega_\nu}\right] &&\text{if}\qquad \omega_\nu=\omega_\mu\\
&\frac{2}{\hbar}\left[\frac{n_\mu+n_\nu+1}{\omega_\mu+\omega_\nu}-\frac{n_\mu-n_\nu}{\omega_\mu-\omega_\nu}\right] &&\text{if}\qquad \omega_\nu\neq\omega_\mu
\end{aligned}
\right.\mathp
\label{eq:def_F0}
\end{align}
The tensor $\Theta_{abcd}$ is the solution of the 
Dyson-like equation
\allowdisplaybreaks
\begin{equation}
\Theta_{abcd}=\overset{\scriptscriptstyle{(4)}}{\Phi}_{abcd}+\sum_{l_1l_2l_3l_4}\overset{\scriptscriptstyle{(4)}}{\Phi}_{abl_1l_2} \Lambda^{l_1l_2l_3l_4}  \Theta_{l_3l_4cd}\mathp
\label{eq:Theta_1}
\end{equation}
Notice that in all these equations the dependence of the quantities on $\bRcal$ is understood.

We have obtained for the second derivative a relation which is different from the one found for the first derivative.
Indeed, as shown in Eq.~\eqref{eq:first_derivative_F},  the first derivative of the SCHA free energy is equal to the average
of the first derivative of the potential. 
On the contrary, the second derivative of the SCHA free energy is equal to the average of the second derivative 
of the potential, the SCHA matrix $\Phi_{ab}$ of Eq.~\eqref{eq:SCHA_matrix_def}, plus two terms depending on the third and fourth order SCHA tensors.
In component-free notation,  we can write Eq.~\eqref{eq:sec_der_F_1} in compact form:
\begin{equation}
\frac{\partial^2F}{\partial \bRcal \partial \bRcal}=\bPhi+\boldsymbol{\overset{\scriptscriptstyle{(3)}}{\Phi}}\bLambda\boldsymbol{\overset{\scriptscriptstyle{(3)}}{\Phi}}
+\boldsymbol{\overset{\scriptscriptstyle{(3)}}{\Phi}}\bLambda\bTheta\bLambda\boldsymbol{\overset{\scriptscriptstyle{(3)}}{\Phi}}\mathv
\label{eq:compact_expr_curv}
\end{equation}
where the contraction on the indices is understood.
Moreover, it is convenient to introduce a `super-index' $A=(pq)$. 
In this way, for example, $\Lambda^{pqhk}=\Lambda^{AB}$, $\Theta^{pqhk}=\Theta^{AB}$
and $\overset{\ssf}{\Phi}{}_{pqhk}=\overset{\ssf}{\Phi}{}_{AB}$ are square symmetric `super-matrices' of order $(3\Nat)^2$,
and the contraction of indices between them can be seen as a matrix product.

As explained in the previous sections, the curvature of the free energy in a high-symmetry phase as a function of temperature
is essential in order to identify and characterize a second order phase transition. Diagonalizing the real symmetric matrix 
$\partial^2F/\partial\Rcal^a\partial\Rcal^b$ we obtain eigenvalues and eigenvectors as a function of temperature. In the presence of a second order phase transition, 
there is at least one eigenvalue that becomes negative at the transition temperature, and the corresponding eigenvector identifies
the instability distortion pattern which reduces the free energy. 
By definition, the SCHA matrix $\bPhi$ is positive-definite (see comment after Eq.~\eqref{eq:def_Upsilon}). On the contrary, 
as shown in Eq.~\eqref{eq:app_def_neg_lambda}, $\bLambda$ is negative-definite  thus 
$\overset{\sst}{\bPhi}\bLambda\overset{\sst}{\bPhi}$ is negative-semidefinite. It is this term, which for reasons that will be clear later we call `bubble'
(see~Sec.~\ref{sec:Diagrammatic_representation}), that allows the second derivative of the free energy to have negative
eigenvalues. The formula obtained for $\partial^2F/\partial\Rcal^a\partial\Rcal^b$ also clarifies in this way
the long-standing debate about the possibility of having 
second order phase transitions within the SCHA~\cite{PhysRevLett.28.895,PhysRevLett.29.369}:
the SCHA can describe a second order phase transition only if 
$\boldsymbol{\overset{\scriptscriptstyle{(3)}}{\Phi}} \ne 0$.

Using the interpretation of the 4th-rank tensors as super-matrices of order $(3\Nat)^2$,
$\bTheta$ is readily obtained by inverting Eq.~\eqref{eq:Theta_1} in matrix form:
\begin{equation}
\bTheta=\Bigl[\mathds{1}-\boldsymbol{\overset{\scriptscriptstyle{(4)}}{\Phi}}\bLambda\Bigr]^{-1}\,\,\boldsymbol{\overset{\scriptscriptstyle{(4)}}{\Phi}}\mathp
\label{eq:btheta_expr}
\end{equation}
Substituting Eq.~\eqref{eq:btheta_expr} into Eq.~\eqref{eq:compact_expr_curv} we obtain the compact expression for the
free energy Hessian:
\begin{equation}
\frac{\partial^2F}{\partial \bRcal \partial \bRcal}=\bPhi+
\boldsymbol{\overset{\scriptscriptstyle{(3)}}{\Phi}}\,\,\bLambda\Bigl[\mathds{1}-\boldsymbol{\overset{\scriptscriptstyle{(4)}}{\Phi}}\bLambda\Bigr]^{-1}\boldsymbol{\overset{\scriptscriptstyle{(3)}}{\Phi}}\mathp
\label{eq:compact_expr_curv_daimpl}
\end{equation}
This is the equation that has been implemented .
It is also interesting to write Eq.~\eqref{eq:compact_expr_curv_daimpl} in a more symmetric fashion:
\begin{equation}
\frac{\partial^2F}{\partial \bRcal \partial \bRcal}=\bPhi-\boldsymbol{\overset{\scriptscriptstyle{(3)}}{\Phi}}\sqrt{-\bLambda}
\,\,
\frac{\mathds{1}}{\mathds{1}-\bXi}
\,\,
\sqrt{-\bLambda}\,\boldsymbol{\overset{\scriptscriptstyle{(3)}}{\Phi}}\mathv
\end{equation}
where $\bXi=-\sqrt{-\bLambda}\,\,\overset{\ssf}{\bPhi}\sqrt{-\bLambda}$ is the adimensional real symmetric matrix that rules the convergence of the geometric series. 
For example, from this formula we clearly see that
in the limiting case where the absolute values of $\bXi$'s eigenvalues are much smaller than one and $\bXi$ can be discarded with respect to the identity, 
the curvature is given by the SCHA matrix plus the bubble only.
\section{Phonons in the SCHA}
\label{sec:phonons_in_the_SCHA}
From the results obtained, it is tempting to use the curvature of the free energy with respect to the centroids to define 
a phonon-like dispersion. To this purpose, for each temperature, we consider the free energy curvature in the corresponding 
equilibrium configuration $\bRcal_{\eq}$, divided by the square root of the masses:
\begin{equation}
D^{\ssF}_{ab}=\left.\frac{1}{\sqrt{M_aM_b}}\frac{\partial^2F}{\partial \Rcal^a \partial \Rcal^b}\right|_{\bRcal_{\eq}}\mathp
\label{eq:SCHA_dyn_mat}
\end{equation}
This matrix can be considered as the temperature-dependent, free energy-based,  generalization of the temperature-independent harmonic dynamical matrix
\begin{equation}
D^{\ssz}_{ab}=\left.\frac{1}{\sqrt{M_aM_b}}\frac{\partial^2V}{\partial R^a \partial R^b}\right|_{\bRcal_{\ssz}}\mathp
\label{eq:Harmonic_dyn_mat}
\end{equation}
Here $\bRcal_{\ssz}$ is the temperature-independent configuration for which the potential $V(\bR)$ has a minimum. 
We associate the `free energy dynamical matrix' $D^{\ssF}_{ab}$ to `free energy phonons', quasi-particles
whose energies $\hbar\Omega_{\mu}$ and polarization vectors $\epsilon^a_{\mu}$ are obtained by diagonalization as
\begin{equation}
\sum_{b}D^{\ssF}_{ab}\epsilon^b_{\mu}=\Omega^2_{\mu}\epsilon^a_{\mu}\mathp
\end{equation}
Since $D^{\ssF}_{ab}$ is positive-definite if and only if $\partial^2F/\partial\Rcal^a\partial\Rcal^b|_{\bRcal_{\eq}}$ is positive-definite, 
an instability in the system corresponds to at least a frequency $\Omega_{\mu}$ becoming imaginary. It is this fact that
justifies the interpretation of $D^{\ssF}_{ab}$ as a temperature-dependent generalized dynamical matrix describing 
temperature-dependent anharmonic phonons. 
It is worthwhile to emphasize that the theory developed so far is `static', in the sense that it is not based on time-dependent properties, but on the variation
of the free energy with respect to a static variation of the centroids position. Moreover, it is important to observe
that we cannot use
\begin{equation}
 D^{\ssS}_{ab}=\frac{1}{\sqrt{M_aM_b}}\Phi_{ab}(\bRcal_{\eq})
\label{eq:def_D}
\end{equation}
to study system instabilities and defines phonon-like particles, even if 
in some cases it has given temperature-dependent anharmonic phonons in 
good agreement with experiments~\cite{PhysRevB.92.140303,PhysRevLett.111.177002}.
Indeed,  $D^{\ssS}_{ab}$ it is not given by the second derivative of the free energy.
Moreover, by definition, $D^{\ssS}_{ab}$ is positive-definite, thus it is impossible to observe any softening in its eigenvalues.

The free energy dynamical matrix $D^{\ssF}_{ab}$ is a particularly important tool when we consider crystals. 
Indeed, in that case we can use the same techniques that are standard for the harmonic theory~\cite{RevModPhys.73.515}. 
Exploiting the translational lattice symmetry,  we define  the SCHA dynamical matrices $\bD^{\ssF}(\bq)$ in the unit cell as
a function of the quasi-momentum  $\bq$.  We can explicitly calculate $\bD^{\ssF}(\bq)$ on a coarse grid 
of the Brillouin zone (BZ) and later Fourier interpolate the result to obtain the matrix
on an arbitrary finer grid or a path. Thus, diagonalizing $\bD^{\ssF}(\bq)$, we obtain the spectrum 
$\Omega^2_{\mu}(\bq)$ and the polarization vectors $\epsilon^a_{\mu}(\bq)$ on a path of the BZ.  An imaginary phonon 
in a point  $\bq$ indicates that  the system is unstable for a distortion with 
modulation $\bq$ that reduces the lattice periodicity. This is, for example, what happens in charge-density wave instabilities. 
Therefore, with moderate workload, it is possible to have a complete picture of the crystal instabilities.
In particular, with calculations on supercells of reasonable size it is possible, in principle, to study lattice instabilities
which are periodic on very large supercells or even incommensurate.
\section{Diagrammatic representation}
\label{sec:Diagrammatic_representation}
In this section we  give a perspicuous diagrammatic description of Eq.~\eqref{eq:compact_expr_curv_daimpl}, in order to
reformulate it in a language familiar to the field theorists. 
The diagrammatic description can also be useful as a basis for further developments of the theory, as we will see
later in Sec.~\ref{sec:Anstaz_for_a_dynamic_theory}.

\begin{figure}
\centering
\includegraphics[width=\columnwidth]{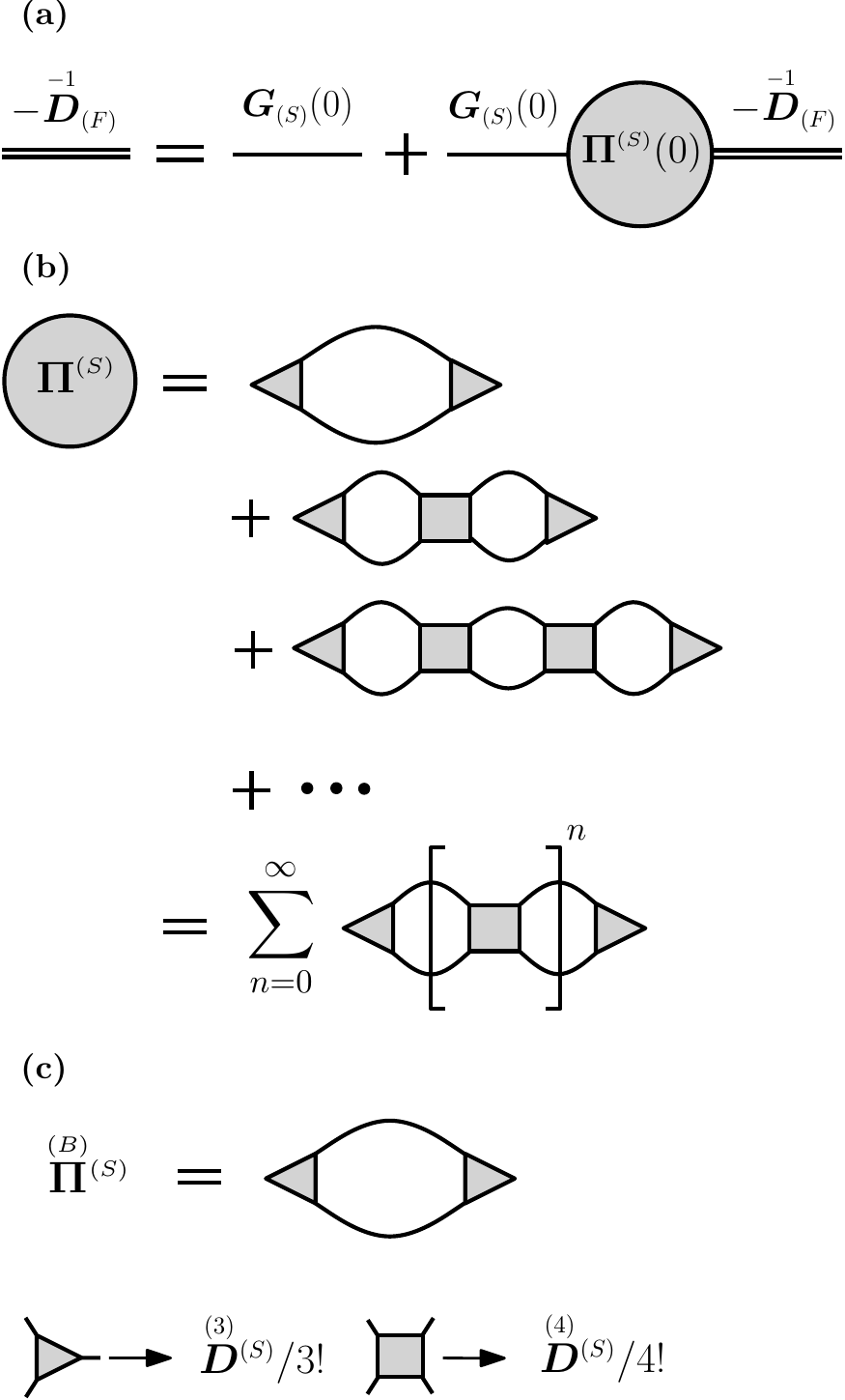}
\caption{Figure $a)$: Diagrammatic representation of Eq.~\eqref{eq:diagrammatic_static_invSCHAdynmat_1}.
Figure $b)$: Diagrammatic representation of the SCHA self-energy $\bPi^{\ssS}$, Eq.~\eqref{eq:diagrammatic_static_selfenergy}. 
Since in that equation only the static value $\bPi^{\ssS}(0)$ is considered, 
the sum over the frequencies of the internal lines is performed, but the total
frequency is kept equal to zero.
Figure $c)$: Diagrammatic representation of $\overset{\ssB}{\bPi}{}^{\ssS}$, the bubble part of the SCHA self-energy,
Eq.~\eqref{eq:diagrammatic_static_selfenergy_bubble}.}
\label{fig:Feyman_diagr_stat}
\end{figure}

Fixed the temperature, with the corresponding $\bRcal_{\eq}$ we define the quadratic `SCHA Hamiltonian'
\begin{equation}
H^{\ssS}=\sum_{a}\frac{{p^2_{a}}}{2M_a}+\frac{1}{2}\sum_{ab}\Phi_{ab}(\bRcal_{\eq})\,(R-\Rcal_{\eq})^a(R-\Rcal_{\eq})^b\mathv
\label{eq:HssS_def}
\end{equation}
and we consider the corresponding SCHA thermodynamic Green function $G^{ab}_{\ssS}(z)$
for the displacements normalized by masses $\sqrt{M_a}(R-\Rcal_{\eq})^a$.
Since $H^{\ssS}$ is quadratic
\begin{equation}
\overset{\ssi}{G}{}^{ab}_{\ssS}(z)=z^2\delta_{ab}-D_{ab}^{\ssS}\mathv
\label{eq:relaz_GS_DS}
\end{equation}
where $\overset{\ssi}{G}{}^{ab}_{\ssS}$ indicates the inverse matrix of $G^{\ssS}_{ab}$ (similar notation for the
inverse will be used later also in other formulas). We also consider $\chi^{abcd}_{\ssS}(0)$, the SCHA `static' loop, i.e. the loop
with $G^{ab}_{\ssS}$ and total frequency equal to zero:
\begin{equation}
\chi_{\ssS}^{abcd}(0)=\frac{1}{\beta}\sum_l\,{G}_{\ssS}^{ac}(i\Omega_l){G}_{\ssS}^{bd}(i\Omega_{-l})\mathv
\label{eq:static_loop}
\end{equation}
where $\Omega_l=2\pi l/\hbar\beta$ is the $l$-th Matsubara frequency. 
With standard techniques for Matsubara frequency summation we obtain~\cite{mahan2000many,PhysRev.128.2589} 
\begin{align}
&\frac{1}{\beta}\sum_lG_{\ssS}^{ac}(i\Omega_{l})G_{\ssS}^{bd}(i\Omega_{-l})=\nonumber\\
&\quad\qquad\frac{\hbar^2}{4}\sum_{\mu\nu}\frac{F(0,\omega_{\mu},\omega_{\nu})}{\omega_{\mu}\omega_{\nu}}\,e^a_{\nu}e^b_{\mu}e^c_{\nu}e^d_{\mu}\mathv
\label{eq:sumMatzub_1}
\end{align}
with $F(0,\omega_{\mu},\omega_{\nu})$ defined in Eq.~\eqref{eq:def_F0}.
From Eq.~\eqref{eq:main_lambda_1}, Eq.~\eqref{eq:static_loop} and Eq.~\eqref{eq:sumMatzub_1}  we obtain a relation between the tensor
$\Lambda^{abcd}$ and the static loop $\chi_{\ssS}^{abcd}(0)$:
\begin{equation}
\chi_{\ssS}^{abcd}(0)=-2\Lambda^{abcd}\sqrt{M_aM_bM_cM_d}\mathp
\label{eq:relazione_lambda_chizero}
\end{equation}
Therefore, using Eq.~\eqref{eq:SCHA_dyn_mat} and Eq.~\eqref{eq:def_D}, 
formula~\eqref{eq:compact_expr_curv_daimpl} divided by the square root of the masses gives
\begin{equation}
{\bD}{}^{\ssF}={\bD}{}^{\ssS}+\bPi^{\ssS}(0)\mathv
\label{eq:diagrammatic_static_invSCHAdynmat_00}
\end{equation}
where, as usual, we have used bold symbols in component-free notation and we have defined
\begin{align}
\bPi^{\ssS}(0)&=\overset{\bsst}{\bD}{}^{\ssS}\left(-\frac{1}{2}\,\bchi_{\ssS}(0)\right)\nonumber\\
 &\qquad\times\left[\mathds{1}-\overset{\bssf}{\bD}{}^{\ssS}\left(-\frac{1}{2}\,\bchi_{\ssS}(0)\right)\right]^{-1}\overset{\bsst}{\bD}{}^{\ssS}\mathp
\label{eq:diagrammatic_static_selfenergy}
\end{align}
Here we have generalized the definition~\eqref{eq:def_D} to the $n$-th order as
\begin{equation}
\overset{\ssn}{D}{}^{\ssS}_{a_1\ldots a_n}=\frac{\overset{\ssn}{\Phi}_{a_1\ldots a_n}(\bRcal_{\eq})}{\sqrt{M_{a_1}\ldots M_{a_n}}}\mathp
\label{eq:def_D_n}
\end{equation}
Notice that we did not use the superscript $\scriptstyle{(2)}$ for the second order tensor defined by Eq.~\eqref{eq:def_D}.
In terms of the SCHA Green function defined in Eq.~\eqref{eq:relaz_GS_DS},
 Eq.~\eqref{eq:diagrammatic_static_invSCHAdynmat_00} is readily  written as
\begin{equation}
-{\bD}{}^{\ssF}=\overset{\ssi}{\bG}{}^{\ssS}(0)-\bPi^{\ssS}(0)\mathv
\label{eq:diagrammatic_static_invSCHAdynmat_0}
\end{equation}
which is equivalent to the Dyson-like equation
\begin{equation}
-\overset{\ssi}{\bD}{}_{\ssF}=\bG_{\ssS}(0)+\bG_{\ssS}(0)\,\bPi^{\ssS}(0)(-\overset{\ssi}{\bD}_{\ssF})\mathv
\label{eq:diagrammatic_static_invSCHAdynmat_1}
\end{equation}
where the matrix product is understood.
If the opportune diagram symmetry factors are taken into account, Eq.~\eqref{eq:diagrammatic_static_invSCHAdynmat_1}
with Eq.~\eqref{eq:diagrammatic_static_selfenergy} have the Feynman diagrams representation
shown in Fig.~\ref{fig:Feyman_diagr_stat}a and Fig.~\ref{fig:Feyman_diagr_stat}b. This is the diagrammatic representation of the curvature formula~\eqref{eq:compact_expr_curv_daimpl}
(divided by the square root of the masses). 
Analogous diagrammatic series has been obtained by G\"otze and Michel in Ref.~\citenum{GotzeZfurphys}.
The first term of the series giving $\bPi^{\ssS}(0)$ is the SCHA `bubble' $\overset{\ssB}{\bPi}{}^{\ssS}(0)$.
It is given by the formula
\begin{equation}
\overset{\ssB}{\bPi}{}^{\ssS}(0)=\overset{\bsst}{\bD}{}^{\ssS}\left(-\frac{1}{2}\,\bchi_{\ssS}(0)\right)\overset{\bsst}{\bD}{}^{\ssS}
\label{eq:diagrammatic_static_selfenergy_bubble}
\end{equation}
and corresponds to the diagram in Fig.~\ref{fig:Feyman_diagr_stat}c. The SCHA `bubble' is 
the term $\overset{\sst}{\bPhi}\bLambda\overset{\sst}{\bPhi}$ 
of Eq.~\eqref{eq:compact_expr_curv}, divided by the square root of the masses. 
This explains the name `bubble' given to that term.

Before concluding this section, it is worthwhile to remark that, in spite of the symbol used, at this level
the $\bPi^{\ssS}(0)$ defined in  Eq.~\eqref{eq:diagrammatic_static_selfenergy} is just an auxiliary quantity, without a specific physical meaning. 
However, the choice of the symbol is not casual because later we will interpreted it as a self-energy.
This will give a deeper meaning to the results obtained.
\section{Stochastic implementation}
\label{sec:Stochastic_implementation}
The stochastic implementation of the SCHA (SSCHA) has demonstrated to be an efficient method
{to analyze thermal properties of solids 
in situations where the harmonic approximation breaks down~\cite{PhysRevLett.111.177002,PhysRevLett.114.157004,PhysRevB.89.064302,Errea81,
PhysRevB.93.174308,PhysRevB.92.140303,0953-8984-28-49-494001}}. The SSCHA is described in Ref.~\citenum{PhysRevB.89.064302} 
and consists in minimizing with {a} conjugate-gradient (CG) method 
the functional $\Fcal[\rhotrial_{\bRcal,\bvarPhi}]$ with respect to $\bRcal$ and $\bvarPhi$. 
The functional and its gradient 
are expressed {as averages taken with $\rhotrial_{\bRcal,\bvarPhi}$
of observables $\Ods(\bR)=\Ods\bigl(V(\bR),\bf(\bR)\bigr)$ that are functions only 
of the potential $V(\bR)$ and  
the forces $\bf(\bR)=-\partial V/\partial \bR$}. 
The method is `stochastic' because these averages are evaluated with the {importance} sampling technique.
Since the observables depend only on the position, Eqs.~\eqref{eq:avg_pos_operator}--\eqref{eq:dens_prob} apply.
The space of configurations is statistically sampled with a (large) population of finite size $N_{\Ical}$,
whose members $\bR_{\ssIcal}$  are distributed according to the probability density $\rhotrial_{\bRcal,\bvarPhi}(\bR)$.  
For each element $\bR_{\ssIcal}=\bRcal+\bu_{\ssIcal}$, {$\bu_{\ssIcal}$} being the displacement  from the
centroids $\bRcal$, 
the forces $\bf(\bRcal+\bu_{\ssIcal})$ 
and the potential energy $V(\bRcal+\bu_{\ssIcal})$ 
are calculated {by any energy-force engine,
i.e., making use of first-principles methods or empirical potentials}. In that way the average integrals can be straightforwardly
computed. However, at each step of the CG minimization algorithm the distribution probability $\rhotrial_{\bRcal,\bvarPhi}(\bR)$ changes. Thus, in principle,
at each minimization step a new population should be generated 
and for its members the energies and the forces should be calculated. In order to reduce the number of calls to the energy-force engine, 
in actual calculations a reweighting procedure is adopted~\cite{PhysRevB.89.064302}.
Energy and forces are computed only once for the population elements that are distributed according to an initially fixed probability density
$\rhotrial{}_{\ssin}(\bR)$. 
The approximate averages for a generic distribution probability $\rhotrial_{\bRcal,\bvarPhi}(\bR)$ are then computed {as}
\begin{align}
\Bavg{\Ods}_{\rhotrial_{\bRcal,\bvarPhi}}&\simeq
\frac{1}{N_{\mathcal{I}}}\sum_{\Ical=1}^{N_{\Ical}}
\frac{\rhotrial_{\bRcal,\bvarPhi}(\bRcal+\bu_{\ssIcal})}{\rhotrial{}_{\ssin}(\bRcal+\bu_{\ssIcal})}\nonumber\\
&\mkern48mu\times\Ods\bigl(V(\bRcal+\bu_{\ssIcal}),\bf(\bRcal+\bu_{\ssIcal})\bigr)\vphantom{\frac{\rhotrial_{\bRcal,\bvarPhi}(\bRcal+\bu_{\ssIcal})}{\rhotrial^{\ssz}(\bRcal+\bu_{\ssIcal})}}
{.}
\label{eq:average_montecarlo}
\end{align}
{Obviously, the equality holds for $N_{\Ical}\rightarrow+\infty$.}

We want to use the stochastic approach also to compute the free energy curvature through Eq.~\eqref{eq:compact_expr_curv_daimpl}.
{Considering} a configuration $\bRcal$, after the SSCHA  minimization of the functional $\Fcal[\rhotrial_{\bRcal,\bvarPhi}]$ 
with respect to $\bvarPhi$, the SCHA matrix $\bPhi$ for that configuration is available{. T}herefore we only  
need to express $\overset{\sst}{\bPhi}$ and $\overset{\ssf}{\bPhi}$ in a form that is suited for the stochastic calculation
(here and in what follows the dependence of the matrices on $\bRcal$ is understood).
{As demonstrated in Appendix~\ref{app:Stochastic_calculation_of_derivative_averages} (see
Eqs.~\eqref{eq:app_phi3_con_fbb},~\eqref{eq:app_SCHA3_mat_fin} and 
Eqs.~\eqref{eq:app_phi4_con_fbb},~\eqref{eq:app_SCHA4_mat_fin} with Eq.~\eqref{eq:app_def_fbb_used_finale}), it can
be shown making use of integration by parts that}
\begin{subequations}\begin{align}
&\overset{\scriptscriptstyle{(3)}}{\Phi}_{abc}=-\sum_{pq}\Upsilon_{ap}\Upsilon_{bq}
\Bavg{u^pu^q\, \fbb_c}_{{\displaystyle\rhotrial_{\bRcal,\bPhi}}}
\label{Eq:stoc_avg_3rd}\\
&\overset{\scriptscriptstyle{(4)}}{\Phi}_{abcd}=-\sum_{pqr}\Upsilon_{ap}\Upsilon_{bq}\Upsilon_{cr}\,
\Bavg{u^pu^qu^r\,\fbb_d}_{{\displaystyle\rhotrial_{\bRcal,\bPhi}}}\mathp
\label{Eq:stoc_avg_4th}
\end{align}\label{Eq:stoc_avg_Vbb}\end{subequations}
{Here 
$\Upsilon_{ab}$ is the matrix obtained from $\Phi_{ab}$ through the definition~\eqref{eq:def_Upsilon}, and 
\begin{equation}
\fbb_i=\f_i-\biggl[\Bavg{\f_i}_{{\displaystyle\rhotrial_{\bRcal,\bPhi}}}-\sum_j\Phi_{ij}\,u^j\biggr]\mathp
\label{eq:def_fbb}
\end{equation}
The equations~\eqref{Eq:stoc_avg_Vbb} express the third and {fourth} order SCHA tensors
in terms of averages of forces and displacements only (in the definition~\eqref{eq:def_fbb}
the term subtracted from the forces $\f_a$ is computed analytically with negligible cost, since $\Bavg{\partial V/\partial \bR}_{{\displaystyle\rhotrial_{\bRcal,\bPhi}}}$ and $\bPhi$
are known). Therefore, they can be calculated through Eq.~\eqref{eq:average_montecarlo}. 

It is interesting to observe that, in the limit of an infinitely large population sampling,
adding to $\fbb_i$ a term odd in the displacements does not change the value of $\overset{\sst}{\bPhi}$ obtained from Eq.~\eqref{Eq:stoc_avg_3rd}.
Therefore, the $\fbb_i$ used in Eq.~\eqref{Eq:stoc_avg_3rd} is actually defined only up to an additive factor that is odd in the displacements. 
Analogously, if we use an infinite sampling, the $\fbb_i$ used in Eqs.~\eqref{Eq:stoc_avg_4th} is defined only up to an additive factor that is even in the displacements. 
However, depending on the actual $\fbb_i$ used, we obtain different results when we use a finite sampling to 
compute the averages.   
The specific choice of Eq.~\eqref{eq:def_fbb}, identical for both equations~\eqref{Eq:stoc_avg_Vbb}, guarantees that if the potential $V$ is quadratic, 
then the SSCHA tensors (i.e. the SCHA tensors calculated stochastically)
$\overset{\sst}{\bPhi}$ and $\overset{\ssf}{\bPhi}$ are correctly zero with any finite sampling used to compute the averages.
Therefore, the definition~\eqref{eq:def_fbb} reduces the stochastic error and accelerates the convergence.
Notice that if we compute the curvature of the free energy in a stationary point, since it is $\partial F/\partial\bRcal=0$
then from Eq.~\eqref{eq:first_derivative_F} the term $\Bavg{\f_i}_{{\displaystyle\rhotrial_{\bRcal,\bPhi}}}$ in Eq.~\eqref{eq:def_fbb} is zero. 
In particular, this is true when we evaluate the curvature in the equilibrium configuration $\bRcal_{\eq}$, which is the relevant case when
we study structural second order phase transitions.

In the limit of a fully converged stochastic calculation, the SSCHA tensors $\overset{\sst}{\bPhi}$ and 
$\overset{\ssf}{\bPhi}$ satisfy 
both acoustic sum rule (ASR) and invariance with respect to permutations of indices and symmetry transformations.
Actually, in Appendix~\ref{app:Stochastic_calculation_of_derivative_averages} it is shown that 
the SSCHA $n$-th order tensor satisfies the ASR with any finite population sampling, 
as long as the total force acting on the system is zero for any population element (as it must be), and the ASR is satisfied by $\bPhi$. 
Therefore, it is not necessary to impose any extra-condition to make $\overset{\sst}{\bPhi}$ and $\overset{\ssf}{\bPhi}$
satisfy the ASR.

For the invariance properties the situation is different. We can distinguish two kind of
operators acting on a tensor $\overset{\ssn}{\bPhi}$:
the $n!$ operators $\Tcal_{\pi}$, which permute the tensor indices according to the permutations $\pi\in\sigma_n$, and
the $\Nsym$ operators $\Tcal_{S}$, whose action corresponds to the symmetry transformations $S\in\Gcal_{\sym}$ 
(excluding lattice translations, if it is a crystal).
If we are considering a crystal, the SSCHA calculation is performed on a supercell made of $\Ncell$ unit cells, with periodic
boundary conditions. In that case we consider also the $\Ncell$ operators $\Tcal_{\bl}$ whose action
corresponds to the translations by lattice vectors 
non commensurate with the supercell $\bl\in\Gcal_{\lat}$. The SSCHA tensors are invariant
with respect to these operations only in the limit $N_{\Ical}\rightarrow +\infty$ (for simplicity
we consider the crystal case):
\begin{subequations}\begin{align}
\Tcal_{\pi}\overset{\ssn}{\bPhi}&=\overset{\ssn}{\bPhi}&&\forall \pi\in\sigma_n            \label{eq:perm_prop}\\
\Tcal_{S}\overset{\ssn}{\bPhi}&=\overset{\ssn}{\bPhi}&&\forall   S\in\Gcal_{\sym}        \label{eq:sym_prop}\\
\Tcal_{\bl}\overset{\ssn}{\bPhi}&=\overset{\ssn}{\bPhi}&&\forall \bl\in\Gcal_{\lat}\mathp          \label{eq:lat_prop}
\end{align}\label{eq:_symmet_prop_gen}\end{subequations}
For  calculations performed with finite-size populations these conditions are not satisfied.
We enforce them by applying the projectors $\Pcal_{\perm}$, $\Pcal_{\sym}$ and $\Pcal_{\lat}$ to the result:
\begin{subequations}\begin{align}
\Pcal_{\perm}&=\frac{1}{n!}\sum_{\pi\in\sigma_n}\Tcal_{\pi}\label{eq:perm_proj}\\
\Pcal_{\sym}&=\frac{1}{\Nsym}\sum_{S\in\Gcal_{\sym}  }\Tcal_{S} \label{eq:sym_proj}\\
\Pcal_{\lat}&=\frac{1}{\Ncell}\sum_{\bl\in  \Gcal_{\lat}}\Tcal_{\bl}\mathp \label{eq:lat_proj}
\end{align}\label{eq:_symmet_proj_gen}\end{subequations}

For calculations with finite sampling the action of the projectors~\eqref{eq:_symmet_proj_gen} has two benefits:
we obtain SSCHA tensors with the correct properties and we reduce the statistical noise and improve, with negligible cost, the rapidity of the statistical convergence
with respect to $N_{\Ical}$.
Indeed, the necessity of imposing the property~\eqref{eq:perm_prop} is due to the
fact that Eqs.~\eqref{Eq:stoc_avg_Vbb} are not symmetric with respect to permutation of indices. That is caused by 
the arbitrariness in the choice of the variables integrated by parts in the derivation of the formulas, shown in appendix~\ref{app:Stochastic_calculation_of_derivative_averages}.
As a consequence, an approximate evaluation of the averages causes spurious asymmetries, which are eliminated by applying the projector 
$\Pcal_{\perm}$ to the result. The necessity of imposing the properties~\eqref{eq:sym_prop} and~\eqref{eq:lat_prop} is instead due to the fact that, in general, 
the population generated to compute the averages is composed of elements whose distribution in configuration space 
does not respect the symmetries of the system. 
This leads to  spurious fluctuations which spoil the symmetry properties of the result and which are eliminated by applying the projectors
$\Pcal_{\sym}$ and $\Pcal_{\lat}$. Applying these projectors to the result corresponds
to computing the averages through Eq.~\eqref{eq:average_montecarlo} using a larger population of $N_{\lat}\times N_{\sym}\times N_{\Ical}$ elements
obtained by applying the $N_{\lat}\times N_{\text{sym}}$ symmetry operations on the $N_{\Ical}$ members of the original population.

In conclusion, the formulas implemented in the SSCHA are:
\allowdisplaybreaks[0]
\begin{subequations}\begin{align}
&\overset{\scriptscriptstyle{(3)}}{\Phi}_{abc}\simeq
\Pcal_{\sym}\Pcal_{\lat}\Pcal_{\perm}
\frac{1}{N_{\Ical}}\sum_{\Ical}\frac{\rhotrial_{\bRcal,\bPhi}(\bRcal+\bu_{\ssIcal})}{\rhotrial{}_{\ssin}(\bRcal+\bu_{\ssIcal})}\nonumber\\
&\mkern80mu\times\biggl[-\sum_{pq}\Upsilon_{ap}\Upsilon_{bq}\,
u_{\ssIcal}^p\,u_{\ssIcal}^q\, \fbb_c(\bRcal+\bu_{\ssIcal})\biggr]\label{Eq:stoc_avg_Vbb_3rd_impl}\mathv\\
\allowdisplaybreaks
\allowdisplaybreaks[0]
&\overset{\scriptscriptstyle{(4)}}{\Phi}_{abcd}\simeq
\Pcal_{\sym}\Pcal_{\lat}\Pcal_{\perm}
\frac{1}{N_{\Ical}}\sum_{\Ical}\frac{\rhotrial_{\bRcal,\bPhi}(\bRcal+\bu_{\ssIcal})}{\rhotrial{}_{\ssin}(\bRcal+\bu_{\ssIcal})}\nonumber\\
&\mkern10mu\times\biggl[-\sum_{pqr}\Upsilon_{ap}\Upsilon_{bq}\Upsilon_{cr}\,
u_{\ssIcal}^pu_{\ssIcal}^qu_{\ssIcal}^r\,\fbb_d(\bRcal+\bu_{\ssIcal})\biggr]
\mathp\label{Eq:stoc_avg_Vbb_4th_impl}
\end{align}\label{Eq:stoc_avg_Vbb_impl}\end{subequations}
\allowdisplaybreaks
\section{Perturbative limit}
\label{sec:Perturbative_limit}
In this section we analyze the lowest perturbative order of the SCHA and of the
free energy dynamical matrix $D^{\ssF}_{ab}$. First we set some definitions. Expanding the potential $V$ around
its minimum $\Rcal^a_{\ssz}$, the Hamiltonian $H$ is written as
\begin{equation}
H=H^{\ssz}+\sum_{n\ge 3}\,\frac{1}{n!}\sum_{a_1\cdots a_n}\,\overset{\ssn}{\phi}_{a_1\ldots a_n}u^{a_1}\ldots u^{a_n}\mathv
\label{eq:Ham_harm+anaharm}
\end{equation}
where $u^a=R^a-\Rcal^a_{\ssz}$ is the displacement with respect to the potential minimum,
\begin{equation}
H^{\ssz}=\sum_a\frac{p_a^2}{2M_a}+V(\bRcal_{\ssz})+\frac{1}{2}\sum_{ab}\phi_{ab}\,u^au^b
\label{eq:Ham_harm}
\end{equation}
is the quadratic harmonic Hamiltonian, and 
\begin{equation}
\overset{\ssn}{\phi}_{a_1\ldots a_n}=\left.\frac{\partial^nV}{\partial R^{a_1}\ldots\partial R^{a_n}}\right|_{\bRcal_{\ssz}}
\label{eq:n-th_order_force_constant}
\end{equation}
is the $n$-th order force constant tensor. Notice that for the second order force constant matrix $\phi_{ab}$ 
we do not use the superscript $\scriptstyle{(2)}$. In order to avoid confusion, it is worthwhile to stress that the $n$-th
force constant $\overset{\ssn}{\phi}_{a_1\ldots a_n}$ is the $n$-th derivative of the potential, evaluated at the potential minimum  $\bRcal_{\ssz}$,
whereas the $n$-th SCHA tensor $\overset{\ssn}{\Phi}_{a_1\ldots a_n}(\bRcal)$, defined in Eq.~\eqref{eq:SCHA_matrix_def_nth}, is the $n$-th derivative of the potential averaged 
with the distribution $\displaystyle\rhotrial_{\bRcal,\bPhi(\bRcal)}$.

The part of the Hamiltonian in Eq.\eqref{eq:Ham_harm+anaharm} not included in $H^{\ssz}$ defines the anharmonic part of the potential,
which we treat as a (small) perturbation of $H^{\ssz}$. With $G^{ab}(z)$ and $G^{ab}_{\ssz}(z)$ we indicate the Green function of 
$H$ and $H^{\ssz}$ for the variable $\sqrt{M_a}(R^a-\Rcal^a_{\ssz})$, respectively. The latter is given as
\begin{equation} 
\overset{\ssi}{G}{}^{ab}_{\ssz}(z)=z^2\delta^{ab}-D^{\ssz}_{ab}\mathv
\label{eq:relaz_G0_D0}
\end{equation}
where $D^{\ssz}_{ab}=\phi_{ab}/\sqrt{M_aM_b}$ is the harmonic dynamical matrix, already defined in Eq.~\eqref{eq:Harmonic_dyn_mat}. 
The relation between the full and harmonic Green functions is given by the Dyson equation
\begin{equation}
\overset{\ssi}{\bG}(z)=\overset{\ssi}{\bG}{}^{\ssz}(z)-\bPi^{\ssz}(z)\mathv
\label{eq:Dyson_harm}
\end{equation}
which is equivalent to
\begin{equation}
{\bG}(z)={\bG}{}_{\ssz}(z)+{\bG}{}_{\ssz}(z)\,\bPi^{\ssz}(z)\,{\bG}(z)\mathv
\label{eq:Dyson_harm_1}
\end{equation}
where, in order to use a consistent notation, we have indicated with $\bPi^{\ssz}(z)$ the harmonic self-energy, i.e.
the self-energy obtained by taking $H^{\ssz}$ as non-interacting unperturbed Hamiltonian. 
At the lowest perturbative order~\cite{PhysRev.128.2589}
\begin{equation}
\bPi^{\ssz}(z)\simeq\overset{\ssT}{\bPi}{}^{\ssz}+\overset{\ssL}{\bPi}{}^{\ssz}+\overset{\ssB}{\bPi}{}^{\ssz}(z)\mathv
\label{eq:perturbative_harmonic_selfenergy}
\end{equation}
where $\overset{\ssL}{\bPi}{}^{\ssz}$, $\overset{\ssT}{\bPi}{}^{\ssz}$ and $\overset{\ssB}{\bPi}{}^{\ssz}(z)$
are the loop, tadpole and bubble harmonic self-energies, respectively, which have the following expressions:
\begin{equation}
\overset{\ssL}{\Pi}{}^{\ssz}_{ab}=-\frac{1}{2}\sum_{c_1c_2}\overset{\ssf}{D}{}^{\ssz}_{abc_1\!c_2}
\left[\frac{1}{\beta}\sum_lG^{c_1\!c_2}_{\ssz}(i\Omega_l)\right]\mathv
\label{eq:pert_loop}
\end{equation}
\begin{align}
\overset{\ssT}{\Pi}{}^{\ssz}_{ab}=&-\frac{1}{2}\sum_{\substack{c_1c_2\\d_1d_2}}\overset{\sst}{D}{}^{\ssz}_{abc_1}\overset{\ssz}{G}{}_{\ssz}^{c_1\!c_2}(0)
\overset{\sst}{D}{}^{\ssz}_{c_2d_1\!d_2}\nonumber\\
&\qquad\qquad\qquad\qquad\times\left[\frac{1}{\beta}\sum_lG^{d_1\!d_2}_{\ssz}(i\Omega_l)\right]\mathv
\label{eq:pert_tadpole}
\end{align}
\begin{align}\overset{\ssB}{\Pi}{}^{\ssz}_{ab}(z)=&-\frac{1}{2}\sum_{\substack{c_1c_2\\d_1d_2}}\overset{\sst}{D}{}^{\ssz}_{ac_1\!c_2}\overset{\sst}{D}{}^{\ssz}_{bd_1\!d_2}\nonumber\\
&\quad\times\left[\frac{1}{\beta}\sum_lG_{\ssz}^{c_1\!c_2}(i\Omega_l)G_{\ssz}^{d_1\!d_2}(z-i\Omega_{l})\right]\mathp
\label{eq:pert_bubble}
\end{align}
Here we have generalized the definition~\eqref{eq:Harmonic_dyn_mat} of the harmonic dynamical matrix to the $n$-th order:
\begin{equation}
\overset{\ssn}{D}{}^{\ssz}_{a_1\cdots a_n}=\frac{ \overset{\ssn}{\phi}{}_{a_1\cdots a_n}}{\sqrt{M_{a_1}\cdots M_{a_n}}}\mathp
\label{eq:def_dyn_harm_nth}
\end{equation}
Notice that loop and tadpole self-energies do not depend on the value of the frequency $z$. In fact they are real symmetric.
On the contrary, the bubble is a complex symmetric matrix depending on $z$.
In Fig.~\ref{fig:Feyman_diagr_pert} the diagrammatic representation of the harmonic perturbative result at the lowest order is shown.
\begin{figure}
\centering
\includegraphics[width=\columnwidth]{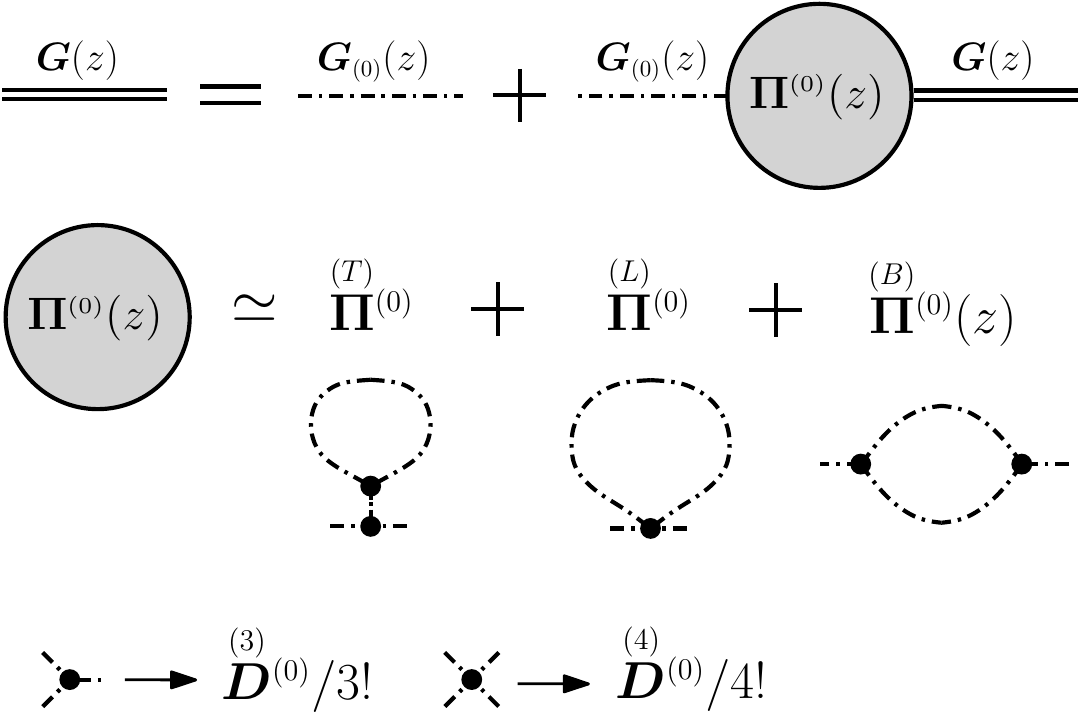}
\caption{Diagrammatic description of the harmonic perturbation theory at the lowest perturbative order, see Eqs.~\eqref{eq:Dyson_harm_1}--\eqref{eq:pert_bubble}. 
The dashed line corresponds to the harmonic propagator. The double solid line corresponds to the full propagator.
Notice that in Fig.~\ref{fig:Feyman_diagr_stat} we have already used a double solid line to indicate $\overset{\ssi}{\bD}_{\ssF}$. This is not casual because later
we will interpret  $\overset{\ssi}{\bD}_{\ssF}$ as the static full propagator (see Eq.~\eqref{eq:staticG_DF}).
Third and fourth order vertices
are associated to $\overset{\sst}{\bD}{}^{\ssz}/3!$ and $\overset{\ssf}{\bD}{}^{\ssz}/4!$, respectively (see definition~\eqref{eq:def_dyn_harm_nth}). 
Sum over internal degrees of freedom is performed.}
\label{fig:Feyman_diagr_pert}
\end{figure}
From the SCHA equations, retaining only the lowest order corrections to the harmonic values $\Rcal^a_{\ssz}$
and $\phi_{ab}$, using the SCHA matrix defined in  Eq.~\eqref{eq:def_D} we obtain (see Eq.~\eqref{eq:pert_DS})
\begin{equation}
\bD^{\ssS}\simeq\bD^{\ssz}+\overset{\ssT}{\bPi}{}^{\ssz}+\overset{\ssL}{\bPi}{}^{\ssz}\mathp
\label{eq:SCHA_Green_perturb_dyn}
\end{equation}
Equivalently, using the SCHA propagator $\bG_{\ssS}(z)$
defined in Eq.~\eqref{eq:relaz_GS_DS}
we can write
\begin{equation}
\overset{\ssi}{\bG}{}^{\ssS}(z)\simeq\overset{\ssi}{\bG}{}^{\ssz}(z)-\overset{\ssT}{\bPi}{}^{\ssz}-\overset{\ssL}{\bPi}{}^{\ssz}\mathv
\label{eq:SCHA_Green_perturb}
\end{equation}
that is
\begin{equation}
{\bG}{}_{\ssS}(z)\simeq{\bG}{}_{\ssz}(z)+{\bG}{}_{\ssz}(z)\left[\overset{\ssT}{\bPi}{}^{\ssz}+\overset{\ssL}{\bPi}{}^{\ssz}\right]
{\bG}{}_{\ssS}(z)\mathp
\label{eq:SCHA_Green_perturb_2}
\end{equation}
At the lowest perturbative order we also have (see Eq.~\eqref{eq:app_per_bubb})
\begin{equation}
\bPi^{\ssS}(0)\simeq\overset{\ssB}{\bPi}{}^{\ssS}(0)\simeq\overset{\ssB}{\bPi}{}^{\ssz}(0)
\label{eq:SCHA_staticself_perturb}
\end{equation}
where $\bPi^{\ssS}(0)$ and $\overset{\ssB}{\bPi}{}^{\ssS}(0)$ are the quantities 
defined in Eq.~\eqref{eq:diagrammatic_static_selfenergy} and Eq.~\eqref{eq:diagrammatic_static_selfenergy_bubble}, respectively.
From Eqs.~\eqref{eq:SCHA_Green_perturb_dyn}--\eqref{eq:SCHA_Green_perturb_2} 
we see that, at the lowest perturbative order,  the SCHA and harmonic propagators
are related through the harmonic loop and tadpole self-energies only~\cite{PhysRevB.91.054304}.  However, from Eq.~\eqref{eq:diagrammatic_static_invSCHAdynmat_00}
 and Eq.~\eqref{eq:SCHA_staticself_perturb} we see that
in order to obtain the SCHA dynamical matrix, defined in Eq.~\eqref{eq:SCHA_dyn_mat}, we need the harmonic static bubble too:
\begin{equation}
{\bD}{}^{\ssF}\simeq{\bD}{}^{\ssz}+\overset{\ssT}{\bPi}{}^{\ssz}+\overset{\ssL}{\bPi}{}^{\ssz}+\overset{\ssB}{\bPi}{}^{\ssz}(0)\mathp
\label{eq:pert_dyna_mat}
\end{equation}
Notice that, in particular, this implies that the term $\overset{\sst}{\bPhi}\bLambda\bTheta\bLambda\overset{\sst}{\bPhi}$ 
in the curvature formula, Eq.~\eqref{eq:compact_expr_curv}, can be discarded at the lowest perturbative order.
In terms of the harmonic propagator defined in Eq.~\eqref{eq:relaz_G0_D0}, the formula~\eqref{eq:pert_dyna_mat} can be written as (Cfr.~Eq.~\eqref{eq:diagrammatic_static_invSCHAdynmat_1})
\begin{align}
&-\overset{\ssi}{\bD}{}_{\ssF}\simeq \bG_{\ssz}(0)\nonumber\\
&\mkern40mu+\bG_{\ssz}(0)\left[\overset{\ssT}{\bPi}{}^{\ssz}+\overset{\ssL}{\bPi}{}^{\ssz}+\overset{\ssB}{\bPi}{}^{\ssz}(0)\right]
(- \overset{\ssi}{\bD}{}_{\ssF})\mathp
\label{eq:static_perturbative}
\end{align}
Equations~\eqref{eq:SCHA_Green_perturb_2} and~\eqref{eq:static_perturbative} are the main SCHA results at the lowest harmonic 
perturbative order. They are represented in diagrammatic form in Fig.~\ref{fig:pert_res}a and in Fig.~\ref{fig:pert_res}b, respectively. 
\begin{figure}
\centering
\includegraphics[width=1.0\columnwidth]{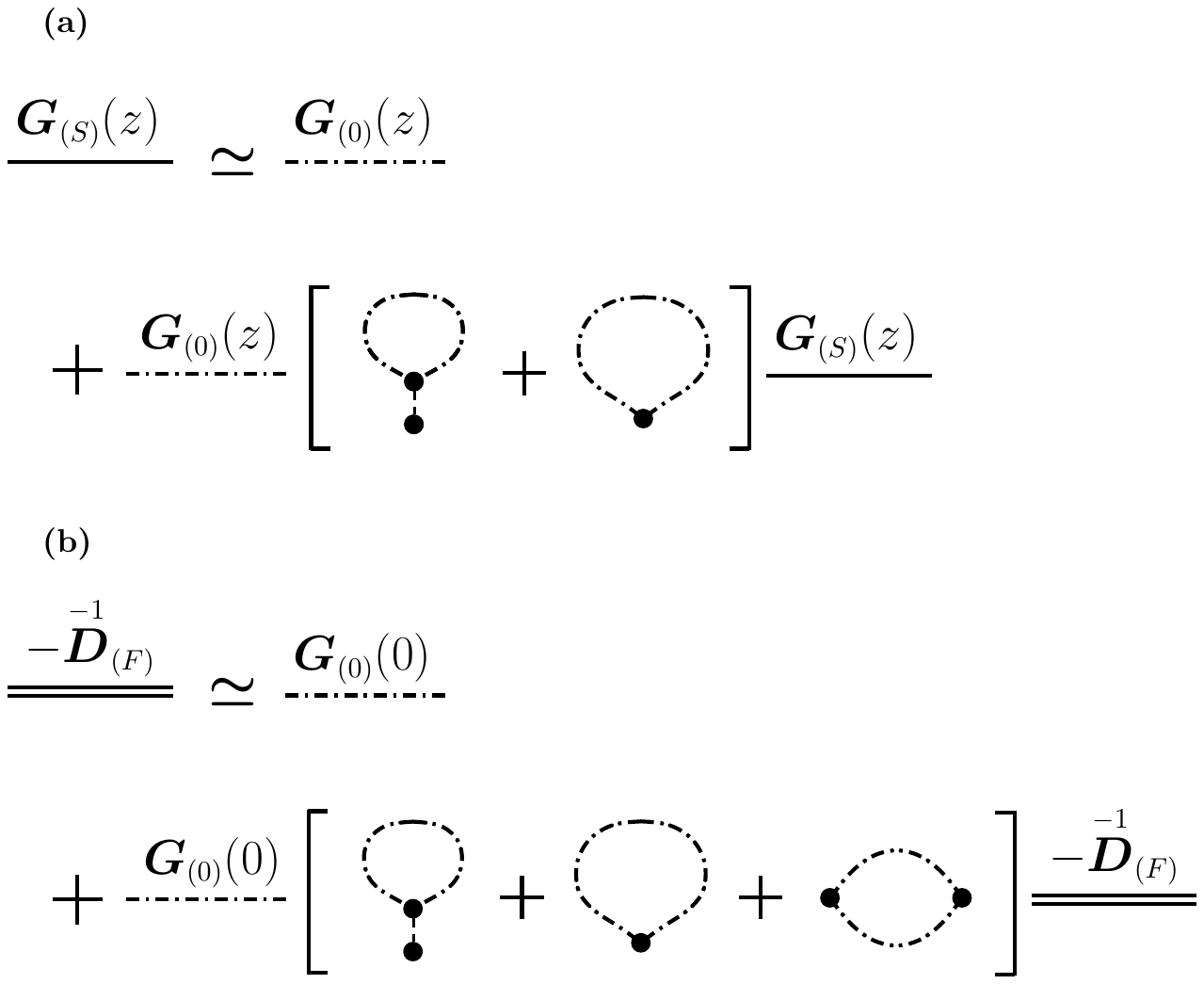}
\caption{Diagrammatic description of the SCHA results at the lowest harmonic perturbative order. Figure a): 
relation between the SCHA and the harmonic propagator, 
Eq.~\eqref{eq:SCHA_Green_perturb_2}.
Figure b): relation between the free energy dynamical matrix within SCHA and the harmonic propagator, Eq.~\eqref{eq:static_perturbative}.}
\label{fig:pert_res}
\end{figure}

It is interesting to observe that, at the lowest perturbative order,
the free energy curvature takes into account only the static harmonic bubble, 
whereas in the full propagator the bubble actually depends on the frequency $z$, as we can see from Eq.~\eqref{eq:perturbative_harmonic_selfenergy}.
This is consistent with the fact that we have developed only a `static' theory 
(obviously, this fact does not have consequences for the tadpole and loop term, because they do not depend on the frequency). 
In the next section we will investigate possible dynamic extensions of the results found thus far. 
%

\section{Ansatz for a dynamic theory}
\label{sec:Anstaz_for_a_dynamic_theory}
In this section we propose a possible `dynamical' extension of the `static' results obtained above. This could be used to interpret
the outcomes of inelastic scattering processes between phonons and external incident particles (typically neutrons)
in the framework of the SCHA approximation. The extension that we are going to present
is reasonable because it returns the expected results in two limits. In the static limit 
it gives results coherent with the ones already obtained for the free energy curvature and
at the lowest perturbative order it gives the correct results already known in literature. 
Nevertheless, it is worthwhile to stress that, at variance with the `static' results, the dynamical extension that we are going
to propose is only an ansatz, reasonable but not based on a rigorous demonstration. 
For that reason it can be considered as a basis for a future rigorous extension of the 
static theory.

Fixed the temperature, and the relative $\Rcal_{\eq}^a$, we consider the full Green function 
$G_{ab}(z)$ for $H$ and the Green function $G^{\ssS}_{ab}(z)$ for $H^{\ssS}$ in the variable $\sqrt{M_a}(R^a-\Rcal_{\eq}^a)$. 
We consider a Dyson-type relation between them:
\begin{equation}
\overset{\ssi}{\bG}(z)=\overset{\ssi}{\bG}{}^{\ssS}(z)-\bPi^{\ssS}(z)\mathv
\label{eq:Green_func_ansatz}
\end{equation}
which is equivalent to
\begin{equation}
{\bG}(z)={\bG}{}_{\ssS}(z)+{\bG}{}_{\ssS}(z)\,\bPi^{\ssS}(z)\,{\bG}(z)\mathv
\label{eq:Green_func_ansatz_1}
\end{equation}
where $\bPi^{\ssS}(z)$ is the SCHA self-energy. The aim of this section is to propose an expression for $\bPi^{\ssS}(z)$.
The first assumption is that its static value, i.e. its value for $z=0$, is given by  Eq.~\eqref{eq:diagrammatic_static_selfenergy}. 
At that level, the symbol used did not have a physical meaning.
Now we are explicitly interpreting it as the static SCHA self-energy. 
Comparing Eq.~\eqref{eq:Green_func_ansatz} to Eq.~\eqref{eq:diagrammatic_static_invSCHAdynmat_0}, this is equivalent to saying that
\begin{equation}
\overset{\ssi}{\bG}(0)=-\bD^{\ssF}\mathp
\label{eq:staticG_DF}
\end{equation}
This is the same kind of relation that exists between the harmonic static Green function and the harmonic dynamical matrix.
Therefore, Eq.~\eqref{eq:staticG_DF} gives a deeper meaning to the consideration in Sec.\ref{sec:phonons_in_the_SCHA}
that $\bD^{\ssF}$ is the anharmonic generalization of the harmonic dynamical matrix. A real pole of the Green function 
corresponds to the energy of a phonon with zero linewidth, i.e. with infinite lifetime. Equation~\eqref{eq:staticG_DF}
means that we observe a phonon with zero energy, i.e. we see a phonon softening and therefore an instability, when 
$\bD^{\ssF}$ has a null eigenvalue. This is exactly the result found in Sec.~\ref{Derivatives_of_F}  and Sec.~\ref{sec:phonons_in_the_SCHA}.
Thus the interpretation of  Eq.~\eqref{eq:diagrammatic_static_selfenergy} as the static SCHA self-energy is 
consistent with the rigorous (static) results obtained for the free energy curvature.

The  subsequent step is to give an expression for the SCHA self-energy at $z$ different from zero.  
As a second part of our hypothesis, we assume for $\bPi^{\ssS}(z)$
the same structure of $\bPi^{\ssS}(0)$, given by Eq.~\eqref{eq:diagrammatic_static_selfenergy} and 
illustrated by the diagrams in Fig.~\ref{fig:Feyman_diagr_stat}b, but readily generalized to any $z$. Therefore it is
\allowdisplaybreaks[0]
\begin{align}
\bPi^{\ssS}(z)&=\overset{\bsst}{\bD}{}^{\ssS}\left(-\frac{1}{2}\,\bchi_{\ssS}(z)\right)\nonumber\\
 &\qquad\times\left[\mathds{1}-\overset{\bssf}{\bD}{}^{\ssS}\left(-\frac{1}{2}\,\bchi_{\ssS}(z)\right)\right]^{-1}\overset{\bsst}{\bD}{}^{\ssS}\mathv
\label{eq:diagrammatic_dynamic_selfenergy}
\end{align}
\allowdisplaybreaks
with
\begin{equation}
\chi_{\ssS}^{abcd}(z)=\frac{1}{\beta}\sum_l\,{G}_{\ssS}^{ac}(i\Omega_l){G}_{\ssS}^{bd}(z-i\Omega_{l})\mathp
\label{eq:dynamic_loop}
\end{equation}
Using standard techniques for Matsubara frequencies summations~\cite{mahan2000many}, 
we obtain an explicit expression for this term:
\begin{align}
&\frac{1}{\beta}\sum_lG_{\ssS}^{ac}(i\Omega_{l})G_{\ssS}^{bd}(z-i\Omega_{l})=\nonumber\\
&\quad\qquad\frac{\hbar^2}{4}\sum_{\mu\nu}\frac{F(z,\omega_{\mu},\omega_{\nu})}{\omega_{\mu}\omega_{\nu}}\,e^a_{\nu}e^b_{\mu}e^c_{\nu}e^d_{\mu}\mathv
\label{eq:sumMatzub_2}
\end{align}
where  $\omega_{\mu}^2$ and $e^a_{\mu}$ are eigenvalues and corresponding eigenvectors of $D^{\ssS}_{ab}$, respectively, and for $z\neq 0$
\begin{align}
F(z,\omega_\nu,\omega_\mu)&=\frac{2}{\hbar}\Biggl[\frac{(\omega_\nu+\omega_\mu)[1+n_\nu+n_\mu]}{(\omega_\nu+\omega_\mu)^2-z^2}\nonumber\\
&\qquad-\frac{(\omega_\nu-\omega_\mu)[n_\nu-n_\mu]}{(\omega_\nu-\omega_\mu)^2-z^2}\Biggr]\mathp
\label{eq:def_F}
\end{align}
The assumption expressed by Eqs.~\eqref{eq:diagrammatic_dynamic_selfenergy},~\eqref{eq:dynamic_loop} is reasonable because at the lowest perturbative limit it gives the correct result. Indeed,
by using the same arguments of Sec.~\ref{sec:Perturbative_limit}, at the lowest perturbative order 
we readily generalize Eq.~\eqref{eq:SCHA_staticself_perturb} to
\begin{equation}
\bPi^{\ssS}(z)\simeq\overset{\ssB}{\bPi}{}^{\ssz}(z)\mathp
\end{equation}
Thus, from Eq.~\eqref{eq:SCHA_Green_perturb} and Eq.~\eqref{eq:Green_func_ansatz} we obtain
\begin{equation}
\overset{\ssi}{\bG}(z)\simeq \overset{\ssi}{\bG}{}^{\ssz}(z)-\overset{\ssT}{\bPi}{}^{\ssz}-\overset{\ssL}{\bPi}{}^{\ssz}-\overset{\ssB}{\bPi}{}^{\ssz}(z)\mathv
\end{equation}
which is the correct perturbative result shown in Eqs.~\eqref{eq:Dyson_harm}  
and~\eqref{eq:perturbative_harmonic_selfenergy}. In conclusion, according to our ansatz, the full Green function $\bG(z)$
is (approximately) given by Eq.~\eqref{eq:Green_func_ansatz_1} with Eqs.~\eqref{eq:diagrammatic_dynamic_selfenergy},~\eqref{eq:dynamic_loop}. 
In that way we obtain a minimal extension of the static theory which reproduces the correct instabilities and gives
the correct results at the lowest perturbative level. By using this formula we can study anharmonic effects in a non perturbative way also for
the dynamic case. In Fig~\ref{fig:dyn_ansatz} we give the diagrammatic expression for our ansatz, the self-energy $\bPi^{\ssS}(z)$ 
being the one in Fig.~\ref{fig:Feyman_diagr_stat}b.
An analogous diagrammatic series has been proposed in Ref.~\citenum{PhysRevLett.24.1424}.
\begin{figure}
\centering
\includegraphics[width=\columnwidth]{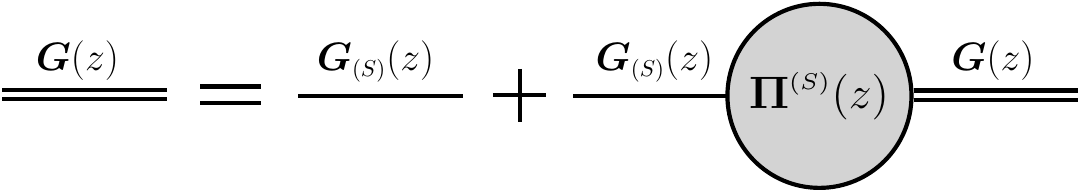}
\caption{Diagrammatic representation of our dynamical conjecture, Eq.~\eqref{eq:Green_func_ansatz_1}. It is the generalization
to $z\neq 0$ of  the static result represented in Fig.~\ref{fig:Feyman_diagr_stat}a. With $\bG$ and $\bG{}_{\ssS}$ we indicate the full Green function and
the SCHA Green function, Eq.~\eqref{eq:relaz_GS_DS}, for the variable $\sqrt{M_a}(\Rcal^a-\Rcal_{\eq}^a)$, respectively.
The SCHA self-energy $\bPi^{\ssS}$ is represented in Fig.~\ref{fig:Feyman_diagr_stat}b.}
\label{fig:dyn_ansatz}
\end{figure}

It is interesting to observe that, inspired by the perturbative result in Eq.~\eqref{eq:perturbative_harmonic_selfenergy},
one could be tempted to naively obtain a dynamic SCHA theory simply by adding 
a dynamic bubble term on top of the standard SCHA results (which, as shown in Eq.~\eqref{eq:SCHA_Green_perturb}, contain only 
tadpole and loop at the lowest perturbative level).
This approach of adding a dynamic bubble has been taken, for example,
in PbTe~\cite{PhysRevLett.112.175501} and PdH~\cite{PhysRevB.87.214303}, 
where the strong anharmonicity induces satellite peaks in the spectral function.
Now we can see that this essentially consists in adopting our ansatz, but discarding
all the terms in $\bPi^{\ssS}(z)$ described by the diagrams of Fig.~\ref{fig:Feyman_diagr_stat}b, except the non-perturbative SCHA dynamic bubble
given in Fig.~\ref{fig:Feyman_diagr_stat}c:
\begin{equation}
\overset{\ssB}{\bPi}{}^{\ssS}(z)=\overset{\bsst}{\bD}{}^{\ssS}\left(-\frac{1}{2}\,\bchi_{\ssS}(z)\right)\overset{\bsst}{\bD}{}^{\ssS}\mathp
\label{eq:sscha_bubble}
\end{equation}
This, in general, is not justified. As long as we consider a non-perturbative situation, there is in principle 
no hierarchy that  allows to discard the other terms. Therefore, the term given by Eq.~\eqref{eq:sscha_bubble} 
has to be considered an incomplete expression for $\bPi^{\ssS}(z)$ and a better choice is to take into account the full 
expression of Eq.~\eqref{eq:diagrammatic_dynamic_selfenergy}. 
Of course, there can be situations in which even if the regime is not perturbative, because the third order is not smaller
than the harmonic term, nevertheless the superior orders are smaller. In that case it would be justified to use Eq.~\eqref{eq:sscha_bubble}
to evaluate $\bPi^{\ssS}(z)$. However, this is a further assumption that, in order to be adopted,
has to be justified case by case.
%
%
\section{Numerical test}
\label{sec:Numerical_test}
In order to give a numerical demonstration of our findings, we apply  
the theory to  a toy model based on the SnTe crystal (an analogous model could be used for GeTe).
SnTe crystallizes at room temperature and ambient pressure in the NaCl-structure (Fm-3m),
called $\beta$-SnTe phase, where two fcc lattices of Sn and Te interpenetrate. 
At low temperature, around $100$ K, it undergoes a phase transition and stabilizes in a rhombohedral structure (R3m),
called $\alpha$-SnTe. The phase transition can be described in terms of a two-step symmetry reduction:
a fixed unit cell polar displacement, between the two fcc, along the [111] cubic direction, which eliminates the inversion center,
and a strain of the unit cell along the cube diagonal~\cite{PSSB:PSSB201248412}. We concentrate on the first distortion.
We define the interatomic potential $V(\bu)$ of the toy-model as a function of the displacements $u^a=R^a-\Rcal_{\ssz}^a$ 
from the equilibrium position of the rock-salt structure $\bRcal_{\ssz}$ and we keep, beyond the quadratic part,
only the anharmonic third and fourth order terms:
\begin{align}
V(\bu)&=\frac{1}{2}\sum_{ab}\phi_{ab}u^a u^b+
\frac{1}{3!}\sum_{abc}\overset{\scriptscriptstyle{(3)}}{\phi}{}_{abc}\,u^a u^bu^c\nonumber\\
&\qquad+\frac{1}{4!}\sum_{abcd}\overset{\scriptscriptstyle{(4)}}{\phi}{}_{abcd}\,u^a u^bu^cu^d\mathp
\end{align}
The harmonic matrix $\phi_{ab}$ has been obtained from first principle calculation 
for SnTe on a 2x2x2 grid of the Brillouin zone (BZ) (details in App.~\ref{app:Toy_model_definition}). 
With the experimental lattice parameter $a_{\scriptscriptstyle{\text{exp}}}=6.312\,\angs$ we do not observe any instability in the total energy 
(i.e. the harmonic matrix is positive-definite). However, a lattice instability appears and increases at $\Gamma\in$ BZ   
as we increase the lattice parameter. Therefore in order to achieve, for explicative purposes, an increased instability at the harmonic level, 
we calculated ab initio the harmonic matrix with a higher lattice parameter: $a_{\scriptscriptstyle{\text{toy}}}=6.562\,\angs$.
Moreover, in order to keep the toy model as simple as possible and focus on the main purpose of the numerical test, 
we ignored the LO-TO splitting at $\Gamma$, which is present in real undoped SnTe samples.
In Fig.~\ref{fig:spec_harm} we show the obtained (harmonic) phonon dispersion along a high-symmetry path of the fcc BZ.
There are imaginary phonons in several points, the optical phonon in $\Gamma$
corresponding to the highest instability. 
\begin{figure}
\centering
\includegraphics[width=\columnwidth]{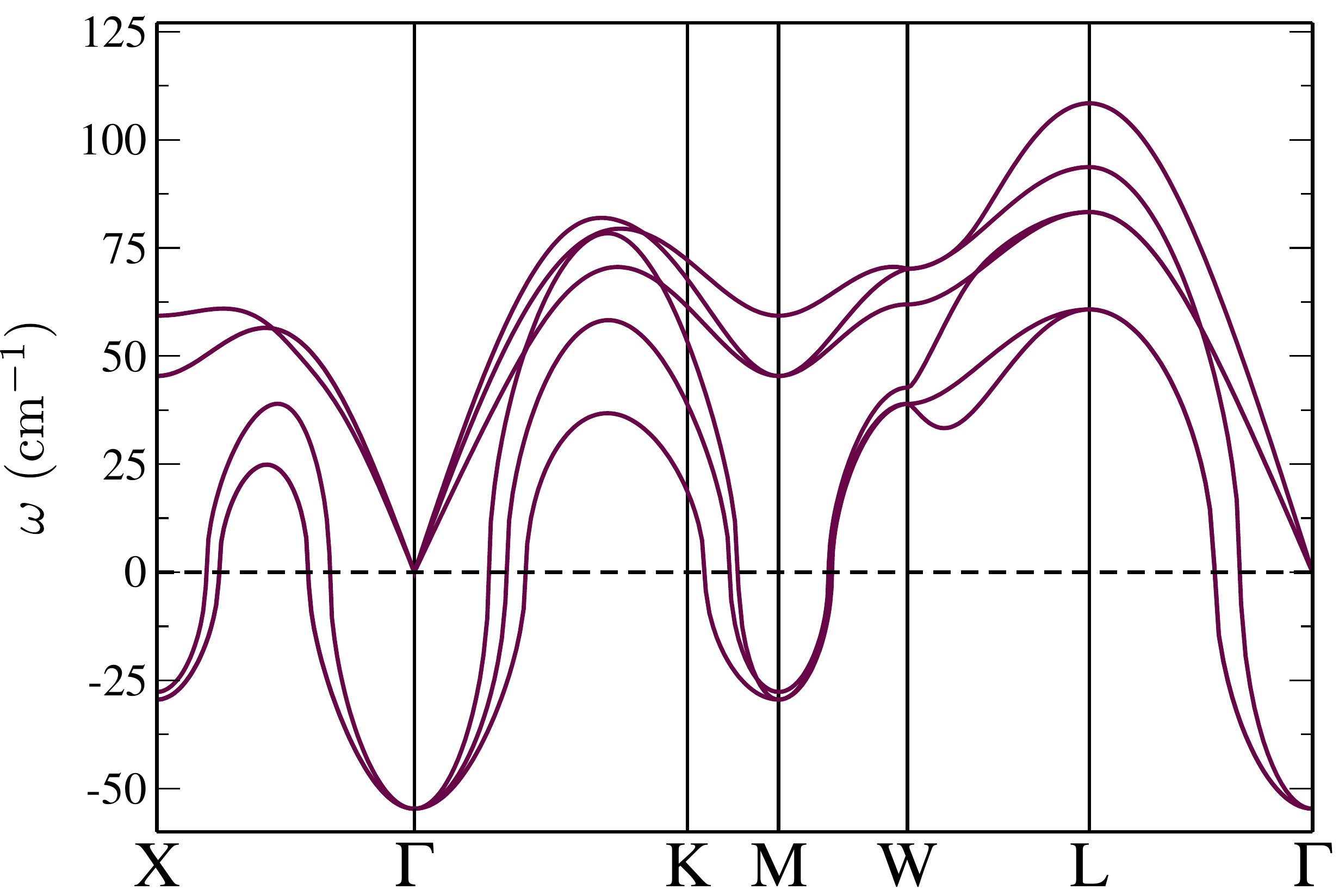}
\caption{Harmonic phonon dispersion for the toy model along an high-symmetry path of the BZ.}
\label{fig:spec_harm}
\end{figure}

For the third and fourth order contributions, we follow the model described in Refs.~\citenum{PhysRevB.90.014308, PhysRevLett.113.105501}.  
We define short-range anharmonic terms by using reciprocal displacements of nearest-neighbor atoms  
(in the rock-salt structure each atom has 6 nearest-neighbors). In particular, as explained in App.~\ref{app:Toy_model_definition},
in our model $\phi^{\sst}_{abc}$ is proportional to a single parameter $p_3$, and $\phi^{\ssf}_{abcd}$
is a linear function of two parameters $p_4$, $p_{4\chi}$. We take
$p_4=7.63\,\text{eV}/\angs^4$, $p_{4\chi}=4.86\,\text{eV}/\angs^4$ and $p_3=6.70\,\text{eV}/\angs^3$.
\subsubsection{Free energy curvature} 
We consider the free energy profile obtained by displacing the atoms in the unit cell along the 
[111] cubic direction. In order to describe this distortion we write the atomic position $\bRcal$ as a function of a
scalar, adimensional parameter $Q$: 
\begin{equation}
\bRcal(Q)=\bRcal_{\ssz}+Q(\bRcal_{\sso}-\bRcal_{\ssz})\mathv
\label{eq:par_Q_def}
\end{equation}
where $\bRcal_{\scriptscriptstyle{(1)}}$ is the configuration corresponding to the minimum of the potential energy 
along the distortion path.
Therefore, $\bRcal(Q)$ is linear, $Q=0$ and $Q=1$ corresponding to the high symmetry phase (Fm-3m)  and
to the low-symmetry energy minimum (R3m), respectively. In Fig.~\ref{fig:F_vs_T} we show $\Delta V(\bRcal(Q))$, the variation 
of the potential (per unit cell) along this distortion path.
This curve depends on $\phi_{ab}$, $p_4$, $p_{4\chi}$. The harmonic term is responsible for the initial decrease whereas
the fourth order term gives the subsequent increase. On the contrary, due to the symmetry of the rock-salt structure, 
the value of $p_3$ is not relevant for the energy pattern (as a matter of fact, the value of $p_3$ does not affect
the energy value of any unit cell configurations).

The free energy along the path has been calculated with the SSCHA on a 2x2x2 supercell.
Fixed $\bRcal$ and the temperature, the GB functional $\Fcal[\displaystyle\rhotrial_{\bRcal,\bvarPhi}]$ has been minimized
with respect to $\bvarPhi$ as described in Ref.~\citenum{PhysRevLett.111.177002,PhysRevB.89.064302}. 
In Fig.~\ref{fig:F_vs_T} we show a complete variation path for the free energy $\Delta F(\bRcal(Q))$ at three temperatures.
For reasons that will be clear in a while, we studied also the case without third order ($p_3=0$).
However, it is interesting to remark that at $Q=0$ the SCHA result is independent from $p_3$. 
\begin{figure}
\centering
\includegraphics[width=\columnwidth]{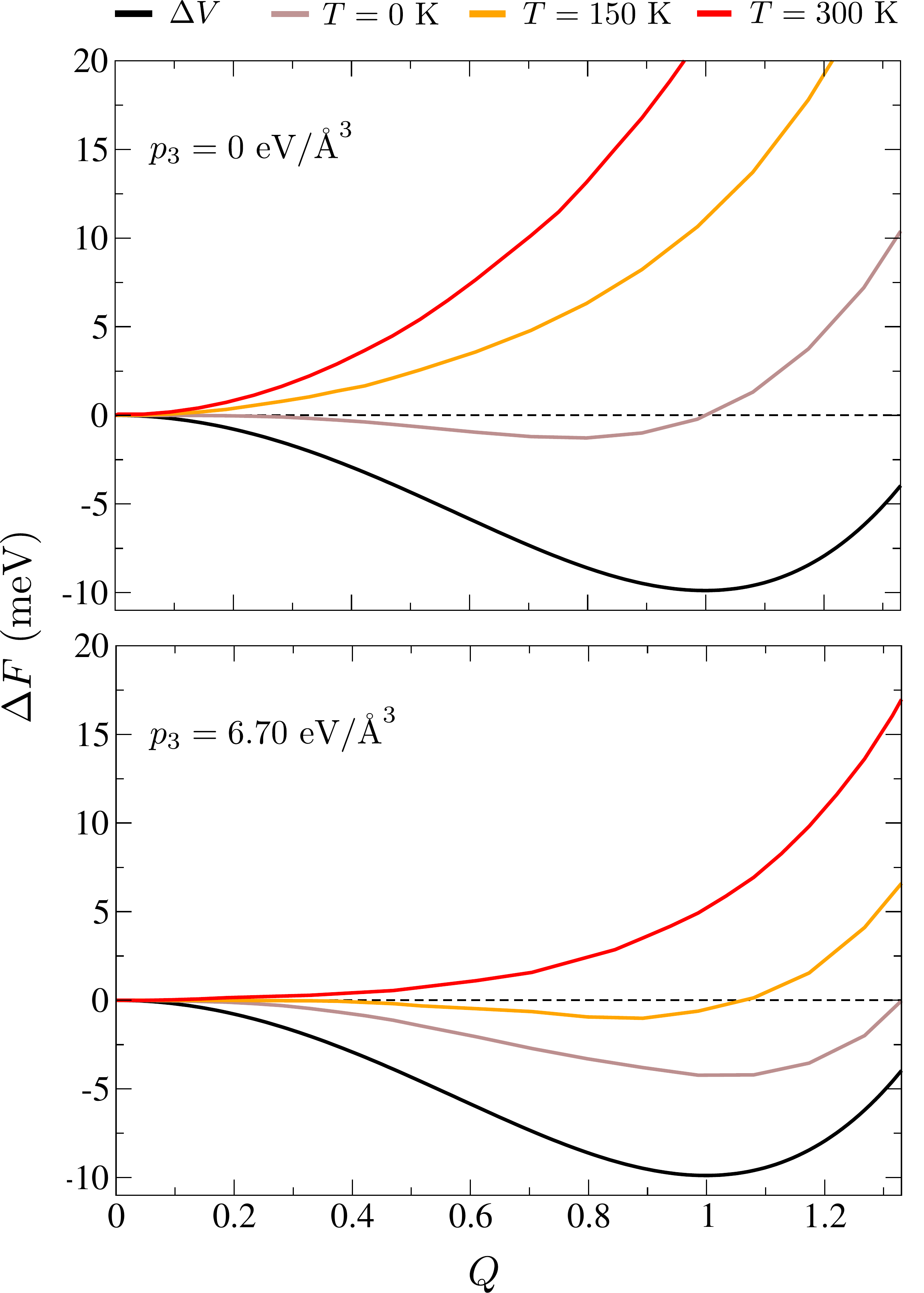}
\caption{Variation of the free energy,
at three temperatures ($0$K, $150$K, $300$K), as a function of the atomic displacement of Eq.~\eqref{eq:par_Q_def}. 
Top panel: without third order. Bottom panel: with third order, $p_3=6.70$ eV/$\angs^3$.
Vertical axis: variation of the free energy (per unit cell) with respect to the value in the undistorted position, 
$\Delta F(Q)=F(Q)-F(0)$, in meV. Horizontal axis: order parameter $Q$.
In the two plots the (temperature independent) variation of the potential energy $\Delta V(Q)=V(Q)-V(0)$ is also shown.}
\label{fig:F_vs_T}
\end{figure}

A first remarkable, somewhat counterintuitive, conclusion can be deduced from the results of Fig.~\ref{fig:F_vs_T}.
While the potential energy path $V(Q)$ is independent from $p_3$, at given temperature
the two free energy paths $F(Q)$ obtained with $p_3=0$ and $p_3=6.70\,\text{eV}/\angs^3$ are considerably different. 
This has important consequences. The presence or not of a second order phase transition and, when there is such a transition, 
the transition temperature $T_c$ and the low-symmetry equilibrium configuration $\bRcal_{\eq}$ for
$T<T_c$ are properties which cannot be inferred from the potential energy profile. 
\begin{figure}
\centering
\includegraphics[width=1.12\columnwidth]{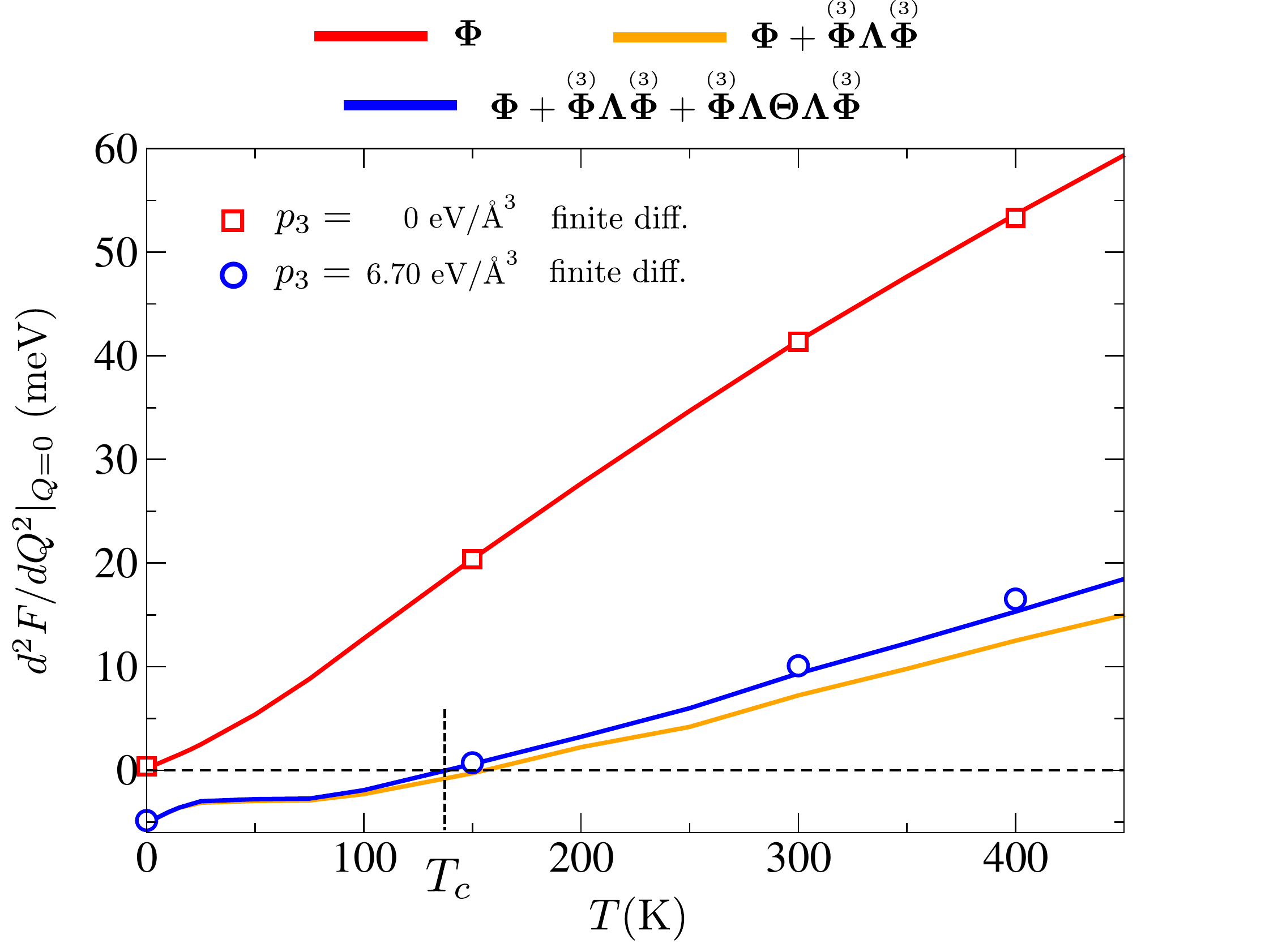}
\caption{Curvature of the free energy in the high-symmetry phase, $d^2F/dQ^2|_{Q=0}$, as a function of temperature $T$.
The transition temperature $T_c$ is around $140$ K. Lines: curvature calculated
with Eq.~\eqref{eq:application_curvature}. Three different quantities, contracted with $d\bRcal/dQ$, are shown (see legend).
Dots: curvature, with and without third order, estimated by finite difference from the values of the free energy
calculated for several configurations around the high-symmetry phase.}
\label{fig:d2FdQ2_finite_diff}
\end{figure}

From the values of the free energy computed near $Q=0$, the curvature in the origin $d^2F/dQ^2|_{Q=0}$ 
has been evaluated by finite difference. The results at four temperatures are shown with dots in Fig.~\ref{fig:d2FdQ2_finite_diff}.
We compare these values with the curvature in $Q=0$ calculated
by contracting $d\bRcal/dQ$  with the formula for $\partial^2\Fcal/\partial\bRcal\partial\bRcal$ of
Eq.~\eqref{eq:compact_expr_curv}:
\allowdisplaybreaks[0]
\begin{align}
\frac{d^2F}{dQ^2}&=\frac{d\bRcal}{dQ}\frac{\partial^2 F}{\partial\bRcal\partial\bRcal}\frac{d\bRcal}{dQ}\\
&=\frac{d\bRcal}{dQ}\bPhi\frac{d\bRcal}{dQ}
+\frac{d\bRcal}{dQ}
   \overset{\sst}{\bPhi}\bLambda\overset{\sst}{\bPhi}                   
 \frac{d\bRcal}{dQ}\nonumber\\
&\mkern+156mu+\frac{d\bRcal}{dQ}
   \overset{\sst}{\bPhi}\bLambda\bTheta\bLambda\overset{\sst}{\bPhi}
  \frac{d\bRcal}{dQ}\mathp
  \label{eq:application_curvature}
\end{align}
\allowdisplaybreaks
This formula is evaluated at $Q=0$.
In order to be consistent with the finite differences result, 
all the ingredients
have been calculated by using the SSCHA on a 2x2x2 supercell.
Once the SSCHA minimization at $Q=0$ has been completed and the converged
value for $\bPhi(\bRcal_{\ssz})$ has been obtained, $\overset{\sst}{\bPhi}(\bRcal_{\ssz})$ and 
$\overset{\ssf}{\bPhi}(\bRcal_{\ssz})$ have been computed using Eq.~\eqref{Eq:stoc_avg_Vbb_impl}. 
For each temperature, we used the converged value of $\bPhi(\bRcal_{\ssz})$ to generate the population used to
compute the averages. Therefore, in this case it is 
$\rhotrial{}_{\ssin}(\bR)=\rhotrial_{\bRcal,\bPhi}(\bR)$.
Notice that, as explained in Sec.~\ref{sec:Stochastic_implementation}, since the calculation has been performed in
a stationary point of the free energy,
the term $\Bavg{\f_i}_{\rhotrial_{\bRcal,\bPhi}}$ on the right-hand side of Eq.~\eqref{eq:def_fbb} is zero.

For explicative purposes we have used Eq.~\eqref{eq:compact_expr_curv} to express the curvature. 
Thus we have three terms in Eq.~\eqref{eq:application_curvature}
and in Fig.~\ref{fig:d2FdQ2_finite_diff} we plot three lines to show their different contributions.
The term  obtained from $\bPhi$ does not depend on the value of $p_3$, whereas the other two terms depend quadratically on $p_3$. 
As a consequence, $[d\bRcal/dQ\,\bPhi\,d\bRcal/dQ]_{Q=0}$ gives the curvature
in the high-symmetry phase when the third order is absent. 
This is confirmed, within the statistical error ($\simeq 2\,\text{meV}$), 
by comparing the red curve and the red dots in Fig.~\ref{fig:d2FdQ2_finite_diff}.
For $p_3=6.70\,\text{eV}/\angs^3$ the other two terms 
$[d\bRcal/dQ\,\overset{\sst}{\bPhi}\bLambda\overset{\sst}{\bPhi}\,d\bRcal/dQ]_{Q=0}$
and
$[d\bRcal/dQ\,\overset{\sst}{\bPhi}\bLambda\bTheta\bLambda\overset{\sst}{\bPhi}\,d\bRcal/dQ]_{Q=0}$
are necessary in order to obtain the curvature. 
This is confirmed by comparing the blue curve and the blue dots in Fig.~\ref{fig:d2FdQ2_finite_diff}.
As explained in Sec.~\ref{sec:Perturbative_limit}, only at the lowest perturbative order it is possible to 
neglect the term $[d\bRcal/dQ\,\overset{\sst}{\bPhi}\bLambda\bTheta\bLambda\overset{\sst}{\bPhi}\,d\bRcal/dQ]_{Q=0}$. 
Indeed, in Fig.~\ref{fig:d2FdQ2_finite_diff} 
we show with a yellow line the curvature computed with only the SCHA matrix and the bubble. In this case the difference with respect 
to the correct value increases with temperature and, even if small, it is already beyond the statistical error at $250$ K. 

In this section we have numerically proved the correctness of Eq.~\eqref{eq:compact_expr_curv}. 
We conclude with a consideration. As already stressed in Sec.~\ref{Derivatives_of_F}, the first term
of Eq.~\eqref{eq:application_curvature} is always positive. Therefore, with $p_3=0$ it is possible to observe 
only a first-order phase transition within the SCHA approximation. 
With $p_3=6.70\,\text{eV}/\angs^3$, the plot in Fig.~\ref{fig:d2FdQ2_finite_diff} shows that the
free energy curvature in $Q=0$ changes sign for $T\simeq140$ K. However, in Fig.~\ref{fig:F_vs_T}
we see that at $T=150\,K$ the free energy has already developed a lower minimum 
in $|Q|\simeq0.9$. As a consequence, the toy model studied undergoes 
a first order phase transition even with $p_3$ different from zero.
\subsubsection{Phonons} 
In this section we apply the concept of free energy dynamical matrix defined in Sec.~\ref{sec:phonons_in_the_SCHA}.
To be precise, fixed the temperature,  we compute the second derivative of the free energy in $\bRcal_{\ssz}$, 
divided by the square root of the masses. Notice that, properly speaking, this is $\bD^{\ssF}$ 
only for $T>T_c$, when $\bRcal_{\eq}$ is equal to $\bRcal_{\ssz}$, because
at temperatures below the transition temperature $\bRcal_{\eq}$ departs from $\bRcal_{\ssz}$. 
Nevertheless, for explicative purposes and having this caveat in mind, we will use the same symbol even at $T<T_c$ .

The matrix $\bD^{\ssF}$ is given by the matrix $\bD^{\ssS}$ plus the static self-energy $\bPi^{\ssS}(0)$ which, in turn,
is made of the bubble term $\overset{\ssB}{\bPi}{}^{\ssS}(0)$ plus other factors, negligible at the lowest perturbative level
(see Eqs.~\eqref{eq:diagrammatic_static_invSCHAdynmat_00},~\eqref{eq:diagrammatic_static_selfenergy},~\eqref{eq:diagrammatic_static_selfenergy_bubble},~\eqref{eq:SCHA_staticself_perturb}). 
Since we are considering a crystal, we exploit the lattice translational symmetry and we write the dynamical matrices
in the unit cell as a function of the quasimomentum. In Fig.~\ref{fig:spectrum} we plot the spectrum of these matrices along
a high-symmetry path of the BZ. We consider two temperatures.  The matrix $\bD^{\ssS}$ coincides with the free energy
dynamical matrix $\bD^{\ssF}$ when the third order is absent. Since $\bD^{\ssS}$ is positive-definite, the spectrum is always positive.
However, with $p_3=6.70\,\text{eV}/\angs^3$  the dynamical matrix $\bD^{\ssF} $ is qualitatively different from $\bD^{\ssS}$.
Below the transition temperature the phonon spectrum becomes imaginary (negative eigenvalue) in $\Gamma$, and only in that point.
The other instabilities that were present in the harmonic phonon spectrum, Fig.~\ref{fig:spec_harm}, have been washed 
out by the zero-point energy and anharmonicity. Notice that in this case the comparison between the harmonic and the free energy dynamical
matrix is particularly meaningful because, for symmetry reasons, both are computed in the same point $\bRcal_{\ssz}$.
In Fig.~\ref{fig:spectrum}  we also show the spectrum obtained by adding only the bubble $\overset{\ssB}{\bPi}{}^{\ssS}(0)$ 
to $\bD^{\ssS}$. From the results shown in Fig.~\ref{fig:d2FdQ2_finite_diff} we expected  in $\Gamma$ a very small 
difference between the full formula and the one
considering only the bubble. However, here we have a more complete picture. As we can see, in other points 
of the BZ the spectrum is more affected by the presence of terms beyond the bubble.
For example, at $400$ K for the 5th mode in $L$, the terms beyond the bubble change the spectrum around $13\,\text{cm}^{-1}$.

\begin{figure*}
\centering
\includegraphics[width=\textwidth]{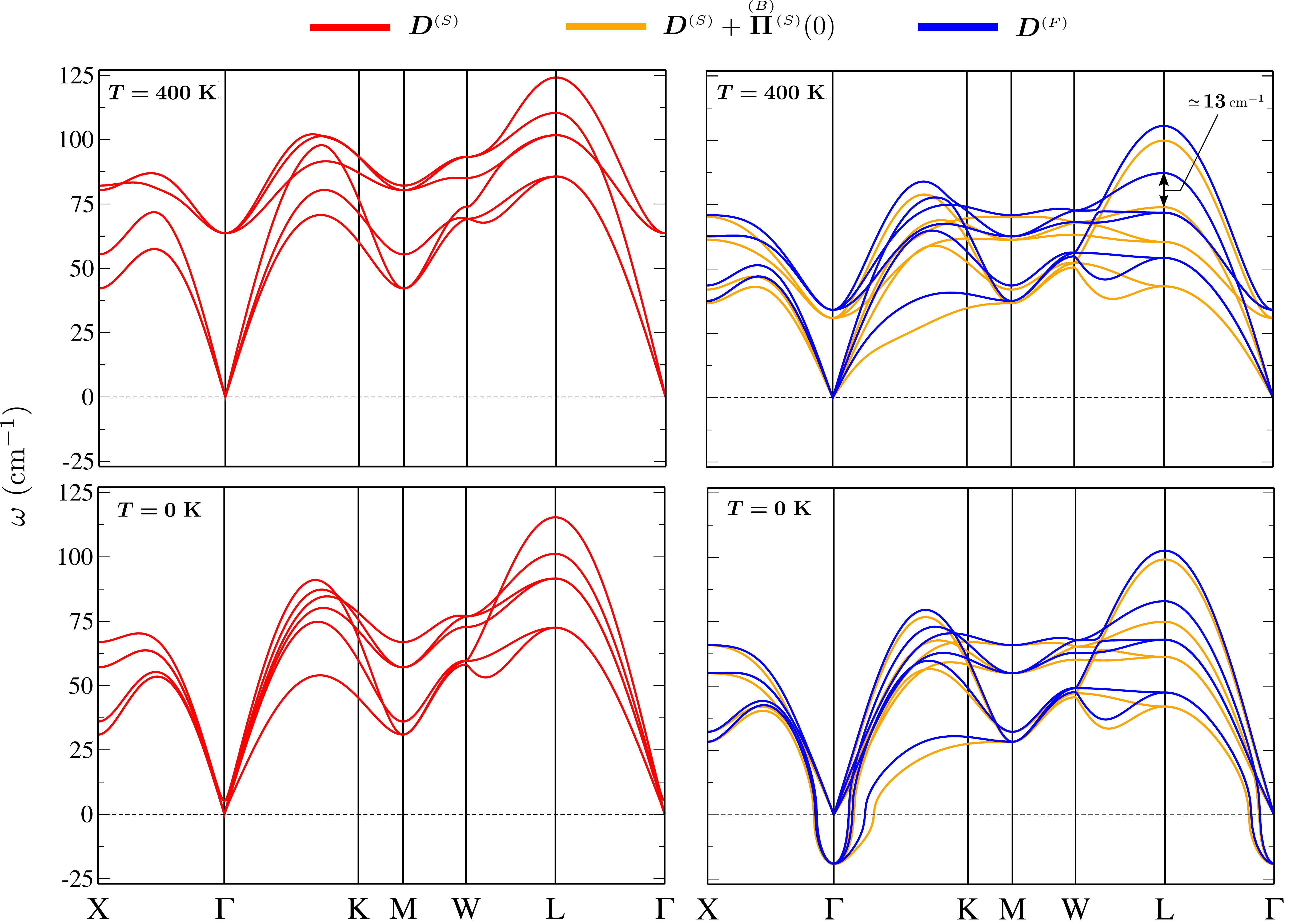}
\caption{(color online) Spectrum along a BZ high-symmetry line of the
matrix $\bD^{\ssS}$, independent of $p_3$, and of the free energy dynamical matrix $\bD^{\ssF}=\bD^{\ssS}+\bPi^{\ssS}(0)$
for $p_3=6.70\,\text{eV}/\angs^3$, at two temperatures. When the third order is equal to zero, $\bD^{\ssF}$ is equal to $\bD^{\ssS}$.
For $p_3=6.70\,\text{eV}/\angs^3$ the system shows phonon softening, i.e. instability, in $\Gamma$. 
The spectrum obtained by adding only the bubble $\overset{\ssB}{\bPi}{}^{\ssS}(0)$ to  $\bD^{\ssS}$ is also shown. 
At $400$ K it is marked the considerable difference, around $13\,\text{cm}^{-1}$, between the energies of the 
5th mode in $L$ obtained with $\bD^{\ssS}+\overset{\ssB}{\bPi}{}^{\ssS}(0)$ and with $\bD^{\ssF}$.}
\label{fig:spectrum}
\end{figure*}
\subsubsection{Convergence} 
Since for our test we used a toy model, i.e. an analytic potential, we could evaluate the averages
using populations of very big size at small computational cost. 
However, in view of first principle applications for realistic materials, 
we carefully performed convergence tests of the curvature formula with respect to the population size $N_{\Ical}$.

First, we tested the convergence of $d^2F/dQ^2|_{Q=0}$ at various temperatures. 
As said, for each temperature we calculated the curvature using the converged value of $\bPhi(\bRcal_{\ssz})$ to generate the population used
to compute the averages in Eq.~\eqref{Eq:stoc_avg_Vbb_impl}. 
As shown in the upper left-hand panel of Fig.~\ref{fig:conv_test}, the convergence can be considered reached with $N_{\Ical}=10^{4}$. 
However, it is worthwhile to say that, in general, fitting the values of the curvature versus temperature with
a polynomial allows to wash out part of the stochastic noise and obtain good estimations for $T_c$ with smaller populations.
In this case, for example, fitting with a $4$th degree polynomial the results obtained with $N_{\Ical}=10^3$ 
gives a value for $T_c$ which is only $9$ K smaller than the converged one. 

As we have seen in Fig.~\ref{fig:d2FdQ2_finite_diff}, for $d^2F/dQ^2|_{Q=0}$ the terms beyond the bubble, which depend on $\overset{\ssf}{\bPhi}$,
have a limited relevance. For that reason,  we performed an analogous convergence test for the frequency of the $5$th mode in $L$ of $\bD^{\ssF}$. Indeed, 
as shown in Fig.~\ref{fig:spectrum}, for that specific mode the terms beyond the bubble play a non negligible role
in the determination of the spectrum. Therefore, this quantity  is particularly significant 
to analyze the convergence of the different terms comprising the curvature formula.
Here, as in the previous paragraph, with $\bD^{\ssF}$ we are indicating the curvature of the free energy in $\bRcal_{\ssz}$ divided by the
square root of masses, even at temperatures below $T_c$. 
As shown in the upper right-hand panel of Fig.~\ref{fig:conv_test}, also in this case the convergence can be considered reached with $N_{\Ical}=10^{4}$.
However, the absolute stochastic error is already smaller than $3\,\text{cm}^{-1}$ with $N_{\Ical}=10^3$.

It is interesting to see how the two terms $\overset{\sst}{\bPhi}$
and $\overset{\ssf}{\bPhi}$ affect the convergence, separately.
To that end, we plot in the the other panels of Fig.~\ref{fig:conv_test} the curvature and the frequency of the chosen mode, versus temperature,
obtained once with $\overset{\sst}{\bPhi}$ computed with different population sizes $N_{\Ical}$
but with $\overset{\ssf}{\bPhi}$ fixed to the converged value (obtained with a population of $10^5$ elements),
and in the other case the inverse. The conclusion is that the total convergence is affected
in a similar way from the two tensors $\overset{\sst}{\bPhi}$ and $\overset{\ssf}{\bPhi}$. 
This could be surprising since the 4th order tensor $\overset{\ssf}{\bPhi}$ is obtained by averaging a quantity that depends three times on the displacements, whereas
for the 3rd order tensors $\overset{\sst}{\bPhi}$ it is averaged a less complicated quantity which depends only two times on the displacements.
However, it has to be considered that in the curvature formula $\overset{\ssf}{\bPhi}$ is fully contracted (at variance with $\overset{\sst}{\bPhi}$).
Indeed, the random fluctuations on the single components tend to cancel each other and, thus,
the convergence of a contracted tensor is expected to be faster than the convergence of a single tensor component. 
\begin{figure*}
\centering
\includegraphics[width=\textwidth]{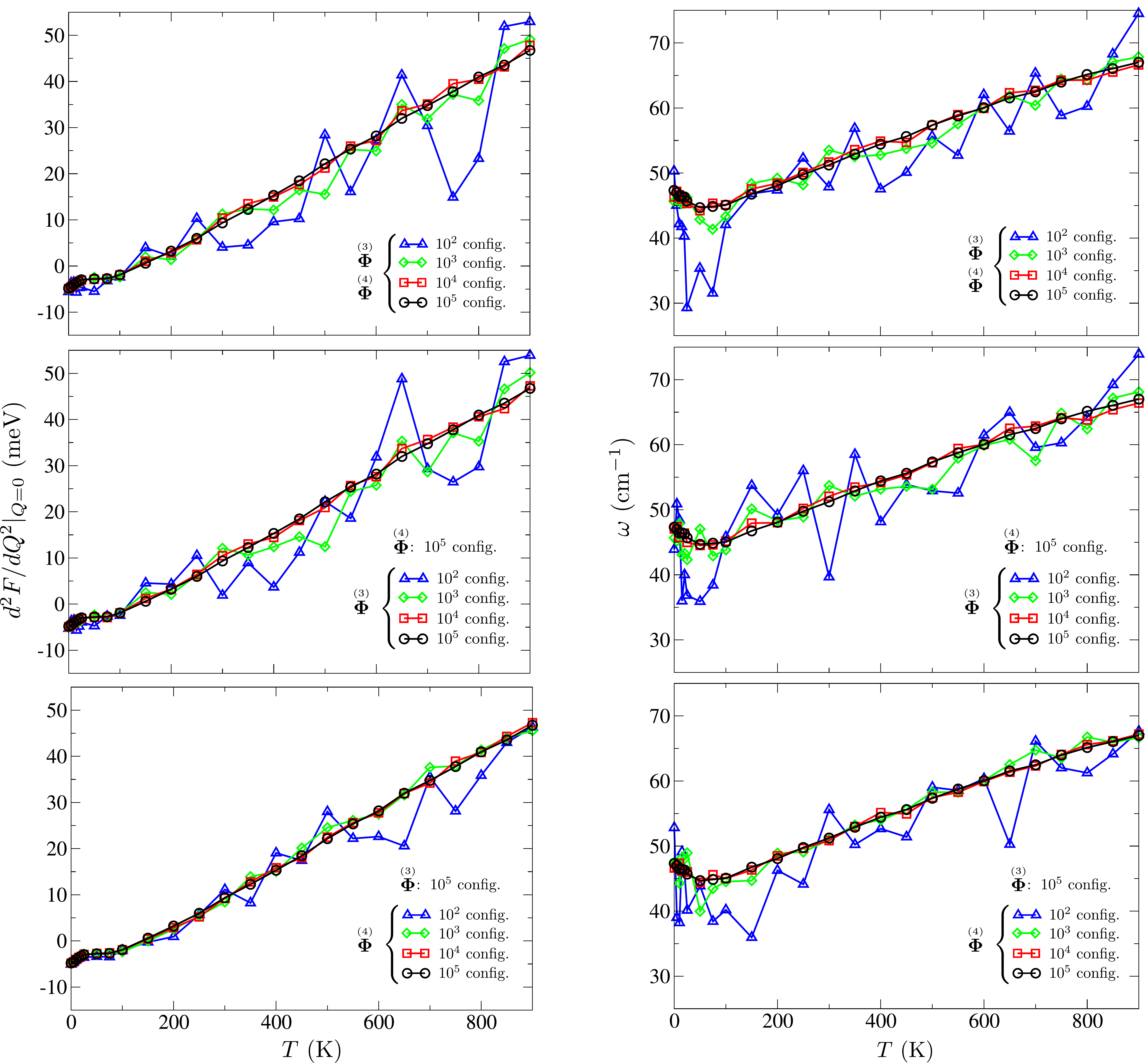}
\caption{(Color online) Convergence test. 
Left-hand column: curvature in the high-symmetry phase, for the distortion considered,
as a function of the temperature. Right-hand column: frequency of the 5th mode in $L$ (also highlighted 
in Fig.~\ref{fig:spectrum}) as a function of temperature. For this mode the effect of the terms beyond the bubble is not negligible.
Populations of different size are used  to compute the tensors $\overset{\sst}{\bPhi}$ and $\overset{\ssf}{\bPhi}$
with Eq.~\eqref{Eq:stoc_avg_Vbb_impl}.
Upper row panels: different population sizes are used to compute both tensors. 
With population of $10^4$ elements the result can be considered converged within the statistical error. 
Second row (third row) bottom panels: different population sizes are used to compute only
$\overset{\sst}{\bPhi}$($\overset{\ssf}{\bPhi}$) whereas $\overset{\ssf}{\bPhi}$($\overset{\sst}{\bPhi}$) 
is computed with $10^5$ elements. In the two cases the convergence trend is similar.}
\label{fig:conv_test}
\end{figure*}

\section{Conclusions}
\label{sec:conclusions}
In this work we present an approach to study structural second order phase transitions in 
molecules and solids within the self-consistent harmonic approximation. 
The developed method  allows to estimate  transition temperature and instability modes. 
It is based on the analytic formula giving the second derivative of the SCHA free energy 
with respect to the average atomic positions. 
The Hessian of the SCHA free energy is also expressed in terms of thermal averages of forces and displacements.
Therefore, the method is suitable for a stochastic implementation in conjunction with any energy-force engine.
{Considering} a configuration, it allows to calculate directly the free energy curvature
{once} the SSCHA calculation has been performed in that point. As a consequence,
it permits to avoid the very computational demanding finite difference approach of computing
the curvature through several SSCHA calculations for different configurations{~\cite{Errea81}}. 
Moreover, the imposition of symmetries on the result reduces the statistical noise and speeds up
statistical convergence with respect to the population size used to compute the averages
with the {importance} sampling technique. 

The efficiency of this method makes it ideal to be used  
in conjunction with first principle energy-force engines to study realistic materials,
{such as ferroelectrics or CDW materials}.
With the curvature formula it is possible to find the instabilities of a general condensed matter
system. In particular, the method is especially convenient for crystals, since
by exploiting the lattice translational symmetry and the 
Fourier interpolation technique it is possible 
to find distortions lowering the free energy with any modulation in space (e.g.
periodic on large supercells or even incommensurate)
with SSCHA {calculations} performed on supercells of moderate size.
In order to demonstrate our findings, numerical tests have been performed on a toy model.
The results confirm both correctness of the theory and numerical efficiency of the implemented method.

In addition to its practical utility, the developed theory sheds light on several fundamental
aspects of the SCHA. In particular, the role of the auxiliary effective quadratic {Hamiltonian} is clarified.
It is shown that the SCHA matrix is only a term of the free energy Hessian and, in general, 
it does not {define} an anharmonic dynamical matrix.
On the contrary, an anharmonic temperature dependent, free energy based, dynamical matrix is obtained
through the free energy curvature. It generalizes the temperature independent harmonic dynamical matrix
and defines temperature dependent anharmonic phonons.

The theory developed for the SCHA free energy curvature is static, as it does not take into account 
any dynamical effects. Inspired by a perspicuous diagrammatic interpretation of the results, 
we propose a tentative minimal dynamic extension of the static theory in order to associate 
{spectral functions with}
anharmonic phonons and interpret the results of scattering processes in a full non-perturbative way.
{Similarly, the dynamic theory allows to calculate phonon lifetimes in the non perturbative 
limit}.
At variance with the curvature formula, the suggested dynamic extension is not based on a rigorous demonstration. 
Nevertheless, it is expected to give good results, because it is correct in both static limit and lowest 
perturbative limit, and thus it opens the way to further theoretical developments and interesting applications.
\section*{ACKNOWLEDGMENTS}
The authors acknowledge support from the Graphene Flagship.
I.E. acknowledges financial support from the
Spanish Ministry of Economy, Industry, and Competitiveness (Grant No. FIS2016-76617-P).
M.C. acknowledges support from Agence Nationale de la Recherche under contract ANR-13-IS10-0003-01, from the Graphene Flagship, 
PRACE for awarding us access to resource on Marenostrum at BSC and the computer facilities provided
by CINES, IDRIS, and CEA TGCC (Grant EDARI No.~2017091202).
\appendix
\section{Proofs of the SCHA method}
\label{app:SCHA_proofs}
In this appendix we give an explicit demonstration of the SCHA self-consistent equation, Eq.~\eqref{eq:SCHA_matrix_def},
and of the expression for the first and second derivative of the SCHA free energy, 
Eq.~\eqref{eq:first_derivative_F} and Eq.~\eqref{eq:sec_der_F_1}, respectively. 
We find convenient to use a notation slightly different from the one used in the main text.
Considering a trial harmonic matrix $\varPhi_{ab}$ we define three matrices from it:
the matrix $\Dcal_{ab}=\varPhi_{ab}/\sqrt{M_aM_b}$;  the matrix $\Ecal^{ab}=\sum_{\mu}\xi^2(\omega^2_\mu)\,e^a_{\mu}e^b_{\mu}$, where
$\omega_{\mu}^2$ and $e^a_{\mu}$ are eigenvalues and eigenvectors of $\Dcal_{ab}$, respectively, 
and $\xi^2(\omega^2_\mu)=\hbar(1+2n_{\mu})/2\omega_{\mu}$, where $n_{\mu}=1/(e^{\beta\hbar\omega_{\mu}}-1)$
is the bosonic average occupation number; and the matrix $\varPsi^{ab}=\Ecal^{ab}/\sqrt{M_aM_b}$.
Introducing the diagonal mass matrix $M_{ab}=\delta_{ab}M_a$, we can summarize these definitions in the compact form
\begin{subequations}\begin{align}
&\bDcal=\bM^{-\frac{1}{2}}\bvarPhi\bM^{-\frac{1}{2}}\label{eq:def_matrici_utili_1}\\
&\bEcal=\xi^2(\bDcal)\label{eq:def_matrici_utili_2}\\
&\bvarPsi=\bM^{-\frac{1}{2}}\bEcal\bM^{-\frac{1}{2}}\label{eq:def_matrici_utili_3}\mathp
\end{align}\label{eq:def_matrici_utili}\end{subequations}
In this notation the matrix $\bUpsilon$ of Eq.~\eqref{eq:def_Upsilon}
coincides with the inverse of $\bvarPsi$, which we indicate with the symbol $\overset{\ssi}{\bvarPsi}$.
Thus the thermal average of an observables  $\mathds{O}(\bR)$ that is  function only of the position
is given by~\cite{doi:10.1063/1.4829836}
\begin{equation}
\Bavg{\Ods}_{{\displaystyle\rhotrial_{\bRcal,\bvarPhi}}}=
\frac{1}{\sqrt{\text{det}\,(2\pi\bvarPsi)}}
\int\Ods(\bRcal+\bu)e^{-\frac{1}{2}\bu\overset{-1}{\bvarPsi}\bu}\,d\bu\mathp
\label{eq:app_avg_formula}
\end{equation}
For the subsequent derivation it is convenient to perform the change of variable
\begin{equation}
u^a=\sum_{\mu}L^a_{\mu}\,y^{\mu}\qquad\text{with }\quad L^a_{\mu}=\frac{e^a_{\mu}}{\sqrt{M_a}}\xi_{\mu}\mathv
\label{eq:change_var}
\end{equation}
where we have introduced the compact notation $\xi^2_{\mu}=\xi^2(\omega_\mu^2)$.
With that change of variable, the average is written as a Gaussian integral:
\begin{equation}
\Bavg{\Ods}_{{\displaystyle\rhotrial_{\bRcal,\bvarPhi}}}=\int\Ods(\bRcal+\bL\by)\,[dy]\mathv
\label{eq:app_int_new_var}
\end{equation}
where
\begin{equation}
[dy]=\prod_{\mu=1}^{3\Nat}\,\frac{e^{-\frac{(y^{\mu})^2}{2}}}{\sqrt{2\pi}}\,dy^{\mu}\mathp
\end{equation}
We now demonstrate the following two relations:
\begin{subequations}
\begin{align}
&\frac{\partial}{\partial \Rcal^a}\Bavg{\Ods}_{{\displaystyle\rhotrial_{\bRcal,\bvarPhi}}}=
		\avg{\frac{\partial\Ods}{\partial R^a}}_{{\displaystyle\rhotrial_{\bRcal,\bvarPhi}}}\label{eq:der_obs_1}\\
&\frac{\partial}{\partial\varPhi_{ab}}\Bavg{\Ods}_{{\displaystyle\rhotrial_{\bRcal,\bvarPhi}}}
=\frac{1}{2}\sum_{cd}\frac{\partial\varPsi^{cd}}{\partial\varPhi_{ab}}\avg{\frac{\partial^2\Ods}{\partial R^c\partial R^d}}_{{\displaystyle\rhotrial_{\bRcal,\bvarPhi}}}\,\,.\label{eq:app_der_obs_2}
\end{align}\label{eq:app_der_obs}\end{subequations}
From the explicit expression of the average, the first identity is trivially obtained. 
In order to demonstrate the second one, we use integration by parts:
\begin{align}
&\frac{\partial}{\partial\varPhi_{ab}}\Bavg{\Ods}_{{\displaystyle\rhotrial_{\bRcal,\bvarPhi}}}=
 \sum_c\int\frac{\partial\Ods}{\partial R^c}(\bRcal+\bL\by)\nonumber\\
 &\mkern210mu\times\sum_{\mu}\frac{\partial L^{c}_{\mu}}{\partial\varPhi_{ab}}y^{\mu}\,[dy]\nonumber\\
&\qquad=-\sum_c\sum_{\mu}\frac{\partial L^{c}_{\mu}}{\partial\varPhi_{ab}}\int\frac{\partial\Ods}{\partial R^c}(\bRcal+\bL\by)\,\frac{\partial [dy]}{\partial y^{\mu}}\nonumber\\
&\qquad=\sum_{cd}\sum_{\mu}\frac{\partial L^{c}_{\mu}}{\partial\varPhi_{ab}} L^{d}_{\mu}\nonumber\\
&\mkern120mu\times\int\frac{\partial^2\Ods}{\partial R^c\partial R^d}(\bRcal+\bL\by)\,[dy]\nonumber\\
&\qquad=\frac{1}{2}\sum_{cd}\frac{\partial}{\partial\varPhi_{ab}} \left(\sum_{\mu}L^{d}_{\mu}L^{c}_{\mu}\right)\nonumber\\
&\mkern120mu\times\int\frac{\partial^2\Ods}{\partial R^c\partial R^d}(\bRcal+\bL\by)\,\,[dy]\nonumber\\
&\qquad=\frac{1}{2}\sum_{cd}\frac{\partial\varPsi^{cd}}{\partial\varPhi_{ab}} \int\frac{\partial^2\Ods}{\partial R^c\partial R^d}(\bRcal+\bL\by)\,\,[dy]\nonumber\\
&\qquad=\frac{1}{2}\sum_{cd}\frac{\partial\varPsi^{cd}}{\partial\varPhi_{ab}}\avg{\frac{\partial^2\Ods}{\partial R^c\partial R^d}}_{{\displaystyle\rhotrial_{\bRcal,\bvarPhi}}}\mathv
\end{align}
the boundary terms at infinity being zero due to the exponential.

Denoting with $\widetilde{F}_{\bRcal,\bvarPhi}$ the free energy of $\displaystyle\rhotrial_{\bRcal,\bvarPhi}$,
we prove next the following relations:
\begin{subequations}
\begin{align}
&\frac{\partial}{\partial\Rcal^a}\left[\Ftrial_{\bRcal,\bvarPhi}-\avg{\Vtrial_{\bRcal,\bvarPhi}}_{\tagmedia}\right]=0\label{eq:app_dadim1}\\
&\frac{\partial}{\partial\varPhi_{ab}}\left[\Ftrial_{\bRcal,\bvarPhi}-\avg{\Vtrial_{\bRcal,\bvarPhi}}_{\tagmedia}\right]=-\frac{1}{2}\sum_{cd}\varPhi_{cd}\frac{\partial\varPsi^{cd}}{\partial\varPhi_{ab}}\mathv
\label{eq:app_dadim2}
\end{align}\label{eq:app_dadim}\end{subequations}
where $\Vtrial_{\bRcal,\bvarPhi}=1/2\sum_{ab}\varPhi_{ab}u^au^b$ is the trial potential. 
Indeed, since the trial Hamiltonian is quadratic, we have the standard result
\begin{equation}
\Ftrial_{\bRcal,\bvarPhi}=\sum_{\mu}\left[\frac{\hbar\omega_{\mu}}{2}-\frac{1}{\beta}\ln\left(1+n_\mu\right)\right]\mathp
\end{equation}
Therefore, since $dn_{\mu}/d\omega_{\mu}=-\beta\hbar n_{\mu}(1+n_{\mu})$,
\begin{equation}
\frac{\partial \Ftrial_{\bRcal,\bvarPhi}}{\partial\omega_{\mu}^2}=\frac{1}{2}\xi^2_{\mu}\mathp
\end{equation}
The matrices $\Dcal_{ab}$ and $\Ecal_{ab}$ 
have same eigenvectors but eigenvalues $\omega_{\mu}^2$ and $\xi^2_{\mu}$, respectively.
Thus,
\begin{align}
\frac{\partial \widetilde{F}_{\bRcal,\bvarPhi}}{\partial D_{ab}}&=\sum_{\mu}  \frac{\partial 
\widetilde{F}_{\bRcal,\bvarPhi}}{\partial \omega^2_{\mu}}  \frac{\partial\omega_{\mu}^2}{\partial \Dcal_{ab}}\\
&=\frac{1}{2}\sum_{\mu}\xi^2_{\mu}   \frac{\partial\omega_{\mu}^2}{\partial \Dcal_{ab}}\\
&=\frac{1}{2}\sum_{cd}\Ecal^{cd}\frac{\partial \Dcal_{cd}}{\partial \Dcal_{ab}}\mathv
\end{align}
or, equivalently,
\begin{equation}
\frac{\partial \Ftrial_{\bRcal,\bvarPhi}}{\partial\varPhi_{ab}}=\frac{1}{2}\sum_{cd}\varPsi^{cd}\frac{\partial\varPhi_{cd}}{\partial\varPhi_{ab}}\mathp
\label{eq:app_1_1}
\end{equation}
Moreover, 
\begin{equation}
\avg{\Vtrial_{\bRcal,\bvarPhi}}_{\tagmedia}=\frac{1}{2}\sum_{\mu}\varPhi_{cd}\varPsi^{cd}\mathp
\label{eq:app_1_2}
\end{equation}
Indeed,
\begin{align}
\avg{\Vtrial_{\bRcal,\bvarPhi}}_{\tagmedia}&=\frac{1}{2}\sum_{ab}\varPhi_{ab}\avg{u^au^b}_{\bRcal,\bvarPhi}\nonumber\\
&=\frac{1}{2}\sum_{ab}\varPhi_{ab}\sum_{\nu\mu}L^a_{\nu}L^b_{\mu}\int y^\nu y^\mu\,[dy]\nonumber\nonumber\\
&=\frac{1}{2}\sum_{ab}\varPhi_{ab}\sum_{\nu\mu}L^a_{\nu}L^b_{\mu}\delta^{\nu\mu}\nonumber\\
&=\frac{1}{2}\sum_{ab}\varPhi_{ab}\varPsi^{ab}.
\end{align}
From Eq.~\eqref{eq:app_1_1} and Eq.~\eqref{eq:app_1_2} the relation~\eqref{eq:app_dadim2} is readily obtained. 
The equation~\eqref{eq:app_dadim1} comes from the observation that
neither $\Ftrial_{\bRcal,\bvarPhi}$ nor $\avg{\Vtrial_{\bRcal,\bvarPhi}}_{\tagmedia}$ depend on $\bRcal$.

The SCHA functional, i.e. the GB functional restricted to quadratic trial Hamiltonians, can be written as~\cite{PhysRevLett.111.177002,PhysRevB.89.064302}
\begin{equation}
\Fcal[\tagmedia]=\Ftrial_{\bRcal,\bvarPhi}-\Bavg{\Vtrial_{\bRcal,\bvarPhi}}_{\tagmedia}+\Bavg{V}_{\tagmedia}\mathp
\end{equation}
From this relation, and Eq.~\eqref{eq:app_der_obs} and Eq.~\eqref{eq:app_dadim} we obtain:
\begin{subequations}\begin{align}
&\frac{\partial}{\partial\Rcal^a}\Fcal[\tagmedia]=\avg{\frac{\partial V}{\partial R^a}}_{\tagmedia}\label{eq:app_derivate_parziali_funzionale_1}\\
&\frac{\partial}{\partial\varPhi_{ab}}\Fcal[\tagmedia]=\frac{1}{2}\sum_{cd}\left[\avg{\frac{\partial^2V}{\partial R^c\partial R^d}}_{\tagmedia}
-\varPhi_{cd}\right]\frac{\partial\varPsi^{cd}}{\partial\varPhi_{ab}}\mathp
\label{eq:app_derivate_parziali_funzionale_2}
\end{align}\label{eq:app_derivate_parziali_funzionale}\end{subequations}
Fixed $\bRcal$, with $\bPhi(\bRcal)$ we indicate the matrix that minimizes $\Fcal[\tagmedia]$ with respect to $\bvarPhi$.
This implies that $\partial \Fcal[\tagmedia]/\partial\bvarPhi$ is equal to zero in $\bPhi(\bRcal)$:
\begin{equation}
\left.\frac{\partial }{\partial \bvarPhi}\Fcal[\tagmedia]\right|_{\bPhi(\bRcal)}=0\mathp
\label{eq:app_annull_der_SCHAmat}
\end{equation}
Therefore, from  Eq.~\eqref{eq:app_derivate_parziali_funzionale_2} we have the self-consistent relation
\begin{equation}
\Phi_{ab}(\bRcal)=\avg{\frac{\partial^2V}{\partial R^a\partial R^b}}_{{\displaystyle\rhotrial_{\bRcal,\bPhi(\bRcal)}}}\mathp
\label{eq:app_SCHA_rel}
\end{equation}
Defining $F(\bRcal)=\Fcal[\displaystyle\rhotrial_{\bRcal,\bPhi(\bRcal)}]$, from Eq.~\eqref{eq:app_derivate_parziali_funzionale_1} and
Eq.~\eqref{eq:app_annull_der_SCHAmat} we have
\begin{equation}
\frac{\partial F}{\partial \Rcal^a}=\avg{\frac{\partial V}{\partial R^a}}_{\tagmediaSCHA}\mathp
\label{eq:app_first_der}
\end{equation}
Deriving one more time and using Eq.~\eqref{eq:app_der_obs}
and Eq.~\eqref{eq:app_SCHA_rel},
\begin{align}
\frac{\partial^2F}{\partial \Rcal^a\partial\Rcal^b}
&=\frac{\partial}{\partial \Rcal^b}
\left[\avg{\frac{\partial V}{\partial R^a}}_{\tagmediaSCHA}\right]\label{eq:app_der_int_0}\\
&=\Phi_{ab}+\sum_{pq}\overset{\sst}{\Phi}{}_{apq}\nonumber\\
&\qquad\qquad\times \sum_{c\le d}\left.\frac{1}{2}\frac{\partial \varPsi^{pq}}{\partial \varPhi_{cd}}\right|_{\bPhi}\frac{\partial \Phi_{cd}}{\partial \Rcal^b}\mathv
\label{eq:app_der_int_1}
\end{align}
where $\overset{\sst}{\bPhi}(\bRcal)$ is defined as a generalization of Eq.~\eqref{eq:app_SCHA_rel} to higher orders
(see Eq.~\eqref{eq:SCHA_matrix_def_nth}) and in the application of the chain rule  we have derived only with respect 
to the independent components of the symmetric matrix $\varPhi_{cd}$. Moreover, using Eq.~\eqref{eq:app_SCHA_rel},
\begin{align}
\frac{\partial \Phi_{ab}}{\partial\Rcal^c}&=\frac{\partial}{\partial \Rcal^c}
\left[\avg{\frac{\partial^2 V}{\partial R^a\partial R^b}}_{\tagmediaSCHA}\right]\nonumber\\
&=\overset{\sst}{\Phi}{}_{abc}+\sum_{lm}\overset{\ssf}{\Phi}{}_{ablm}\nonumber\\
&\qquad\qquad\times\sum_{p\le q}\left.\frac{1}{2}\frac{\partial \varPsi^{lm}}{\partial \varPhi_{pq}}\right|_{\bPhi}\frac{\partial \Phi_{pq}}{\partial \Rcal^c}\mathp
\label{eq:app_der_int_2}
\end{align}
In the next section we will give the explicit expression of the 4th order tensor $\Lambda^{abcd}(\bRcal)$
satisfying the relation
\begin{equation}
\sum_{p\le q}\left.\frac{1}{2}\frac{\partial \varPsi^{lm}}{\partial \varPhi_{pq}}\right|_{\bPhi}\frac{\partial \Phi_{pq}}{\partial \Rcal^c}
=\sum_{pq}\Lambda^{lmpq}\frac{\partial \Phi_{pq}}{\partial \Rcal^c}\mathp
\label{eq:app_tens_lambda}
\end{equation}
Using it we rewrite  Eq.~\eqref{eq:app_der_int_1} and Eq.~\eqref{eq:app_der_int_2} in the following way:
\begin{align}
&\frac{\partial^2F}{\partial \Rcal^a\partial\Rcal^b}=\Phi_{ab}+\sum_{lmpq}\overset{\sst}{\Phi}{}_{alm}\Lambda^{lmpq}\frac{\partial \Phi_{pq}}{\partial \Rcal^b}\label{eq:app_der_int_1_2}\\
&\frac{\partial \Phi_{ab}}{\partial\Rcal^c}=\overset{\sst}{\Phi}{}_{abc}+
\sum_{lmpq}\overset{\ssf}{\Phi}{}_{ablm}\Lambda^{lmpq}\frac{\partial \Phi_{pq}}{\partial \Rcal^c}\mathp
\label{eq:app_der_int_2_2}
\end{align}
The second equation can be solved by iteration, in the hypothesis that the resulting
series converges:
\allowdisplaybreaks[0]
\begin{align}
&\frac{\partial \Phi_{ab}}{\partial\Rcal^c}=\overset{\sst}{\Phi}{}_{abc}+
\sum_{l_1l_2l_3l_4}\overset{\ssf}{\Phi}{}_{abl_1l_2}\Lambda^{l_1l_2l_3l_4}\overset{\sst}{\Phi}{}_{l_3l_4c}\nonumber\\
&+\sum_{\substack{l_1l_2l_3l_4\\j_1j_2j_3j_4}}
\overset{\ssf}{\Phi}{}_{abl_1l_2}\Lambda^{l_1l_2l_3l_4}\overset{\ssf}{\Phi}{}_{l_3l_4j_1j_2}\Lambda^{j_1j_2j_3j_4}\overset{\sst}{\Phi}{}_{j_3j_4c}
\nonumber\\
&+\ldots
\end{align}
\allowdisplaybreaks
Substituting this solution into Eq.~\eqref{eq:app_der_int_1_2} we obtain
\begin{align}
&\frac{\partial^2 F}{\partial \Rcal^a \partial \Rcal^b}=\Phi_{ab}+\sum_{c_1c_2c_3c_4}\overset{\scriptscriptstyle{(3)}}{\Phi}_{ac_1c_2}\Lambda^{c_1c_2c_3c_4}\overset{\scriptscriptstyle{(3)}}{\Phi}_{c_3c_4b}\nonumber\\
&\,\,\,+\sum_{\substack{c_1c_2c_3c_4\\d_1d_2d_3d_4}}\overset{\scriptscriptstyle{(3)}}{\Phi}_{ac_1c_2}\Lambda^{c_1c_2c_3c_4}\Theta_{c_3c_4d_1d_2}\Lambda^{d_1d_2d_3d_4}\overset{\scriptscriptstyle{(3)}}{\Phi}_{d_3d_4b}\mathv
\label{eq:app_sec_der_F_1}
\end{align}
where $\Theta_{abcd}(\bRcal)$ is the 4th order tensor given by the series
\begin{align}
\Theta_{abcd}&=\overset{\ssf}{\Phi}_{abcd}
+\sum_{l_1l_2l_3l_4}\overset{\ssf}{\Phi}_{abl_1l_2}\Lambda^{l_1l_2l_3l_4}\overset{\ssf}{\Phi}_{l_3l_4cd}\nonumber\\
&+\sum_{\substack{l_1l_2l_3l_4\\j_1j_2j_3j_4}}\overset{\ssf}{\Phi}_{abl_1l_2}\Lambda^{l_1l_2l_3l_4}\overset{\ssf}{\Phi}_{l_3l_4j_1j_2}
\Lambda^{j_1j_2j_3j_4}\overset{\ssf}{\Phi}_{j_3j_4cd}\nonumber\\
&+\ldots
\end{align}
Thus, $\Theta_{abcd}(\bRcal)$ solves the Dyson-like equation
\begin{equation}
\Theta_{abcd}=\overset{\scriptscriptstyle{(4)}}{\Phi}_{abcd}+\sum_{l_1l_2l_3l_4}\overset{\scriptscriptstyle{(4)}}{\Phi}_{abl_1l_2} \Lambda^{l_1l_2l_3l_4}  \Theta_{l_3l_4cd}
\label{eq:app_Theta_equation}
\end{equation}
already introduced in Eq.~\eqref{eq:Theta_1}.
\section{The matrix  $\bLa$}
\label{app:the_matrix_L}
In this section we derive the explicit expression of the tensor $\Lambda^{abcd}$ used in Eq.~\eqref{eq:app_tens_lambda}. 
We break the derivation into several intermediate steps.
\subsection{Derivatives of eigenvalues and eigenvectors with respect to matrix elements}
Let us consider a real symmetric matrix $M_{ab}$ with distinct eigenvalues $\lambda_\mu$ and eigenvectors $e_\mu^a$.
From non degenerate perturbation theory, if $M_{ab}$ depends on a parameter $\epsilon$ we have
\begin{align}
&\frac{\partial \lambda_\mu}{\partial \epsilon}=\sum_{ab} M'_{ab}e^a_\mu e^b_\mu\\
&\frac{\partial e^a_\mu}{\partial \epsilon}=\sum_{\nu,\nu\neq\mu}\frac{\sum_{pq}M'_{pq}e^p_{\nu}e^q_{\mu}}{\lambda_\mu-\lambda_\nu}e^a_\nu\mathv
\end{align}
where $M'_{ab}=d M_{ab}/d\epsilon$.
In particular, for $\epsilon=M_{ij}$,
\begin{equation}
\frac{\partial M_{ab}}{\partial M_{ij}}=
\left\{
\begin{aligned}
&\delta_{ai}\delta_{bj}+\delta_{bi}\delta_{aj}&& i\neq j\\
&\delta_{ai}\delta_{bj}&& i=j
\end{aligned}
\right.
\mathp
\end{equation}
Thus,
\begin{equation}
\frac{\partial \lambda_\mu}{\partial M_{ij}}=
\left\{
\begin{aligned}
&2e^i_\mu e^j_\mu&& i\neq j\\
&e^i_\mu e^j_\mu&& i=j
\end{aligned}
\right.
\label{eq:app_deriv_eigval}
\end{equation}
and
\begin{equation}
\frac{\partial e^a_\mu}{\partial M_{ij}}=
\left\{
\begin{aligned}
& \sum_{\nu,\nu\neq\mu} \frac{e^i_{\nu}e^j_{\mu}+e^j_{\nu}e^i_{\mu}}{\lambda_\mu-\lambda_\nu}e^a_\nu&&i\neq j\\
& \sum_{\nu,\nu\neq\mu} \frac{e^i_{\nu}e^j_{\mu}}{\lambda_\mu-\lambda_\nu}e^a_\nu &&i=j
\end{aligned}
\right.
\mathp
\label{eq:app_deriv_eigvec}
\end{equation}
\subsection{Calculation of $\boldsymbol{\partial [F(M)]_{ab}/\partial M_{cd}}$}\label{sec:PertEig}
The matrix $M_{ab}$ can be written as
\begin{equation}
M_{ab}=\sum_\mu\lambda_{\mu}\,e_\mu^a e_\mu^b\mathp
\end{equation}
Given a regular function $F(x)$, $[F(M)]_{ab}$ is a real symmetric matrix having the same eigenvectors $e_\mu^a$ and eigenvalues $F(\lambda_\mu)$:
\begin{equation}
[F(M)]_{ab}=\sum_\mu F(\lambda_{\mu})\,e_\mu^a e_\mu^b\mathp
\end{equation}
Then, we can write
\begin{equation}
\frac{\partial [F(M)]_{ab}}{\partial M_{cd}}=X^{abcd}+Y^{abcd}
\end{equation}
with
\begin{align}
X^{abcd}&=\sum_{\mu}\frac{\partial [F(M)]_{ab}}{\partial \lambda_\mu}\frac{\partial \lambda_\mu}{\partial M_{cd}}\nonumber\\
&=\sum_\mu F'(\lambda_\mu)e^a_{\mu}e^b_{\mu}\frac{\partial \lambda_\mu}{\partial M_{cd}}
\end{align}
and 
\begin{align}
Y^{abcd}&=\sum_{m\mu}  \frac{\partial [F(M)]_{ab}}{\partial e^m_{\mu}}\frac{\partial e^m_{\mu}}{\partial M_{cd}}\nonumber\\
&=\sum_{m\mu}  F(\lambda_{\mu})\left[\delta^{ma}e^b_\mu+\delta^{mb}e^a_\mu\right]   \frac{\partial e^m_{\mu}}{\partial M_{cd}}\nonumber\\
&=\sum_{\mu}  F(\lambda_{\mu})\left[e^b_\mu \frac{\partial e^a_{\mu}}{\partial M_{cd}}+e^a_\mu \frac{\partial e^b_{\mu}}{\partial M_{cd}}\right] \mathp
\end{align}
Therefore, from Eq.~\eqref{eq:app_deriv_eigval}
\allowdisplaybreaks[0]
\begin{align}
X^{abcd}&=\sum_\mu F'(\lambda_\mu)e^a_{\mu}e^b_{\mu}\nonumber\\
&\qquad\qquad\times\left\{
\begin{aligned}
&2e^c_{\mu}e^d_{\mu} &&\text{if }c\neq d\\
&e^c_{\mu}e^d_{\mu} &&\text{if }c=d\\
\end{aligned}
\right.
\end{align}
\allowdisplaybreaks
and from  Eq.~\eqref{eq:app_deriv_eigvec}
\begin{align}
Y^{abcd}&=\sum_{\mu\neq\nu}  F(\lambda_{\mu})\,(e^a_\nu e^b_\mu+e^a_\mu e^b_\nu)\nonumber\\
&\qquad\qquad\times\left\{
\begin{aligned}
&\frac{e^c_{\nu}e^d_{\mu}+e^c_{\mu}e^d_{\nu}}{\lambda_\mu-\lambda_\nu}
&&\text{if }c\neq d\\
&\frac{e^c_{\nu}e^d_{\mu}}{\lambda_\mu-\lambda_\nu}
&&\text{if }c=d\\
\end{aligned}
\right.\mathp
\end{align}
\subsection{Contraction of $\boldsymbol{\partial [F(M)]_{ab}/\partial M_{cd}}$ with a symmetric matrix $\boldsymbol{\mathds{S}_{cd}}$}\label{sec:PertEig}
We are interested in calculating the quantity
\begin{equation}
\sum_{c\le d}\frac{\partial [F(M)]_{ab}}{\partial M_{cd}}\Sds_{cd}=\sum_{d\le c}\frac{\partial [F(M)]_{ab}}{\partial M_{cd}}\Sds_{cd}
\end{equation}
with $\Sds_{cd}=\Sds_{dc}$ a symmetric matrix. First, we define the tensor
\begin{equation}
\widetilde{\Lambda}{}^{abcd}=\widetilde{X}^{abcd}+\widetilde{Y}^{abcd}\mathv
\end{equation}
with
\begin{equation}
\widetilde{X}^{abcd}=\sum_\mu F'(\lambda_\mu)\,e^a_\mu e^b_\mu  e^c_\mu e^d_\mu 
\end{equation}
and
\begin{equation}
\widetilde{Y}^{abcd}=\sum_{\nu\neq\mu} \frac{F(\lambda_\mu)}{\lambda_\mu-\lambda_\nu}\Bigl(e^a_\nu e^b_\mu   + e^a_\mu e^b_\nu \Bigr)\,e^c_\nu e^d_\mu\mathp
\end{equation}
In terms of the tensor $\widetilde{\Lambda}{}^{abcd}$
\begin{equation}
\frac{\partial \left[F(M)\right]_{ab}}{\partial M_{cd}}=
\left\{
\begin{aligned}
&\widetilde{\Lambda}{}^{abcd}+\widetilde{\Lambda}{}^{abdc}&&\text{if }c\neq d\\
&\widetilde{\Lambda}{}^{abcd}&&\text{if }c= d
\end{aligned}
\right.\mathp
\end{equation}
Therefore,
\begin{equation}
\sum_{c\le d}\frac{\partial [F(M)]_{ab}}{\partial M_{cd}}\Sds_{cd}=\sum_{cd}\widetilde{\Lambda}{}^{abcd}\,\Sds_{cd}\mathp
\end{equation}
\iftoggle{draft}{
\begin{widetext}
\begin{align*}
\sum_{c\le d}\frac{\partial [F(M)]^{ab}}{\partial M^{cd}}S^{cd}&=
\sum_{c=d}\frac{\partial [F(M)]^{ab}}{\partial M^{cd}}S^{cd}+\sum_{c<d}\frac{\partial [F(M)]^{ab}}{\partial M^{cd}}S^{cd}\\
&=\sum_{c=d}\left[\widetilde{\frac{dF(M)}{dM}}\right]^{ab}_{cd}S^{cd}+
  \sum_{c<d}\left(\left[\widetilde{\frac{dF(M)}{dM}}\right]^{ab}_{cd}+\left[\widetilde{\frac{dF(M)}{dM}}\right]^{ab}_{dc}\right)S^{cd}\\
&=\sum_{c=d}\left[\widetilde{\frac{dF(M)}{dM}}\right]^{ab}_{cd}S^{cd}+
 \sum_{c<d}\left[\widetilde{\frac{dF(M)}{dM}}\right]^{ab}_{cd}S^{cd}+
 \sum_{c<d}\left[\widetilde{\frac{dF(M)}{dM}}\right]^{ab}_{dc}S^{dc}\\
&=\sum_{cd}\left[\widetilde{\frac{dF(M)}{dM}}\right]^{ab}_{cd}S^{cd}    
\end{align*}
\end{widetext}
Moreover, notice that
\begin{align*}
\widetilde{Y}^{ab}_{cd}S^{cd}&=
\sum_{cd}\sum_{\nu\neq\mu}\frac{F(\lambda_\mu)}{\lambda_\mu-\lambda_\nu}
\Bigl(e^a_\nu e^b_\mu  e^c_\nu e^d_\mu + e^a_\mu e^b_\nu  e^d_\mu e^c_\nu\Bigr)S^{cd}\nonumber\\
&=\sum_{cd}\sum_{\nu\neq\mu}\frac{F(\lambda_\mu)}{\lambda_\mu-\lambda_\nu}
\Bigl(e^a_\nu e^b_\mu  e^c_\nu e^d_\mu\Bigr)S^{cd}+\sum_{cd}\sum_{\nu\neq\mu}\frac{F(\lambda_\mu)}{\lambda_\mu-\lambda_\nu}\Bigl(e^a_\mu e^b_\nu  e^d_\mu e^c_\nu\Bigr)S^{cd}\nonumber\\
&=\sum_{cd}\sum_{\nu\neq\mu}\frac{F(\lambda_\mu)}{\lambda_\mu-\lambda_\nu}
\Bigl(e^a_\nu e^b_\mu  e^c_\nu e^d_\mu\Bigr)S^{cd}+\sum_{cd}\sum_{\nu\neq\mu}\frac{F(\lambda_\mu)}{\lambda_\mu-\lambda_\nu}\Bigl(e^a_\mu e^b_\nu  e^c_\mu e^d_\nu\Bigr)S^{cd}\nonumber\\
&=\sum_{cd}\sum_{\nu\neq\mu}\frac{F(\lambda_\mu)}{\lambda_\mu-\lambda_\nu}
\Bigl(e^a_\nu e^b_\mu  e^c_\nu e^d_\mu\Bigr)S^{cd}+\sum_{cd}\sum_{\nu\neq\mu}\frac{F(\lambda_\nu)}{\lambda_\nu-\lambda_\mu}\Bigl(e^a_\nu e^b_\mu  e^c_\nu e^d_\mu\Bigr)S^{cd}\nonumber\\
&=\sum_{cd}\sum_{\nu\neq\mu}\frac{F(\lambda_\mu)-F(\lambda_\nu)}{\lambda_\mu-\lambda_\nu}
\Bigl(e^a_\nu e^b_\mu  e^c_\nu e^d_\mu\Bigr)S^{cd}\mathp
\end{align*}
\end{widetext}  }
Moreover, notice that
\begin{align}
&\sum_{cd}\widetilde{Y}^{abcd}\,\Sds_{cd}\nonumber\\
&\quad=\sum_{cd}\sum_{\nu\neq\mu}\frac{F(\lambda_\mu)}{\lambda_\mu-\lambda_\nu}\Bigl(e^a_\nu e^b_\mu  e^c_\nu e^d_\mu + e^a_\mu e^b_\nu  e^d_\mu e^c_\nu\Bigr)\Sds_{cd}\nonumber\\
&\quad=\sum_{cd}\sum_{\nu\neq\mu}\frac{F(\lambda_\mu)-F(\lambda_\nu)}{\lambda_\mu-\lambda_\nu}\,e^a_\nu e^b_\mu  e^c_\nu e^d_\mu\,\Sds_{cd}\mathp
\end{align}
Therefore, given a symmetric tensor $\Sds_{cd}$ we have
\begin{equation}
\sum_{c\le d}\frac{\partial [F(M)]^{ab}}{\partial M^{cd}}\,\Sds_{cd}=\sum_{cd}\,\Lambda^{abcd}\,\Sds_{cd}
\label{eq:contract_symm_matrix}
\end{equation}
with $\Lambda^{abcd}$ defined by
\begin{equation}
\Lambda^{abcd}=\sum_{\nu\mu}F_{\mu\nu}\,e^a_\nu e^b_\mu  e^c_\nu e^d_\mu\mathv
\label{eq:dFdM_1}
\end{equation}
where
\begin{equation}
F_{\mu\nu}=
\left\{
\begin{aligned}
&\left.\frac{dF}{d\lambda}\right|_{\lambda_\mu}&&\text{if}\quad\mu=\nu\\
&\frac{F(\lambda_\mu)-F(\lambda_\nu)}{\lambda_\mu-\lambda_\nu}&&\text{if}\quad\mu\neq\nu
\end{aligned}
\right.\mathp
\label{eq:dFdM_2}
\end{equation}
Notice the following two symmetries:
\begin{align}
&\Lambda^{abcd}=\Lambda^{badc}\label{eq:app_symmdF_1}\\
&\Lambda^{abcd}=\Lambda^{cdab}\label{eq:app_symmdF_2}\mathp
\end{align}
\subsection{Degeneracy}
Up to now, we have exclusively considered the non-degenerate case. However, if we
write $F_{\mu\nu}$ in the form
\begin{equation}
F_{\mu\nu}=
\left\{
\begin{aligned}
&\left.\frac{dF}{d\lambda}\right|_{\lambda_\mu}&&\text{if}\quad\lambda_\mu=\lambda_\nu\\
&\frac{F(\lambda_\mu)-F(\lambda_\nu)}{\lambda_\mu-\lambda_\nu}&&\text{if}\quad\lambda_\mu\neq\lambda_\nu
\end{aligned}
\right.\mathv
\label{eq:dFdM_3}
\end{equation}
Eq.~\eqref{eq:dFdM_1} is well defined even in case of degeneracies, and
gives the same tensor $\Lambda^{abcd}$ regardless of the gauge chosen, 
i.e. regardless of the basis set chosen in the degenerate spaces.
In order to see that, let us consider two eigenvectors basis sets $\{e^a_{\mu}\}$ and  $\{\bare^a_\mu\}$ related as
\begin{equation}
e^a_\mu=\sum_\nu \bare^a_\nu \,U^\nu_\mu
\end{equation}
where $U^\nu_\mu$ is an orthogonal matrix, i.e. it satisfies 
\begin{equation}
\sum_{\nu}U_\nu^{\mu_2} U_\nu^{\mu_1}=\sum_{\nu}U^\nu_{\mu_2} U^\nu_{\mu_1}=\delta^{\mu_1\mu_2}\mathv
\end{equation}
and it does not mix eigenvectors with different eigenvalues:
\begin{equation}
\lambda_\nu\neq\lambda_\mu\,\Longrightarrow\, U^\mu_\nu=0\mathp
\end{equation}
Notice that an immediate consequence is that
\begin{equation}
F_{\mu\nu}U^{\mu}_{\rho}=F_{\rho\nu}U^{\mu}_{\rho}\mathp
\end{equation}
Then,
\begin{align}
&\sum_{\nu\mu}F_{\mu\nu}\,e^a_\nu e^b_\mu e^c_\nu e^d_\mu=\nonumber\\ 
&\,=\sum_{\rho_1 \rho_2 \rho_3 \rho_4}
  \sum_{\nu\mu}F_{\mu\nu}\,
  \bare^a_{\rho_1}\bare^b_{\rho_2}\bare^c_{\rho_3}\bare^d_{\rho_4} 
  U^{\rho_1}_{\nu}U^{\rho_2}_{\mu}U^{\rho_3}_{\nu}U^{\rho_4}_{\mu}\nonumber\\
&\,=\sum_{\rho_1 \rho_2 \rho_3 \rho_4}
  \sum_{\nu\mu}F_{\rho_1\rho_2}\,
  \bare^a_{\rho_1}\bare^b_{\rho_2}\bare^c_{\rho_3}\bare^d_{\rho_4}
  U^{\rho_1}_{\nu}U^{\rho_2}_{\mu}U^{\rho_3}_{\nu}U^{\rho_4}_{\mu}\nonumber\\
&\,=\sum_{\rho_1 \rho_2 \rho_3 \rho_4} 
  F_{\rho_2\rho_1}\,
  \bare^a_{\rho_1}\bare^b_{\rho_2}\bare^c_{\rho_3}\bare^d_{\rho_4}
  \delta^{\rho_1\rho_3}\delta^{\rho_2\rho_4}\nonumber\\
&\,=\sum_{\rho_1 \rho_2} 
  F_{\rho_2\rho_1}\,
  \bare^a_{\rho_1}\bare^b_{\rho_2}\bare^c_{\rho_1}\bare^d_{\rho_2}\mathp
\end{align}
Therefore, Eq.~\eqref{eq:dFdM_1} and Eq.~\eqref{eq:dFdM_3}  are well defined even in the degenerate case, because they give the same result, regardless of
the specific basis of eigenvectors chosen. This essentially proves that Eq.~\eqref{eq:contract_symm_matrix}, 
is valid even if $M_{ab}$ is degenerate. In fact, given a matrix $M_{ab}$ degenerate, 
we can consider a real symmetric matrix function $M_{ab}{(\epsilon)}$ continuously depending on a parameter $\epsilon\in[0,1]$,
\begin{equation}
M_{ab}(\epsilon)=\sum_{\mu}\lambda_{\mu}(\epsilon)e^a_{\mu}(\epsilon)e^b_{\mu}(\epsilon)\mathv
\end{equation}
such that  $M_{ab}(0)=M_{ab}$ but with $M_{ab}(\epsilon)$ non degenerate for $\epsilon\neq 0$. Notice that, given $M_{ab}$, the arbitrariness 
in the choice of the function $M_{ab}(\epsilon)$ reflects in the arbitrariness of the eigenvectors $e^b_{\mu}(0)$ for $M_{ab}$, i.e. in the choice of a specific gauge.
We can apply Eq.~\eqref{eq:contract_symm_matrix},  Eq.~\eqref{eq:dFdM_1}, and  Eq.~\eqref{eq:dFdM_2} with $\epsilon\neq 0$ and consider the limit of the result for $\epsilon$
that goes to zero. That gives  Eq.~\eqref{eq:contract_symm_matrix},  Eq.~\eqref{eq:dFdM_1}, and  Eq.~\eqref{eq:dFdM_3} for the specific set $e^b_{\nu}(0)$. 
However, as observed, the formula is gauge invariant, i.e. the specific choice of $e^a_{\nu}(0)$ in the degenerate subspaces is immaterial.
Therefore, the procedure is well defined, as it gives a unique result.
\subsection{The matrix $\boldsymbol{\Lambda}^{abcd}$}
We apply the general results found to  Eq.~\eqref{eq:app_tens_lambda}. We can write:
\allowdisplaybreaks[0]
\begin{align}
\sum_{c\le d}\frac{1}{2}\left.\frac{\partial \varPsi^{ab}}{\partial \varPhi_{cd}}\right|_{\bPhi}&\frac{\partial \Phi_{cd}}{\partial R^m}=\nonumber\\
&=\frac{1}{\sqrt{M_aM_bM_cM_d}}\sum_{c\le d}\frac{1}{2}\left.\frac{\partial \Ecal^{ab}}{\partial \Dcal_{cd}}\right|_{\bDcal(\bPhi)}\frac{\partial \Phi_{cd}}{\partial R^m}
\mathp
\end{align}
\allowdisplaybreaks
Since $\Ecal_{ab}=\xi^2(\Dcal_{ab})$ and the matrix $\partial \Phi_{cd}/\partial R^m$ is symmetric in $(cd)$,
applying Eq.~\eqref{eq:contract_symm_matrix} we obtain
\begin{equation}
\sum_{c\le d}\frac{1}{2}\left.\frac{\partial \varPsi^{ab}}{\partial \varPhi_{cd}}\right|_{\bPhi}\frac{\partial \Phi_{cd}}{\partial R^m}=\sum_{cd}\Lambda^{abcd}\,\frac{\partial \Phi_{cd}}{\partial R^m}\mathv
\end{equation} 
where
\begin{equation}
\Lambda^{abcd}=\frac{1}{2}\sum_{\nu\mu}F_{\mu\nu}\,\frac{e^a_{\nu}}{\sqrt{M_a}}\frac{e^b_{\mu}}{\sqrt{M_b}}\frac{e^c_{\nu}}{\sqrt{M_c}}\frac{e^d_{\mu}}{\sqrt{M_d}}
\label{eq:lambda_1}
\end{equation}
and
\begin{equation}
F_{\mu\nu}=
\left\{
\begin{aligned}
&\frac{d\xi^2_\mu}{d\omega^2_\mu}\quad&&\text{if}\quad\omega_\mu^2=\omega_\nu^2\\
&\frac{\xi^2_\mu-\xi^2_\nu}{\omega^2_\mu-\omega^2_\nu}\quad&&\text{if}\quad\omega_\mu^2\neq\omega_\nu^2
\end{aligned}
\right.\mathp
\label{eq:lambda_2}
\end{equation}
Here $\omega_{\mu}^2$ and $e^a_{\mu}$ are eigenvalues and eigenvectors of $\Dcal_{ab}=\Phi_{ab}/\sqrt{M_aM_b}$, respectively.
Since $\xi_{\mu}^2=\hbar(1+2n_{\mu})/2\omega_{\mu}$ and $dn_{\mu}/d\omega_{\mu}=-\beta\hbar n_{\mu} (1+n_{\mu})$, we can write
\begin{equation}
F_{\mu\nu}=-\frac{\hbar^2}{4\omega_\mu\omega_\nu}F(0,\omega_\nu,\omega_\mu)
\end{equation}
with
\begin{align}
&F(0,\omega_\nu,\omega_\mu)=\phantom{\Biggl\{}\nonumber\\
&\quad\left\{
\begin{aligned}
&\frac{2}{\hbar}\left[\frac{2n_\nu+1}{2\omega_\nu}-\frac{dn_\nu}{d\omega_\nu}\right] &&\text{if}\qquad \omega_\nu=\omega_\mu\\
&\frac{2}{\hbar}\left[\frac{n_\mu+n_\nu+1}{\omega_\mu+\omega_\nu}-\frac{n_\mu-n_\nu}{\omega_\mu-\omega_\nu}\right] &&\text{if}\qquad \omega_\mu\neq\omega_\nu
\end{aligned}
\right.\mathp
\label{app:eq:def_F0}
\end{align}
Therefore, we obtain the final expression:
\begin{align}
\Lambda^{abcd}=-\frac{\hbar^2}{8}\sum_{\mu\nu}&\frac{F(0,\omega_{\mu},\omega_{\nu})}{\omega_{\mu}\omega_{\nu}}\nonumber\\
&\times\frac{e^a_{\nu}}{\sqrt{M_a}}\frac{e^b_{\mu}}{\sqrt{M_b}}\frac{e^c_{\nu}}{\sqrt{M_c}}\frac{e^d_{\mu}}{\sqrt{M_d}}
\mathp
\label{eq:lambd}
\end{align}

For some derivations it is convenient to express the 4th-rank tensor $\Lambda^{abcd}$
as a square super-matrix $\Lambda^{AB}$, with $A=(ab)$ and $B=(cd)$.
From~Eq.\eqref{eq:app_symmdF_2}  we have that $\bLambda$ is real symmetric (we are using bold symbol in component free notation).
Moreover, it is negative-definite.
In fact, $\xi^2(\omega_{\mu}^2)=\hbar(1+2n_{\mu})/2\omega_{\mu}=\hbar\coth(\beta\hbar\omega_{\mu}/2)/2\omega_{\mu}$ is  monotonically decreasing. Thus 
from Eq.~\eqref{app:eq:def_F0} $F_{\mu\nu}<0$, and, considering a vector $\bT=T_A=T_{ab}$ we have
from Eq.~\eqref{eq:lambda_1}
\begin{equation}
\bT\bLambda\bT=\frac{1}{2}\sum_{\mu\nu}\left(\sum_{ab}\frac{T_{ab}}{\sqrt{M_aM_b}}e^a_\mu e^b_\nu\right)^2F_{\mu\nu}<0\mathp
\label{eq:app_def_neg_lambda}
\end{equation}

\section{Stochastic calculation of SCHA matrices}
\label{app:Stochastic_calculation_of_derivative_averages}
Given an observable $\Ods(\bR)$, it can be proved that
\begin{equation}
\avg{\frac{\partial\Ods}{\partial R^a}}_{\tagmedia}=\sum_b\varPsiinv_{ab}\,\Bavg{u^b\,\Ods}_{\tagmedia}\mathv
\label{eq:app_formula_average_er}
\end{equation}
where $\bu=\bR-\bRcal$. In order to demonstrate this formula we use the change of variable in Eq.~\eqref{eq:change_var}
and the inverse matrix of $\bL$:
\begin{equation}
\Linv{}_a^{\mu}=\frac{e^a_{\mu}}{\xi_{\mu}}\sqrt{M_a}\mathp
\end{equation}
The demonstration is obtained with integration by parts:
\allowdisplaybreaks[0]
\begin{align}
\avg{\frac{\partial\Ods}{\partial R^a}}_{\tagmedia}&=\int \frac{\partial\Ods}{\partial\Rcal^a}\left(\bRcal+\bL\by\right)\,[dy]\nonumber\\
&\mkern-60mu=\sum_{\mu}\Linv^{\mu}_a\int\frac{\partial}{\partial y_\mu}\Ods\left(\bRcal+\bL\by\right)\,[dy]\nonumber\\
&\mkern-60mu=-\sum_{\mu}\Linv^{\mu}_a\int\Ods\left(\bRcal+\bL\by\right)\,\frac{\partial [dy]}{\partial y^\mu}\nonumber\\
&\mkern-60mu=\sum_{\mu}\Linv^{\mu}_a\int y^{\mu}\,\Ods\left(\bRcal+\bL\by\right)\,[dy]\nonumber\\
&\mkern-60mu=\sum_b\sum_{\mu}\Linv^{\mu}_a\Linv^{\mu}_b\int u^b\Ods\left(\bRcal+\bL\by\right)\,[dy]\nonumber\\
&\mkern-60mu=\sum_b\varPsiinv_{ab}\int u^b\Ods\left(\bRcal+\bL\by\right)\,[dy]\nonumber\\
&\mkern-60mu=\sum_b\varPsiinv_{ab}\Bavg{u^b\,\Ods}_{\tagmedia}\mathp
\end{align}
\allowdisplaybreaks
We apply Eq.~\eqref{eq:app_formula_average_er} to find expressions for $\overset{\sst}{\bPhi}(\bRcal)$
and $\overset{\ssf}{\bPhi}(\bRcal)$ suited for a stochastic calculation. The goal is to express them 
as averages of functions of positions and forces only. In what follows the dependence of the matrices on $\bRcal$ is understood.
\subsubsection{Stochastic formula for $\protect\overset{\sst}{\bPhi}$}  
Applying Eq.~\eqref{eq:app_formula_average_er} we obtain 
\begin{align}
\overset{\sst}{\Phi}{}_{abc}&=\avg{\frac{\partial^3V}{\partial R^a\partial  R^b\partial R^c}}_{\tagmediaSCHAnoR}\nonumber\\
&=\sum_p\varPsiinv_{ap}\avg{u^p\frac{\partial^2V}{\partial R^b \partial R^c}}_{\tagmediaSCHAnoR}\nonumber\\
&=\sum_p\varPsiinv_{ap}\left[\avg{\frac{\partial}{\partial R^{b}}\left(u^p\frac{\partial V}{\partial R^{c}}\right)}_{\tagmediaSCHAnoR}\right.\nonumber\\
&\mkern+170mu -\left.\delta^{pb}\avg{\frac{\partial V}{\partial R^c}}_{\tagmediaSCHAnoR}\right]\nonumber\\
&=\sum_{pq}\varPsiinv_{ap}\varPsiinv_{bq}\avg{u^pu^q\frac{\partial V}{\partial R^c}}_{\tagmediaSCHAnoR}\nonumber\\
&\mkern+170mu-\varPsiinv_{ab}\avg{\frac{\partial V}{\partial R^c}}_{\tagmediaSCHAnoR}\nonumber\\
&=\sum_{pq}\varPsiinv_{ap}\varPsiinv_{bq}\avg{u^pu^q\left[\,\,\frac{\partial V}{\partial R^c}-\avg{\frac{\partial V}{\partial R^c}}_{\tagmediaSCHAnoR}\right]}_{\tagmediaSCHAnoR}\mathv
\label{eq:app_SCHA3_mat_ant}
\end{align}
where in the last line we used that $\Bavg{u^pu^q}_{\tagmediaSCHAnoR}=\varPsi^{pq}$.
In terms of the forces $\f_c=-\partial V/\partial R^c$, we write
\begin{equation}
\overset{\sst}{\Phi}{}_{abc}=-\sum_{pq}\varPsiinv_{ap}\varPsiinv_{bq}\avg{u^pu^q\Biggl[\,\,\f_c-
\Bavg{\f_c}_{\tagmediaSCHAnoR}\Biggr]}_{\tagmediaSCHAnoR}\mathp
\end{equation}
Since the average of a function that is odd in the displacements is zero, we can write
\begin{equation}
\overset{\sst}{\Phi}{}_{abc}=-\sum_{pq}\varPsiinv_{ap}\varPsiinv_{bq}\,\Bavg{u^pu^q\,\fbb_c}_{\tagmediaSCHAnoR}\mathv
\label{eq:app_phi3_con_fbb}
\end{equation}
with
\begin{equation}
\fbb_c=\f_c-\biggl[\Bavg{\f_c}_{{\displaystyle\rhotrial_{\bRcal,\bPhi}}}+\Fscr_c^{\odd}(\bu)\biggr]\mathv
\label{eq:app_def_fbb_3_generic}
\end{equation}
where $\Fscr_c^{\odd}(\bu)$  are generic odd functions. However, this is true only in the limit of an infinite 
population sampling. 
In actual calculations we compute the averages with populations of finite size.
In that case, the value of $\overset{\sst}{\bPhi}$ obtained from Eq.~\eqref{eq:app_phi3_con_fbb} depends on the specific 
$\Fscr_c^{\odd}(\bu)$ chosen, i.e. on the
specific expression of $\fbb_c$. In order to reduce the statistical noie and speed up the convergence, we found convenient to
utilize
\begin{equation}
\fbb_c=\f_c-\biggl[\Bavg{\f_c}_{{\displaystyle\rhotrial_{\bRcal,\bPhi}}}-\sum_i\Phi_{ci}\,u^i\biggr]\mathp
\label{eq:app_defusata_fbb}
\end{equation} 
Indeed, this corresponds to $\fbb_c=-\partial\Vbb/\partial R^c$,
where $\Vbb(\bR)=V(\bR)-V^ {\rref}(\bR)$ is the difference between the true potential and the quadratic ``reference'' potential
\allowdisplaybreaks[0]
\begin{align}
V^{\rref}(\bR)&=V(\bRcal)+\sum_a\avg{\frac{\partial V}{\partial R^a}}_{\tagmediaSCHAnoR}(R-\Rcal)^a\nonumber\\
&+\frac{1}{2}\sum_{ab}\Phi_{ab}(R-\Rcal)^a(R-\Rcal)^b\mathp
\label{eq:app_refpot_3}
\end{align}
\allowdisplaybreaks
If $V(\bR)$ is quadratic, then it coincides with $V^{\rref}(\bR)$, thus $\Vbb(\bR)=0$ and $\fbb_c=0$. Therefore, considering Eq.~\eqref{eq:app_phi3_con_fbb} with Eq.~\eqref{eq:app_defusata_fbb} 
implies that if $V(\bR)$ is quadratic then $\overset{\sst}{\bPhi}$ is identically zero, as it must be, with any finite sampling.
\subsubsection{Stochastic formula for $\protect\overset{\ssf}{\bPhi}$} 
With passages analogous to the ones used in the previous demonstration, we obtain:
\allowdisplaybreaks[0]
\begin{align}
\overset{\ssf}{\Phi}{}_{abcd}&=\avg{\frac{\partial^4 V}{\partial R^a \partial R^b \partial R^c \partial R^d}}_{\tagmediaSCHAnoR}\nonumber\\
&=\sum_{pqr}\varPsiinv_{ap}\varPsiinv_{bq}\varPsiinv_{cr}\,\avg{u^pu^qu^r\,\frac{\partial V}{\partial R^d}}_{\tagmediaSCHAnoR}\nonumber\\
&\qquad-\sum_p\Bigl[\varPsiinv_{ac}\varPsiinv_{bp}+\varPsiinv_{bc}\varPsiinv_{ap}+\varPsiinv_{ab}\varPsiinv_{cp}\Bigr]\nonumber\\
&\mkern+220mu\times\avg{u^p\,\frac{\partial V}{\partial R^d}}_{\tagmediaSCHAnoR}\nonumber\\
&=\sum_{pqr}\varPsiinv_{ap}\varPsiinv_{bq}\varPsiinv_{cr}\,\Biggl\{u^pu^qu^r\nonumber\\
&\mkern+50mu\times\Biggl[\frac{\partial V}{\partial R^d}-\sum_iu^i\avg{\frac{\partial^2V}{\partial R^i\partial R^d}}_{\tagmediaSCHAnoR}\Biggr]\Biggr\}_{\tagmediaSCHAnoR}\mathp
\label{eq:app_SCHA4_mat_ant}
\end{align}
\allowdisplaybreaks
Therefore, we can write
\begin{equation}
\overset{\ssf}{\Phi}{}_{abcd}=-\sum_{pqr}\varPsiinv_{ap}\varPsiinv_{bq}\varPsiinv_{cr}
\avg{u^pu^qu^r\Biggl[\,\,\f_d+\sum_i\,\Phi_{di}u^i\Biggr]}_{\tagmediaSCHAnoR}\mathp
\end{equation}
Again, since the average of a function that is odd in the displacements is zero, we can write
\begin{equation}
\overset{\ssf}{\Phi}{}_{abcd}=-\sum_{pqr}\varPsiinv_{ap}\varPsiinv_{bq}\varPsiinv_{cr}\,\Bavg{u^pu^qu^r\,\fbb_d}_{\tagmediaSCHAnoR}\mathv
\label{eq:app_phi4_con_fbb}
\end{equation}
with
\begin{equation}
\fbb_d=\f_d-\biggl[\Fscr_d^{\even}(\bu)-\sum_i\,\Phi_{di}u^i\biggr]\mathv
\label{eq:app_def_fbb_4_generic}
\end{equation}
where $\Fscr_d^{\even}(\bu)$  are generic even functions. 

In order to reduce the statistical noise and accelerate the convergence, 
in our simulations we calculated $\overset{\ssf}{\bPhi}$ with Eq.~\eqref{eq:app_phi4_con_fbb} 
and we defined $\fbb_d$ as in Eq.~\eqref{eq:app_defusata_fbb}, since it is compatible with Eq.~\eqref{eq:app_def_fbb_4_generic}.
the definition~\eqref{eq:app_defusata_fbb} for $\fbb_d$.
In that way, if $V(\bR)$ is quadratic, we obtain $\overset{\ssf}{\bPhi}=0$ for any finite sampling, as it must be. 
Once $\overset{\sst}{\bPhi}$ has been calculated, another possibility, in principle better, would be to define 
\begin{align}
\fbb_d=\f_d-\biggl[\Bavg{\f_d}_{{\displaystyle\rhotrial_{\bRcal,\bPhi}}}+&\frac{1}{2}\sum_{ij}\overset{\sst}{\Phi}_{dij}\varPsi^{ij}\nonumber\\
&-\sum_i\Phi_{di}\,u^i-\frac{1}{2}\sum_{ij}\overset{\sst}{\Phi}_{dij}u^iu^j\biggr]\mathp
\label{eq:app_defnonusata_fbb}
\end{align} 
With this choice it is sufficient that $V(\bR)$ is cubic in order to obtain 
$\overset{\ssf}{\bPhi}=0$, as it must be, with any finite sampling.
Indeed, Eq.~\eqref{eq:app_defnonusata_fbb} corresponds to $\fbb_d=-\partial\Vbb/\partial R^d$,
where $\Vbb(\bR)$ is the difference between $V(\bR)$ and the cubic ``reference'' potential
\begin{align}
V^{\rref}(\bR)&=V(\bRcal)\nonumber\\
&\mkern-40mu+\sum_a\left[\avg{\frac{\partial V}{\partial R^a}}_{\tagmediaSCHAnoR}-\frac{1}{2}\sum_{hk}
\overset{\sst}{\Phi}_{ahk}\varPsi^{hk}\right](R-\Rcal)^a\nonumber\\
&\mkern-40mu+\frac{1}{2}\sum_{ab}\Phi_{ab}(R-\Rcal)^a(R-\Rcal)^b\nonumber\\
&\mkern-40mu+\frac{1}{3!}\sum_{abc}\overset{\sst}{\Phi}_{abc}(R-\Rcal)^a(R-\Rcal)^b(R-\Rcal)^c\mathp
\label{eq:app_refpot_4}
\end{align} 
If $V(\bR)$ is cubic it coincides with this $V^{\rref}(\bR)$,
thus $\overset{\ssf}{\bPhi}$ calculated with Eqs.~\eqref{eq:app_phi4_con_fbb} and~\eqref{eq:app_defnonusata_fbb} is equal to zero with any finite sampling.
However, for the simulations of this paper we did not use the definition~\eqref{eq:app_defnonusata_fbb}.
\subsubsection{Acoustic sum rule}
The SCHA tensor $\overset{\ssn}{\bPhi}{}_{a_1\ldots a_n}$ satisfies the acoustic sum rule (ASR)
if the sum over any atomic index vanishes. Considering that we are using a
double (cartesian,atom) index $a=(\alpha,s)$, in our notation this means that
\begin{equation}
\sum_{a_i}\overset{\ssn}{\bPhi}{}_{a_1\ldots a_i\ldots a_n}\,t^{a_i}=0\mathv
\label{eq:app_def_ASR}
\end{equation}
with $t^{\alpha,s}=t^{\alpha}$ a global translation of the system by the 3D vector $t^{\alpha}$.
The averages in Eq.~\eqref{eq:app_phi3_con_fbb} and Eq.~\eqref{eq:app_phi4_con_fbb}  are evaluated stochastically with a finite-size population
through Eq.~\eqref{eq:average_montecarlo}. We demonstrate that if the matrix $\Phi_{ab}$ satisfies the sum rule and the total force
on the center of mass of the system is zero for any population member (as it must be), then  
the approximate SCHA tensors given by Eqs.~\eqref{eq:app_phi3_con_fbb} and~\eqref{eq:app_phi4_con_fbb} 
with Eq.~\eqref{eq:app_defusata_fbb} satisfy the ASR
with any finite population sampling.
The two above mentioned conditions are expressed by the relations
\begin{align}
&\sum_a \f_a\,t^a=0\label{eq:app_CMF_zero}\\
&\sum_{b}\Phi_{ab}\,t^b=0\mathv\label{eq:app_ASR_Phi}
\end{align}
thus from Eq.~\eqref{eq:app_defusata_fbb} we also have
\begin{equation}
\sum_a \fbb_a\,t^a=0\mathp
\label{eq:app_CMFF_zero}
\end{equation}
This proves that $\sum_c\overset{\sst}{\bPhi}{}_{abc}\,t^c=0$ and
$\sum_d\overset{\sst}{\bPhi}{}_{abcd}\,t^d=0$. 
The proof that the SCHA tensors given by Eq.~\eqref{eq:app_phi3_con_fbb}  and Eq.~\eqref{eq:app_phi4_con_fbb} satisfy the ASR
also on the other indices is obtained by considering that 
\begin{equation}
\sum_b\varPsiinv_{ab}\,t^b=0\mathp
\label{eq:app_ASR_Psi}
\end{equation}
Indeed, Eq.~\eqref{eq:app_ASR_Phi} means that $t^b$ is null eigenvector for $\Phi_{ab}$,
which is equivalent to saying that $\sqrt{M_b}\,t^b$ is null eigenvector for $\Dcal_{ab}$.
However, indicating with $\overset{\ssi}{E}_{ab}$ the inverse of $E^{ab}$ defined in Eq.~\eqref{eq:def_matrici_utili_2}, it is $\overset{\ssi}{\Ecal}_{ab}=\xi^{\scriptscriptstyle{-2}}\bigl(\Dcal_{ab}\bigr)$ and $\xi^{-2}(0)=0$. Therefore,
$\sqrt{M_b}\,t^b$ is null eigenvector for $\overset{\ssi}{\Ecal}{}_{ab}$, i.e. $t^b$ is null eigenvector of 
$\overset{\ssi}\varPsi_{ab}=\sqrt{M_aM_b}\overset{\ssi}{\Ecal}_{ab}$, which proves Eq.~\eqref{eq:app_ASR_Psi}.

Notice that $\overset{\ssf}{\bPhi}$ computed with Eq.~\eqref{eq:app_phi4_con_fbb} satisifes the ASR with any finite sampling even
if we use the definition~\eqref{eq:app_defnonusata_fbb} for $\fbb_a$.
In that case, in order to demonstrate the condition~\eqref{eq:app_CMFF_zero}, we just 
have to consider that $\overset{\sst}{\bPhi}$ satisfies the ASR.

%

In conclusion, and using the matrix $\bUpsilon=\overset{\ssi}{\bvarPsi}$  (see Eq.~\eqref{eq:def_Upsilon})
in order to ease the connection with the main text, the formulas that we used to compute the SSCHA 3rd and 4th order tensors
are:
\allowdisplaybreaks[0]
\begin{subequations}\begin{align}
&\overset{\scriptscriptstyle{(3)}}{\Phi}_{abc}=-\sum_{pq}\Upsilon_{ap}\Upsilon_{bq}
\Bavg{u^pu^q\, \fbb_c}_{{\displaystyle\rhotrial_{\bRcal,\bPhi}}}
\label{eq:app_SCHA3_mat_fin} \\
&\overset{\scriptscriptstyle{(4)}}{\Phi}_{abcd}=-\sum_{pqr}\Upsilon_{ap}\Upsilon_{bq}\Upsilon_{cr}\,
\Bavg{u^pu^qu^r\,\fbb_d}_{{\displaystyle\rhotrial_{\bRcal,\bPhi}}}\mathv
\label{eq:app_SCHA4_mat_fin} 
\end{align}\end{subequations}
\allowdisplaybreaks
where
\begin{equation}
\fbb_i=\f_i-\biggl[\Bavg{\f_i}_{{\displaystyle\rhotrial_{\bRcal,\bPhi}}}-\sum_j\Phi_{ij}\,u^j\biggr]\mathp
\label{eq:app_def_fbb_used_finale}
\end{equation}
\section{perturbative limit of SCHA}
\label{app:PertLimSCHA}
In this section we explicitly calculate the lowest perturbative order of the SCHA results.
The formalism used is the one introduced in Sec.~\ref{sec:Perturbative_limit}.
We treat the anharmonic potential as a (small) perturbation of the harmonic Hamiltonian $H^{\ssz}$,
with $\phi^{\ssn}_{\mu_1\ldots\mu_n}$ of order $\Ocal(\lambda^{n-2})$, $\lambda \ll 1$ being the 
adimensional perturbative expansion parameter (which can be estimated as the ratio between the mean harmonic displacement
and the nearest-neighbor atomic distance~\cite{PhysRev.128.2589}).

We consider the SCHA equilibrium position $\bRcal_{\eq}$, Eq.~\eqref{eq:defin_Req}, and the corresponding 
SCHA square matrix $\bPhi=\bPhi(\bRcal_{\eq})$, Eq~\eqref{eq:SCHA_matrix_def}. 
From Eq.~\eqref{eq:first_derivative_F_in_eq} we have
\begin{align}
0&=\avg{\frac{\partial V}{\partial R^{a}}}_{\rhotrial_{\bRcal_{\eq},\bPhi}}\\
\fcmtrial_{ab}&=\avg{\frac{\partial V}{\partial R^a \partial R^b}}_{\rhotrial_{\bRcal_{\eq},\bPhi}}\mathp
\end{align}
Using the explicit expression Eq.~\eqref{eq:Ham_harm+anaharm} and Eq.~\eqref{eq:Ham_harm} we obtain
\allowdisplaybreaks[0]
\begin{align}
0&=\sum_{c_1}\fcm_{ac_1}\dR^{c_1}+\frac{1}{2}\sum_{c_1c_2}\fct_{ac_1\!c_2}\left(\dR^{c_1}\dR^{c_2}+\varPsi^{c_1\!c_2}\right)
\phantom{\avg{\frac{\partial V}{\partial R^{a}}}_{\rhotrial_{\bRcal_{\eq},\bPhi}}}\nonumber\\
&\quad+\frac{1}{3!}\sum_{c_1c_2c_3}\fcq_{ac_1\!c_2c_3}\left(\dR^{c_1}\dR^{c_2}\dR^{c_3}+3\dR^{c_1}\varPsi^{c_2c_3}\right)
\phantom{\avg{\frac{\partial V}{\partial R^{a}}}_{\rhotrial_{\bRcal_{\eq},\bPhi^{\eq}}}}\nonumber\\
&\qquad+\ldots\phantom{\avg{\frac{\partial V}{\partial R^{a}}}_{\rhotrial_{\bRcal_{\eq},\bPhi}}}
\end{align}
\allowdisplaybreaks
and
\begin{align}
\fcmtrial_{ab}&=\fcm_{ab}+\sum_{c_1}\fct_{abc_1}\dR^{c_1}\phantom{\avg{\frac{\partial V}{\partial R^a \partial R^b}}}\nonumber\\
&\qquad+\frac{1}{2}\sum_{c_1c_2}\fcq_{abc_1\!c_2}\left(\dR^{c_1}\dR^{c_2}+\varPsi^{c_1\!c_2}\right)\nonumber\\
&\qquad\qquad+\ldots\phantom{\avg{\frac{\partial V}{\partial R^a \partial R^b}}}\mathv
\end{align}
where $\delta R^a=\Rcal^a_{\eq}-R^a_{\ssz}=\mathcal{O}(\lambda)$ and 
$\varPsi^{ab}$ is the matrix obtained from $\Phi_{ab}$ according to the definitions~\eqref{eq:def_matrici_utili}.
At the lowest order
\allowdisplaybreaks[0]
\begin{align}
0&=\sum_{c_1}\fcm_{ac_1}\dR^{c_1}\nonumber\\
&\qquad+\frac{1}{2}\sum_{c_1c_2}\fct_{ac_1\!c_2}\psi^{c_1\!c_2}+\mathcal{O}(\lambda^3)
\label{app:eq_dim_pert_1}
\end{align}
and
\begin{align}
\Phi_{ab}&=\phi_{ab}+\sum_{c_1}\fct_{abc_1}\dR^{c_1}\nonumber\\
&\qquad+\frac{1}{2}\sum_{c_1c_2}\fcq_{abc_1\!c_2}\psi^{c_1\!c_2}+\Ocal(\lambda^3)\mathv
\label{app:eq_dim_pert_2}
\end{align}
where $\psi^{ab}$ is the matrix related to $\phi_{ab}$ 
in the same way as $\varPsi^{ab}$ is related to $\Phi_{ab}$ according to the definitions~\eqref{eq:def_matrici_utili}.
\allowdisplaybreaks
Inverting Eq.~\eqref{app:eq_dim_pert_1} we obtain
\begin{equation}
\dR^{a}=-\frac{1}{2}\sum_{c_1c_2c_3}\fcmi{}^{ac_1}\fct_{c_1\!c_2c_3}\psi^{c_2c_3}+\mathcal{O}(\lambda^3)\mathv
\label{app:eq_dim_pert_3}
\end{equation}
where $\fcmi{}^{ab}$ is the inverse matrix of $\fcm_{ab}$.
By substituting Eq.~\eqref{app:eq_dim_pert_3} into Eq.~\eqref{app:eq_dim_pert_2}, we obtain
\begin{align}
\Phi_{ab}&=\phi_{ab}+\frac{1}{2}\sum_{c_1c_2}\fcq_{abc_1\!c_2} \psi^{c_1\!c_2}\nonumber\\
&\quad-\frac{1}{2}\sum_{c_1c_2c_3c_4}\fct_{abc_1}\fcmi{}^{c_1\!c_2}\fct_{c_2c_3c_4}\psi^{c_3c_4}\phantom{\frac{1}{2}\sum_{c_1c_2c_3c_4}}\nonumber\\
&\quad\quad+\Ocal(\lambda^3)\phantom{frac{1}{2}\sum_{c_1c_2c_3c_4}}\phantom{\frac{1}{2}\sum_{c_1c_2c_3c_4}}\mathp
\label{eq:pert_SCHA_2}
\end{align}
Denoting with $\omega^2_{\mu}$ and $e^a_{\mu}$ eigenvalues and eigenvectors of $D^{\ssz}_{ab}=\phi_{ab}/\sqrt{M_aM_b}$, respectively,
with standard techniques for Matsubara frequency summation we obtain~\cite{mahan2000many}
\begin{equation}
-\frac{1}{\beta}\sum_lG_{\ssz}^{ab}(i\Omega_l)=\frac{\hbar}{2}\sum_{\mu}\frac{1+2n_{\mu}}{\omega_{\mu}}\,e^a_{\mu}e^b_{\mu}=\sqrt{M_aM_b}\psi^{ab}\mathv
\label{eq:sumMatz_1}
\end{equation}
where $G_{\ssz}^{ab}(z)$ is the harmonic Green function for the variable $\sqrt{M_a}(R^a-\Rcal^a_{\ssz})$
(see Eq.~\eqref{eq:relaz_G0_D0}).
Therefore, dividing Eq.~\eqref{eq:pert_SCHA_2} by the square root of masses 
and considering the definition of $D_{ab}^{\ssS}$, Eq.~\eqref{eq:def_D}, and Eqs.\eqref{eq:pert_loop}--\eqref{eq:pert_bubble}
we obtain 
\begin{equation}
D^{\ssS}_{ab}=D^{\ssz}_{ab}+\overset{\ssT}{\Pi}{}^{\ssz}_{ab}+\overset{\ssL}{\Pi}{}^{\ssz}_{ab}+\Ocal(\lambda^3)\mathp
\label{eq:pert_DS}
\end{equation}
Moreover from Eq.~\eqref{eq:SCHA_matrix_def_nth} we readily have
\begin{equation}
\overset{\sst}{\Phi}_{abc}=\overset{\sst}{\phi}_{abc}+\Ocal(\lambda^3)
\label{eq:pert_SCHA_3}
\end{equation}
and
\begin{equation}
\overset{\ssf}{\Phi}_{abcd}=\overset{\ssf}{\phi}_{abcd}+\Ocal(\lambda^3)\mathp
\label{eq:pert_SCHA_4}
\end{equation}
Thus, from Eq.~\eqref{eq:diagrammatic_static_selfenergy}, Eq.~\eqref{eq:diagrammatic_static_selfenergy_bubble}, and 
Eq.~\eqref{eq:pert_bubble}, we have
\begin{equation}
\overset{}{\Pi}{}^{\ssS}_{ab}(0)=\overset{\ssB}{\Pi}{}^{\ssS}_{ab}(0)+\Ocal(\lambda^3)
=\overset{\ssB}{\Pi}{}^{\ssz}_{ab}(0)+\Ocal(\lambda^3)\mathp
\label{eq:app_per_bubb}
\end{equation}
Therefore, from Eq.~\eqref{eq:diagrammatic_static_invSCHAdynmat_00} and 
Eq.~\eqref{eq:pert_DS}
we obtain
\begin{equation}
D^{\ssF}_{ab}=D^{\ssz}_{ab}+\overset{\ssT}{\Pi}{}^{\ssz}_{ab}+\overset{\ssL}{\Pi}{}^{\ssz}_{ab}+\overset{\ssB}{\Pi}{}^{\ssz}_{ab}(0)+\Ocal(\lambda^3)\mathp
\end{equation}
\section{Toy model definition}
\label{app:Toy_model_definition}
In this section we define the toy model used in our numerical tests, which we
rewrite here as:
\begin{equation}
V(\bR)=\frac{1}{2}\sum_{ab}\phi_{ab}\,u^a u^b+V_A^{(3)}(\bu)+V_A^{(4)}(\bu)\mathv
\end{equation}
where $\bu=\bR-\bR_{\ssz}$, $\bR_{\ssz}$ being the equilibrium configuration of the rock-salt structure.
The harmonic matrix $\phi_{ab}$ has been obtained with ab initio calculations for SnTe, on a 2x2x2 mesh of the BZ,
performed within Density Functional Perturbation Theory (DFPT)~\cite{RevModPhys.73.515} as implemented in Quantum ESPRESSO~\cite{QE-2009}.
In order to have increased harmonic instability, calculations have been performed with lattice parameter $a_{\scriptscriptstyle{\text{toy}}}=6.562\,\angs$, 
which is higher than the experimental value $a_{\scriptscriptstyle{\text{exp}}}=6.312\,\angs$. We used Perdew-Burke-Ernzerhof (PBE)~\cite{PhysRevLett.77.3865}, projector augmented wave (PAW)~\cite{PhysRevB.50.17953} potentials. We calculated $40$ Khon-Sham states, with a cutoff of 28 Ry and 280 Ry for the wave functions
and the charge density, respectively. The BZ integration has been performed with a Monkhorst-Pack grid~\cite{PhysRevB.13.5188} of 
16x16x16 $\bk$ points. 
The self-consistent solution of the Kohn-Sham equations was obtained when the total energy changed by less than $5\times 10^{-12}$ Ry.

In order to describe the anharmonic terms we follow the model described in Ref.~\citenum{PhysRevB.90.014308,PhysRevLett.113.105501}
and we define short-range anharmonic terms by using reciprocal displacements of nearest-neighbor atoms 
(in the rock-salt structure each atom has 6 nearest-neighbor). The third and fourth order terms are given by
\begin{equation}
V^{\sst}(\bu)=p_3\sum_{s=1}^{\Nat}\sum_{\alpha=x,y,z}\Bigl[\Acal_{s,\alpha_{+}}^3-\Acal_{s,\alpha_{-}}^3\Bigr]
\end{equation}
and
\begin{align}
V^{\ssf}(\bu)=&p_4\sum_{s=1}^{\Nat}\sum_{\alpha=x,y,z}\Bigl[\Acal_{s,\alpha_{+}}^4+\Acal_{s,\alpha_{-}}^4\Bigr]\nonumber\\
+&p_{4_{\chi}}\sum_{s=1}^{\Nat}\sum_{\alpha=x,y,z}
\Bigl[\Acal_{s,\alpha_{+}}^2
  \Bigl((\Ecal^{\scriptscriptstyle{(1)}}_{s,\alpha_{+}})^2+
   (\Ecal^{\scriptscriptstyle{(2)}}_{s,\alpha_{+}})^2\Bigr)\nonumber\\
&\mkern+48mu+\Acal_{s,\alpha_{-}}^2
 \Bigl((\Ecal^{\scriptscriptstyle{(1)}}_{s,\alpha_{-}})^2+
   (\Ecal^{\scriptscriptstyle{(2)}}_{s,\alpha_{-}})^2\Bigr)\Bigr]
\end{align}
with, for example,
\begin{subequations}\begin{align}
\Acal_{s,x_{\pm}}&=\frac{1}{\sqrt{2}}\left(u^{x_{\scriptscriptstyle{\pm}}\!(s),x}- u^{s,x}\right)\\
\Ecal_{s,x_{\pm}}^{\scriptscriptstyle{(1)}}&=\frac{1}{\sqrt{2}}\left(u^{x_{\scriptscriptstyle{\pm}}\!(s),y}- u^{s,y}\right)\\
\Ecal_{s,x_{\pm}}^{\scriptscriptstyle{(2)}}&=\frac{1}{\sqrt{2}}\left(u^{x_{\scriptscriptstyle{\pm}}\!(s),z}- u^{s,z}\right)\mathv
\end{align}\end{subequations}
where $x_{\scriptscriptstyle{+}}(s)$ and $x_{\scriptscriptstyle{-}}(s)$ are
the nearest-neighbor of the atom $s$, along the cartesian direction $+x$ and
$-x$, respectively. Similar notation is used for the directions $\pm y$ and $\pm z$.
According to this definition, the third-order is proportional to the parameter $p_3$  
and the fourth-order is linear function of the parameters $p_4$ and $p_{4\chi}$. 
In order to set reasonable values, $p_4$ and $p_{4\chi}$ have been fixed
by fitting the energy curve obtained ab initio for displacements with
unit cell periodicity (the third order does not affect the value of the
potential for configurations having unit cell periodicity):
$p_4=7.63\,\text{eV}/\angs^4$ and $p_{4\chi}=4.86\,\text{eV}/\angs^4$.
For the third order we set $p_3=6.70\,\text{eV}/\angs^3$
a value larger than the one reported in Ref.~\citenum{PhysRevB.90.014308} for PbTe. 
This has been done to magnify the effect of the third order in the 2x2x2 supercell 
used for the SSCHA calculation. 
\bibliography{bibliography.bib}
\end{document}